\documentclass[journal]{IEEEtran}
\ifCLASSINFOpdf
\else
\fi

\usepackage{color}
\usepackage{amsmath,amssymb, amsthm,bm}
\usepackage{cite}
\usepackage{multirow}
\usepackage{url}
\usepackage{graphicx}

\usepackage{slashbox}
\usepackage{epstopdf}
\usepackage{algorithm}
\usepackage{algorithmic}
\usepackage{subcaption}
\usepackage{cite}

\newcommand*\xbar[1]{%
  \hbox{%
    \vbox{%
      \hrule height 0.5pt 
      \kern0.3ex
      \hbox{%
        \kern-0.05em
        \ensuremath{#1}%
        \kern-0.05em
      }%
    }%
  }%
}
\newcommand{\argmax}{\mathop{\rm argmax}\limits}
\newcommand{\argmin}{\mathop{\rm argmin}\limits}
\newtheorem{Ex}{Example}
\theoremstyle{definition}
\theoremstyle{Rem}
\newtheorem{Rem}{Remark}
\newtheorem{definition}{Definition}
\newtheorem{theorem}{Theorem}
\newtheorem{prop}{Proposition}
\newtheorem{corol}{Corollary}

\def\thline{\noalign{\hrule height 1pt}}

\def\thline{\noalign{\hrule height 1pt}}

\def\l({\left(}
\def\r){\right)}
\begin{document}
%
\title{Epigraphical Relaxation for Minimizing \\Layered Mixed Norms}

\author{Seisuke Kyochi,
        Shunsuke Ono,~\IEEEmembership{Member,~IEEE}, and
        Ivan Selesnick,~\IEEEmembership{Fellow,~IEEE}
\thanks{S. Kyochi is with the Department of Information and Systems Engineering, The University of Kitakyushu, Fukuoka 808-0135, Japan (e-mail: s-kyochi@kitakyu-u.ac.jp).}
\thanks{S. Ono is with the Department of Communications and
Computer Engineering, Tokyo Institute of Technology, Tokyo 152-8550, Japan.}
\thanks{I. Selesnick with the Department of Electrical and Computer Engineering, Tandon School of Engineering, New York University, Brooklyn, NY 10003 USA}
\thanks{Manuscript received 2020.}
}

\markboth{}%
{Kyochi: Epigraphical Relaxation for Minimizing Layered Mixed Norms}
\maketitle

\begin{abstract}
This paper proposes an epigraphical relaxation (ERx) technique for non-proximable mixed norm minimization. Mixed norm regularization methods play a central role in signal reconstruction and processing, where their optimization relies on the fact that the proximity operators of the mixed norms can be computed efficiently. To bring out the power of regularization, sophisticated layered modeling of mixed norms that can capture inherent signal structure is a key ingredient, but the proximity operator of such a mixed norm is often unavailable (\textit{non-proximable}). Our ERx decouples a layered non-proximable mixed norm into a norm and multiple epigraphical constraints. This enables us to handle a wide range of non-proximable mixed norms in optimization, as long as both the proximal operator of the outermost norm and the projection onto each epigraphical constraint are efficiently computable. Moreover, under mild conditions, we prove that ERx does not change the minimizer of the original problem despite relaxing equality constraints into inequality ones. We also develop new regularizers based on ERx: one is decorrelated structure-tensor total variation for color image restoration, and the other is amplitude-spectrum nuclear norm for low-rank amplitude recovery. We examine the power of these regularizers through experiments, which illustrates the utility of ERx.
\end{abstract}


\IEEEpeerreviewmaketitle

\section{Introduction}
\label{sec:intro}
\IEEEPARstart{C}{onvex} optimization with sparsity and low-rankness promoting norms, needless to say, has been a fundamental tool for many tasks in modern signal processing and machine learning\footnote{Recently, deep neural networks have become state-of-the-arts in many applications, including image denoising and super-resolution. However, optimization-based methods and regularizations are still important in inverse problems because they can explicitly take models into account, do not require training data set, and are robust to noise level change.}, e.g., signal/data recovery, regression, classification, and so on \cite{Theodoridis2015, Sra2011, Bach2012, Combettes2008,Goldstein2009,Pustelnik2011,Afonso2011}. Although such norms, for instance, the $\ell_1$-norm and the nuclear norm, often fall into a class of non-smooth functions, recent advances of proximal methods in convex optimization \cite{Bauschke2011,Beck2017} enable us to handle a wide range of minimization problems including non-smooth norms. 

To accurately recover or estimate signals, we should carefully design a suitable convex regularizer that models the desired properties of target signals. Some conventional regularizers are characterized by composite norm functions, called mixed norms. Well-known examples are a class of total variation regularizers (TV) modeling piecewise-smoothness of images\cite{Rudin1992,Bresson2008,Chan2010,Bayram2012,Ono2014DVTV} and its extensions \cite{Bredies2010, Lefkimmiatis2015,Chierchia2014,Lefkimmiatis2015a}. 

Most of the conventional mixed norms are characterized as \textit{2-layered composite functions}. For example, TV is formulated by using the $\ell_{2,1}$-norm \cite{Kowalski2009} in which the $\ell_2$-norm-based vector-valued function is used as the first layer function (the inner function) and the $\ell_1$-norm as the second layer function (the outer function). Such 2-layered mixed norms successes tempt us to model various aspects of the group sparsity or low-rankness of signals by introducing an involved mixed norm with more layer functions (we refer to it as \textit{a layered mixed norm}). In fact, Kowalski \textit{et al.} successfully applied a 3-layered function ($\ell_{2,1,2}$-norm) \cite{gramfort2009, Kowalski2013} to, for example, Magnetoencephalography inverse problems \cite{gramfort2009}. However, exploring the efficient computation of the proximity operator of other layered mixed norms is, in general, a very challenging task, and this difficulty motivates us to develop and use \textit{non-proximable} mixed norms\footnote{In this paper, we refer to a \textit{non-proximable} function as a function that does not have a closed-form proximity operator.} forming a multiple-layered composite function in regularization.

To circumvent this dilemma, we propose a new methodology named epigraphical relaxation (ERx). ERx decouples a layered mixed norm function into an outermost norm and multiple epigraphical constraints via the relaxation of equality constraint into an inequality one. By this manipulation, we can tackle the original minimization problem involving a layered non-proximable mixed norm, as long as both the proximity operator for the outermost norm and the projection onto each epigraphical constraint (called \textit{epigraphical projection}) are available. Since ERx changes the original problem, one may think that the minimizers to the relaxed problem are not guaranteed to be identical to the ones of the original problem. Fortunately, however, we can prove that, for certain types of mixed norms, the minimizers to the problem after ERx are the same as the original ones.

\subsection{Related Work}
Epigraphical techniques have already been successfully applied to handling involved norm ball constraints, e.g., multi-class SVM \cite{Chierchia2014SVM}, randomized data-fidelity constraints \cite{Ono2019}, (non-local) TV, total generalized variation (TGV), and (non-local) structure-tensor TV (STV) constraints \cite{Chierchia2015, Ono2014, Chierchia2013}. In these methods, a mixed norm constraint is decoupled into an epigraphical constraint and a half-space constraint (this technique is termed as \textit{epigraphical splitting}). Inspired by the epigraphical splitting, our ERx converts a non-proximable mixed norm into a proximable norm and epigraphical constraints, where their epigraphical projections can be computed in closed-form. What distinguishes ERx from these existing methods is that ERx targets non-proximable norms in objective function while they focus on non-projectable constraints. 

Some conventional convex optimization methods use ways similar to the proposed method \cite{Grant2006,Lobo1998}. For example, in Sec. 6.4 in \cite{Grant2006}, a cost 2-layered composite function is converted into the outer function with an inequality constraint. In \cite{Lobo1998}, the quadratic cost function is converted to a variable with a \textit{second-order cone constraint}. The proposed method differs in two points. First, ERx extensively generalizes conventional approaches: \textit{various types of layered} composite functions with an arbitrary number of layers (each layer function can be a nonsmooth function, e.g., $\ell_1$-norm, the nuclear norm) can be converted into the outer function with epigraphical constraints. Second, the conventional methods do not rely on proximal approaches, but traditional convex programming, as employed in off-the-shelf modeling software called, e.g., CVX, while the proposed method uses a proximal algorithm, i.e., the PDS algorithm \cite{Chambolle2010,Condat2013,Vu2013,Komodakis2015,Bot2015}. In modern convex optimization for signal processing and machine learning, proximal algorithms have become standard tools taking the place of non-proximal approaches due to the following reasons \cite{Komodakis2015}:
\begin{enumerate}
\item proximal algorithms can find an optimal solution more efficiently (e.g., time to converge) even though the dimensionality is large (as long as proximity operators are efficiently computed). On the other hand, non-proximal algorithms cannot handle large-scale optimization problems. Since ERx expands the dimension of a variable, it is difficult for non-proximal algorithms to handle deeply-layered composite functions efficiently.
\item various types of nonsmooth norms (e.g., the nuclear norm) can be supported in proximal algorithms.
\end{enumerate}
Moreover, among conventional proximal algorithms, the PDS algorithm \cite{Chambolle2010,Condat2013,Vu2013,Komodakis2015,Bot2015} has been attractive recently. Since it can circumvent the inverse matrix calculation required in other proximal algorithms such as ADMM (thus it provides comparable or even less time to converge \cite{Komodakis2015}), a wide range of large-scale optimization problems in signal processing can be handled. Consequently, the PDS-based algorithm proposed in this work is a promising way to solve layered mixed norm minimization.
\subsection{Contributions}
The contributions of this paper are in the following. 
\begin{enumerate}
\item We introduce a generic computational procedure of ERx for layered mixed norm minimization problems. The procedure decouples a layered mixed norm into the outermost norm and the multiple epigraphical constraints. Then, the relaxed problem is solved by the PDS algorithm \cite{Chambolle2010,Condat2013,Vu2013,Komodakis2015,Bot2015}. ERx with the PDS algorithm can handle a very wide range of layered mixed norm minimization problems as long as the closed-form proximity operators for the outermost norm and the multiple epigraphical constraints are available.
\item We thoroughly investigate sufficient conditions under which ERx keeps the minimizer of the original problem. We start by discussing a multi-layered composite function (not necessarily norm) minimization problem. As will be clarified, a key observation is that, whereas the innermost function requires the convexity, the other functions should satisfy not only the convexity but also a strictly increasing property (see Definition \ref{def:sifunc}). Then we analyze whether the $\ell_p$-type vector/matrix norms ($p\in [1,\infty]$) and the Schatten-$p$ norms ($p=1,2,\infty$) satisfy the strictly increasing property or not.
\item We construct two effective regularizers as practical usages of ERx. The first one is termed as decorrelated structure-tensor TV (DSTV) that is STV defined on a luminance-chrominance (luma-chroma) color space, whereas the original STV \cite{Lefkimmiatis2015,Chierchia2014,Lefkimmiatis2015a} is defined on the RGB space, for effective image restoration. We also present a closed-form computation of the epigraphical projection of the $\ell_1$-norm for solving optimization problems with DSTV.

The second regularizer is named as amplitude spectrum nuclear norm (ASNN) for promoting the low-rankness of the amplitude spectra of signals. ASNN is integrated into the robust PCA framework (RPCA) \cite{Candes2011} that extracts principal components from a data matrix corrupted by outliers. In general, a low-rank signal and its shifted version have an identical amplitude spectrum, meaning that both have the same low-rank amplitude spectra in the frequency domain. However, the standard nuclear norm cannot capture this inherent structure when the signal is shifted because it is performed in the signal domain. By replacing the nuclear norm by the ASNN in RPCA, which is termed as frequency-domain robust ($\mathcal{F}$-RPCA), we can robustly extract principal components from a data matrix even in the presence of misalignment (and outliers).
\end{enumerate} 

The preliminary work of this paper is presented in \cite{Kyochi2020}, where we only discuss the sufficient condition to keep the minimizer by ERx in the case of the $\ell_p$-vector norms ($p\in [1,\infty)$). In contrast, this paper treats a much wider class of functions (i.e., strictly increasing vector/matrix functions) and thoroughly investigates a variety of types of the $\ell_p$-vector/matrix norms ($p\in [1,\infty]$) and whether or not the Schatten-$p$ norms ($p=1,2,\infty$) keep the minimizer. Moreover, this paper newly presents the ASNN.

The rest of this paper is organized as follows. Sec. \ref{sec:Prelim} reviews basic tools for convex optimization and the PDS algorithm. Then, ERx for involved layered mixed norms is explained in Sec. \ref{sec:Proposed}. Sec. \ref{sec:AppER} introduces the proposed regularizers, i.e., the DSTV and ASNN regularization functions, and the algorithms for minimizing those layered mixed norms. The proposed regularizers are evaluated in the experiments of compressed image sensing and signal decomposition in Sec. \ref{sec:Experimental}. Finally, this paper is concluded in Sec. \ref{sec:Conc}.   
\subsection{Notations} \label{subsec:notation}
Bold-faced lower-case letters and upper-case letters are vectors and matrices, respectively. The other mathematical notations are summarized in Table \ref{tab:Notations}. 
\begin{table}[t]
\caption{Basic notations}
\vspace{-0.2cm}
\label{tab:Notations}
\begin{center}
\scalebox{0.75}{
\begin{tabular}{c|c}
\thline
Notation & Terminology\\ \thline
$\mathbb{N}$, $\mathbb{R}$, and $\mathbb{R}_+$& Natural, real, nonnegative real numbers\\\hline 
$\mathbb{C}$, $\mathrm{j}$ & Complex numbers and complex unit $\sqrt{-1}$ \\\hline
\begin{tabular}{c} $(a,b),\ (a,b],\ [a,b),\ [a,b]\ (\subset \mathbb{R})$ \\ ($-\infty \leq a<b \leq \infty$) \end{tabular}& Open, half-closed, and closed intervals\\\hline
$A^N$ and $A^{M\times N}$ ($A \subset \mathbb{C}$)& \begin{tabular}{c} $N$- and $M\times N$-dimensional \\ vectors/matrices with elements in $A$
 \end{tabular}\\\hline
$\mathbf{I}$, $\mathbf{J}$, $\mathbf{O}$& Identity, reversal identity, and zero matrix\\\hline
$\mathbf{X}^{\top}$& Transpose of $\mathbf{X}$\\\hline
$x_n$ and $[\mathbf{x}]_n$& $n$-th element of a vector $\mathbf{x}$\\\hline
$X_{m,n}$ and $[\mathbf{X}]_{m,n}$& $(m,n)$-th element of a matrix $\mathbf{X}$\\\hline
$\mathbf{x}_{n} \in \mathbb{R}^{N_0}$ &\begin{tabular}{c}$n$-th subvector of $\mathbf{x} \in \mathbb{R}^{LN_0}$\  ($1\leq n \leq L$)\end{tabular}\\\hline
$\mathbf{X}_{m,n} \in \mathbb{R}^{M_0 \times N_0}$ &\begin{tabular}{c}$(m,n)$-th subblock of $\mathbf{X} \in \mathbb{R}^{L_vM_0 \times L_hN_0}$\\ ($1\leq m \leq L_v$, $1\leq n \leq L_h$)\end{tabular}\\\hline
\begin{tabular}{c} 
$|\mathbf{x}| \in \mathbb{R}^N$, $|\mathbf{X}| \in \mathbb{R}^{M\times N}$ \\
($\mathbf{x} \in \mathbb{C}^N$, $\mathbf{X} \in \mathbb{C}^{M\times N}$)
\end{tabular} 
 & \begin{tabular}{c} Element-wise absolute value, \\ $[|\mathbf{x}|]_n = |x_n|$, $[|\mathbf{X}|]_{m,n} = |X_{m.n}|$\end{tabular}\\\hline
$\mathrm{vec}(\mathbf{X}) \in \mathbb{R}^{M N}$ & \begin{tabular}{c} Vectorization of $\mathbf{X}\in \mathbb{R}^{M \times N}$,\\ $x_{M n + m} = X_{m,n}$\end{tabular} \\\hline
\begin{tabular}{c}$\mathrm{bvec}_{(M_b,N_b)}(\mathbf{X})$\\
$\in \mathbb{R}^{L_v M_b L_h N_b}$\end{tabular} & \begin{tabular}{c}
$M_b\times N_b$-block-wise vectorization of \\$\mathbf{X}\in \mathbb{R}^{L_v M_b \times L_h N_b}$,\\$[ \mathrm{vec}(\mathbf{X}_{1,1})^{\top}\  \ldots\ \mathrm{vec}(\mathbf{X}_{L_v,L_h})^{\top} ]^{\top}$
\end{tabular}
\\ \hline
$\mathrm{mat}(\mathbf{X}),\ \mathrm{bmat}(\mathbf{X})  \in \mathbb{R}^{M\times N}$ & \begin{tabular}{c}
Reverse operation of \\$\mathrm{vec}(\cdot)$ and $\mathrm{bvec}_{(M_b,N_b)}(\cdot)$,\\
$\mathbf{X} = \mathrm{mat}(\mathrm{vec}(\mathbf{X}))$,\\ $\mathbf{X} = \mathrm{bmat}_{(M_b,N_b)}(\mathrm{bvec}_{(M_b,N_b)}(\mathbf{X}))$
\end{tabular}
\\ \hline
 \begin{tabular}{c}$\mathrm{diag}(a_1, \ldots , a_{N})$, \\ 
 $\mathrm{diag}( \mathbf{A}^{(\mathrm{1})}, \ldots , \mathbf{A}^{({N})})$
 \end{tabular}& Diagonal/block-diagonal matrices.\\\hline
$\|\mathbf{x}\|_{p}$ ($p \in [1 , \infty )$)  & \begin{tabular}{c} $\ell_p$-norm, $\|\mathbf{x}\|_{p}=\left(\sum^{N}_{n=1} |x_n|^p\right)^{\tfrac{1}{p}}$
 \end{tabular}\\\hline
 $\|\mathbf{x}\|_{\infty }$  & \begin{tabular}{c} $\ell_\infty$-norm,  $\|\mathbf{x}\|_{\infty}=\max\{ |x_n|\}$
 \end{tabular}\\\hline
 $\mathcal{B}_p(\mathbf{y},\epsilon)$ $p\in [1,\infty]$  & \begin{tabular}{c} $\mathbf{y}$-centered $\ell_p$-norm ball with the radius of $\epsilon$,\\$\mathcal{B}_p(\mathbf{y},\epsilon) = \{ \mathbf{x} \in \mathbb{R}^N \  |\  \|\mathbf{x} - \mathbf{y}\|_p \leq \epsilon\}$
 \end{tabular}\\\hline
 \begin{tabular}{c} $\|\mathbf{x}\|_{p,q}$ ($p, q \in [1 , \infty)$)  \end{tabular} &  \begin{tabular}{c} $\ell_{p,q}$-mixed norm, \\$\|\mathbf{x}\|_{p,q} = \left(\sum^{N}_{n=1} \|\mathbf{x}_{n}\|_p^q\right)^{\tfrac{1}{q}}$ \end{tabular}
\\\hline
$\|\mathbf{X}\|_{p}$ \ \ ($p \in [1 , \infty )$) & \begin{tabular}{c} $\ell_p$-matrix norm, \\ $\|\mathbf{X}\|_{p}=\left(\sum^{M}_{m=1}\sum^{N}_{n=1} |X_{m,n}|^p\right)^{\tfrac{1}{p}}$
 \end{tabular}\\\hline
\begin{tabular}{c} $\|\mathbf{X}\|_{p,q}$ \ \ ($p ,q \in [1 , \infty )$) \end{tabular}  &  \begin{tabular}{c} $\ell_{p,q}$-mixed matrix norm, \\$\|\mathbf{X}\|_{p,q}=\left(\sum^{M}_{m=1}\sum^{N}_{n=1} \|\mathbf{X}_{m,n}\|_p^q\right)^{\tfrac{1}{q}}$ \end{tabular}
\\\hline
  \begin{tabular}{c} $\|\mathbf{X}\|_{S_p}$  \ \ ($p \in [1 , \infty )$) \end{tabular} & \begin{tabular}{c} Schatten $p$-matrix norm, \\$\|\mathbf{X}\|_{S_p}  =\left( \sum_{r=1}^{R} \sigma_r(\mathbf{X})^p \right)^{\tfrac{1}{p}}$ \\
  ($\sigma_r(\mathbf{X}) $ : $r$-th singular value of $\mathbf{X}$)
\end{tabular} \\\hline
  $\|\mathbf{X}\|_{\ast}$  & \begin{tabular}{c} Nuclear norm (Schatten $1$-matrix norm), \\ $\|\mathbf{X}\|_{\ast} = \|\mathbf{X}\|_{S_1}$ 
\end{tabular} \\\hline
  $\|\mathbf{X}\|_{\mathrm{op}}$  & \begin{tabular}{c} Induced norm (Schatten $\infty$-matrix norm), \\ $\|\mathbf{X}\|_{\mathrm{op}} = \|\mathbf{X}\|_{S_\infty} = \max_{\|\mathbf{v}\|_2 \leq 1} \|\mathbf{X}\mathbf{v}\|_2$ 
\end{tabular} \\\hline
$\mathbf{x} \leq \mathbf{y}$ ($\mathbf{x}$, $\mathbf{y}$ $\in \mathbb{R}^{N}$) & $x_n \leq y_n\  (1 \leq {}^\forall n \leq N)$.
\\\hline
 $f^{(k_1,k_2)}$ $(k_1 < k_2)$ & \begin{tabular}{c} $(k_2-k_1+1)$-layered \\ composite function of $\{f^{(k)}\}_{k=k_1}^{k_2}$, \\ $f^{(k_1,k_2)} := f^{(k_2)} \circ f^{(k_2-1)} \circ \cdots \circ f^{(k_1)}$ \\ $f^{(k)}$: $k$-th layer function. \end{tabular} \\\hline
\end{tabular}
}
\end{center}
\vspace{-0.4cm}
\end{table}
\section{Preliminaries}
\label{sec:Prelim}
\subsection{Convex Scalar/Vector-valued Function}
A function $f:\mathbb{R}^{N} \rightarrow \mathbb{R}$ is said to be a convex (scalar-valued) function \cite{Boyd2004} if and only if, for $^\forall \mathbf{x},\ \mathbf{y} \in \mathbb{R}^{N}$ and $^\forall \alpha \in [0,1]$, 
\begin{align}\label{eq:conv_vecconv}
f(\alpha \mathbf{x} + (1-\alpha)\mathbf{y}) \leq \alpha f(\mathbf{x}) + (1-\alpha)f(\mathbf{y}).
\end{align}
In addition, a vector-valued function (vector function, for short) $f:\mathbb{R}^{N} \rightarrow \mathbb{R}^M$ is said to be a convex vector function if and only if the equation \eqref{eq:conv_vecconv} holds for $^\forall \mathbf{x},\ \mathbf{y} \in \mathbb{R}^{N}$ and $^\forall \alpha \in [0,1]$  \cite{Dattorro2005}, where the definition for vector inequality is defined in Table \ref{tab:Notations}.

As a special class of vector functions, for a vector with non-overlapped $K$ blocks $\mathbf{x} = \begin{bmatrix}
\mathbf{x}_1^\top & \ldots & \mathbf{x}_{K}^\top \end{bmatrix}^\top \in \mathbb{R}^{N}$ ($\mathbf{x}_k \in \mathbb{R}^{N_k}$, $1 \leq k \leq K$, $N = \sum_{k=1}^{K} N_k$) we define \textit{block-wise vector function} $f:\mathbb{R}^{N} \rightarrow \mathbb{R}^K$ with functions $f_k:\mathbb{R}^{N_k} \rightarrow \mathbb{R}$ (called block function in this paper) as 
\begin{align*}
f(\mathbf{x}) =  \begin{bmatrix}
f_1(\mathbf{x}_1) & \ldots & f_{K}(\mathbf{x}_{K})
\end{bmatrix}^\top.
\end{align*}
If a block-wise vector function satisfies convexity, we refer to it as \textit{convex block-wise vector function}. Obviously, if a block-wise vector function $f$ is convex if and only if each block function $f_k$ of $f$ is convex.
\subsection{Epigraph}\label{subsec:epi}
Epigraph of a function $f \in \Gamma_0(\mathbb{R}^N)$ ($\Gamma_0(\mathbb{R}^N)$ is the set of proper lower semicontinuous convex functions on $\mathbb{R}^N$) is a subset of the product space $\mathrm{epi}_f \subset \mathbb{R}^N \times \mathbb{R}$ defined by 
$\mathrm{epi}_f := \{ (\mathbf{x},\xi) \in \mathbb{R}^N \times \mathbb{R}\  |\  f(\mathbf{x}) \leq \xi \}$ \cite{Beck2017}. 
\subsection{Basic Tools in Convex Optimization}
This section reviews some basic tools used in this paper (for more detailed information, see \cite{Bauschke2011}).
\subsubsection{Indicator function} $\iota_C$ is the indicator function of $C$ defined by $
\iota_{C}(\mathbf{x}) := 0\ (\mathbf{x} \in C),\  \iota_{C}(\mathbf{x}) := \infty\  (\mathbf{x} \notin C)$.
\subsubsection{Proximity operator and projection onto convex set} The proximity operator, denoted as $\mathrm{prox}_{\gamma f}:\mathbb{R}^{N} \rightarrow \mathbb{R}^{N}$, is defined for a function $f \in \Gamma_0(\mathbb{R}^N)$ and an index $\gamma \in (0, \infty)$ by 
$
	\mathrm{prox}_{\gamma f}(\mathbf{x}) := \argmin_{\mathbf{y}\in \mathbb{R}^N} \gamma f(\mathbf{y}) + \frac{1}{2} \|\mathbf{x}-\mathbf{y}\|^2_2.
$

The projection onto a convex set $C \subset \mathbb{R}^N$, denoted as $\mathcal{P}_C(\mathbf{x}):\mathbb{R}^{N} \rightarrow \mathbb{R}^{N}$, is one the special cases of the proximity operator, which can be represented as
$
	\mathrm{prox}_{\gamma \iota_C}(\mathbf{x}) := \argmin_{\mathbf{y}\in \mathbb{R}^N} \gamma \iota_C(\mathbf{y}) + \frac{1}{2} \|\mathbf{x}-\mathbf{y}\|^2_2 = \argmin_{\mathbf{y} \in C} \|\mathbf{x}-\mathbf{y}\|_2^2 =: \mathcal{P}_C(\mathbf{x}).
$
Examples of proximity operators are listed in Table \ref{tab:complexity} (detailed information for examples of the projection onto epigraph, see Appendix \ref{app:epi_proj}).

\begin{table}[t]
\caption{Proximity operator and computational complexity}
\vspace{-0.2cm}
\label{tab:complexity}
\begin{center}
\scalebox{0.77}{
\begin{tabular}{c|c}
\thline
Function $f$ & \begin{tabular}{c} Proximity operator $\widetilde{\mathbf{x}} = \mathrm{prox}_{\gamma f}(\mathbf{x})$ \\ and computational complexity for \\ $N$- and $M \times N$-dimensional vectors/matrices\end{tabular}\\ \thline
$\|\cdot\|_{2}$  & \begin{tabular}{c} $\widetilde{\mathbf{x}} = \left(1 - \frac{\gamma}{\max\{\|\mathbf{x}\|_2,\gamma\}}\right)\mathbf{x}$ \cite{Beck2017}, $\mathcal{O}(N)$ \end{tabular}\\\hline
 $\iota_{\mathcal{B}_2(\mathbf{y},\epsilon)}(\cdot)$ & \begin{tabular}{c} $\widetilde{x}  = 
\begin{cases}
\mathbf{x} & \mathbf{x} \in \mathcal{B}(\mathbf{y}, \epsilon) \\
\mathbf{y} + \epsilon \frac{\mathbf{x} - \mathbf{y}}{\|\mathbf{x} - \mathbf{y} \|_2} & (\mathrm{otherwise})
\end{cases}$ \cite{Beck2017}, $\mathcal{O}(N)$.
 \end{tabular}\\\hline
 $\|\cdot\|_{1}$  & \begin{tabular}{c} Soft-thresholding $\widetilde{\mathbf{x}} = \mathcal{T}_\gamma(\mathbf{x})$ \cite{Beck2017}, $\mathcal{O}(N)$ \\ ($\widetilde{\mathbf{x}}_n = [\mathcal{T}_\gamma(\mathbf{x})]_n = \mathrm{sign}(x_n)\max\{|x_n|-\gamma, 0\}$ )\end{tabular}\\\hline
$\iota_{\mathcal{B}_1(\mathbf{y},\epsilon)}(\cdot)$ & \begin{tabular}{c} Using projection onto the simplex \cite{Duchi2008}, $\mathcal{O}(N\log N)$.
 \end{tabular}\\\hline
 $\|\cdot\|_{\infty}$  & \begin{tabular}{c} $\mathbf{x}-\mathrm{prox}_{\iota_{\mathcal{B}_1(\mathbf{0},1)}}(\mathbf{x}/\gamma)$ \cite{Beck2017}, $\mathcal{O}(N\log N)$.
 \end{tabular}\\\hline
 \begin{tabular}{c}  $\|\mathbf{x}\|_{2,1}$ \end{tabular} &  \begin{tabular}{c} Block-wise prox. of the $\ell_{2}$-norm, $\mathcal{O}(N)$ \\ $\mathbf{x} = \begin{bmatrix} \mathbf{x}_1^\top & \cdots & \mathbf{x}_{N}^\top \end{bmatrix}^\top$, $\widetilde{\mathbf{x}}_n = \mathrm{prox}_{\gamma \|\cdot\|_2} (\mathbf{x}_n)$  \end{tabular}
\\\hline
  $\|\mathbf{X}\|_{\ast}$  & \begin{tabular}{c} Soft-thresholding on singular values \\ of $\mathbf{X} \in \mathbb{R}^{M\times N}$ \cite{Beck2017}, $\mathcal{O}(MN \min\{M,N\})$\\ ${\bm \sigma} = \begin{bmatrix} \sigma_1\ \ldots \ \sigma_r \end{bmatrix}^\top$ ($\mathbf{X}=\mathbf{U}\mathrm{diag}({\bm \sigma})\mathbf{V}^\top$), \\ $\widetilde{\mathbf{X}} = \mathbf{U}\mathrm{diag}(\mathcal{T}_\gamma({\bm \sigma}))\mathbf{V}^\top$. 
\end{tabular} \\\hline
 $\iota_{[a,b]^N}(\cdot)$ & $\widetilde{x}_n = \min\{ \max\{ x_n , a \} , b \}$ \cite{Beck2017}, $\mathcal{O}(N)$.\\\hline
  $\iota_{\mathrm{epi}_{\tau\|\cdot\|_2}}(\cdot)$ & See \eqref{eq:epil2} \cite{Beck2017}, $\mathcal{O}(N)$.\\\hline
    $\iota_{\mathrm{epi}_{\|\cdot\|_\infty}}(\cdot)$ & See \eqref{eq:epiinf} \cite{Beck2017}, $\mathcal{O}(N\log N)$.\\\hline
      $\iota_{\mathrm{epi}_{\|\cdot\|_1}}(\cdot)$ & See \eqref{eq:phi_lambda} and Proposition \ref{prop:lambda_solution}, $\mathcal{O}(N\log N)$.\\\hline
  $\iota_{\mathrm{epi}_{\|\cdot\|_\ast}}(\cdot)$ & \begin{tabular}{c} See \eqref{eq:phi_lambda} and \eqref{eq:S_p_lambda}, $\mathcal{O}(MN \min\{M,N\})$  \end{tabular} \\\hline
\end{tabular}
}
\end{center}
\vspace{-0.4cm}
\end{table}
\subsubsection{Fenchel-Rockafellar conjugate function and its proximity operator}
For $^\forall f \in \Gamma_0(\mathbb{R}^N)$, the conjugate function $f^\ast$ of $f$ is defined as
$
f^\ast({\bm \xi}) := \sup_{\mathbf{x} \in \mathbb{R}^N} \langle \mathbf{x}, {\bm \xi } \rangle - f(\mathbf{x}),
$
and the proximity operator of the conjugate function is calculated as:
$
\mathrm{prox}_{\gamma f^\ast}(\mathbf{x}) = \mathbf{x} - \gamma\mathrm{prox}_{\frac{1}{\gamma}f} \left(\frac{1}{\gamma}\mathbf{x}\right)
$ \cite{Bauschke2011}.
\subsection{Primal-dual Splitting}\label{subsec:pds}
Consider the following convex optimization problem to find
\begin{align}
	\label{eq:pds}
	\mathbf{x}^{\star} \in \argmin_{\mathbf{x}\in \mathbb{R}^N} G(\mathbf{x}) + H(\mathbf{F}\mathbf{x}) \ (\neq \emptyset),
\end{align}
where $G\in \Gamma_0(\mathbb{R}^N)$, $H \in \Gamma_0(\mathbb{R}^M)$, and $\mathbf{F} \in \mathbb{R}^{M\times N}$, respectively. We assume that the solution set of \eqref{eq:pds} is nonempty. Then the PDS algorithm \cite{Chambolle2010,Condat2013,Vu2013,Komodakis2015,Bot2015} for solving \eqref{eq:pds} is given as follows: 
\begin{align}
	\label{eq:pdsalg}
	\begin{cases}
		\mathbf{x}^{(n+1)} = \mathrm{prox}_{\gamma_1 G} [\mathbf{x}^{(n)} - \gamma_1 \mathbf{F}^{\top}\mathbf{z}^{(n)}] \\
		\mathbf{z}^{(n+1)} = \mathrm{prox}_{\gamma_2 H^{\ast}} [\mathbf{z}^{(n)} + \gamma_2 \mathbf{F}(2\mathbf{x}^{(n+1)} - \mathbf{x}^{(n)})]
	\end{cases},
\end{align}
where prox denotes the proximity operator and $H^{\ast}$ is the conjugate function of $H$. 

According to Theorem 3.3 in \cite{Condat2013}, for given $\gamma_1 > 0$ and $\gamma_2 > 0$ satisfying $\gamma_1\gamma_2\|\mathbf{F}\|_\mathrm{op} \leq 1$ (the definition of the induced norm, see Table \ref{tab:Notations}), the sequence of $\{\mathbf{x}^{(n)}\}_{n=0}^{\infty}$ in \eqref{eq:pdsalg} converges to a minimizer $\mathbf{x}^\star$ of \eqref{eq:pds}. Thanks to the convexity nature, the initial value of $\mathbf{x}^{(0)}$ and $\mathbf{z}^{(0)}$ can be chosen arbitrarily. The convergence rate of the algorithm \eqref{eq:pdsalg} is proven to be $\mathcal{O}(1/n)$ \cite{Chambolle2010,Bot2015}.
\section{Epigraphical Relaxation}\label{sec:Proposed}
This section introduces a minimization approach for the layered mixed-norm regularization that does not have a closed-form proximity operator by using ERx. Sec. \ref{subsec:prob} states the general problem formulation. Sec. \ref{sec:episplit}, Sec. \ref{subsec:linf}, and Sec. \ref{subsec:MMN} present the procedure for ERx and discuss a theoretical guarantee on keeping the minimizer for each vector/matrix norm.
\subsection{Problem Formulation}\label{subsec:prob}
Let us consider the following minimization problem including a $K$-layered composite function $f^{(1,K)}: \mathbb{R}^{\textcolor{black}{N_1}} \rightarrow \mathbb{R}_+$ (see its definition in Table \ref{tab:Notations}, \textcolor{black}{the first layer function $f^{(1)}$ is the innermost function and the $K$-th layer function $f^{(K)}$ is the outermost one}) with a coercive function $g \in \Gamma_0(\mathbb{R}^N)$ (i.e., if $\|\mathbf{x}\| \rightarrow \infty$, $g(\mathbf{x})\rightarrow \infty$) consisting of the sum of proximable functions $g(\mathbf{x}) := \sum^{M}_{m = 1}g_m (\mathbf{B}_m\mathbf{x})$ ($g_m : \mathbb{R}^{N_m^{(B)}} \rightarrow \mathbb{R}_+$, $\mathbf{B}_m \in \mathbb{R}^{N_m^{(B)} \times N}$) as
\begin{align}\label{eq:genform}
\mathcal{S}_{\mathbf{x}} := \argmin_{\mathbf{x}\in \mathbb{R}^{N}} f^{(1,K)}(\mathbf{Ax}) + g(\mathbf{x})\ (\neq \emptyset),
\end{align}
where $\mathbf{A}\in \mathbb{R}^{\textcolor{black}{N_1} \times N}$, and the composite function $f^{(1,K)}$ follows the two assumptions (A1) and (A2) below.
\begin{itemize}
\item[(A1)] \textcolor{black}{
The $k$-th layer function $f^{(k)}:\mathbb{R}^{N_k} \rightarrow \mathbb{R}_+^{N_{k+1}}$ $(1 \leq {}^\forall k \leq K)$ $(\mathbb{R}_+^{N_{K+1}} = \mathbb{R}_+)$ is a convex function ($k=K$) or convex block-wise vector functions ($1\leq k \leq K-1$).}
\item[(A2)] \textcolor{black}{
The $k$-th layer function $f^{(k)}$ ($2 \leq {}^\forall k \leq K$) is \textit{strictly increasing (scalar/vector) function} on $\mathbb{R}_+^{N_{k}}$. On the other hand, it is not necessary that $f^{(1)}$ is strictly increasing. The definition of \textit{strictly increasing functions} in this paper is given as follows.}
\end{itemize}
\begin{definition}\label{def:sifunc}
A function $f:C \rightarrow \mathbb{R}$ ($C \subset \mathbb{R}^N$) is said to be a strictly increasing function on $C$ if for any $\mathbf{x} , \mathbf{y} \in C$ satisfying $\mathbf{x} \leq \mathbf{y}$ and $x_{n_0} < y_{n_0}$ for some $1\leq n_0 \leq N$, then $f(\mathbf{x}) < f(\mathbf{y})$. A \textcolor{black}{block-wise vector} function $f:\mathbb{R}^N \rightarrow \mathbb{R}^M$, 
$
f(\mathbf{x}) =  \begin{bmatrix}
f_1(\mathbf{x}_1) & \ldots & f_{M}(\mathbf{x}_{M})
\end{bmatrix}^\top,
$ ($\mathbf{x} = \begin{bmatrix}
\mathbf{x}_1^\top & \ldots & \mathbf{x}_{{M}}^\top
\end{bmatrix}^\top \in \mathbb{R}^{N}$, $\mathbf{x}_m \in \mathbb{R}^{\textcolor{black}{N_m}}$, \textcolor{black}{$N = \sum_{m=1}^{M}N_m$}) is said to be a strictly increasing (block-wise vector) function if all the block functions $f_{\textcolor{black}{m}} : \mathbb{R}^{\textcolor{black}{N_m}} \rightarrow \mathbb{R}$ are strictly increasing.
\end{definition}
\begin{Ex}
The $\ell_{p,q}$ norm $\|\cdot\|_{p,q}: \mathbb{R}^{N_1 N} \rightarrow \mathbb{R}_+$ ($p ,q \in [1 , \infty )$) is an example of a 2-layered composite function whose setup in the assumption (A1) is
\begin{align}
\textcolor{black}{\mathbf{x} = }&\ \textcolor{black}{\begin{bmatrix}
\mathbf{x}_1^\top & \ldots & \mathbf{x}_{N_1}^\top
\end{bmatrix}^\top,\ \mathbf{x}_n \in \mathbb{R}^{N}} \nonumber\\
\textcolor{black}{ f^{(1)} = }&\ \textcolor{black}{\begin{bmatrix}
\|\mathbf{x}_1\|_p & \ldots & \|\mathbf{x}_{N_1}\|_p
\end{bmatrix}^\top :  \mathbb{R}^{N_1N} \rightarrow \mathbb{R}_+^{N_1},}\nonumber\\
\textcolor{black}{f^{(2)} =}&\ \textcolor{black}{\|\cdot\|_{q}:  \mathbb{R}^{N_1} \rightarrow \mathbb{R}_+}.
\end{align}
\end{Ex}
\begin{Rem}\upshape 
\textcolor{black}{Note that $f^{(K)}$ and each block function $f^{(k)}_n$ of $f^{(k)}$ ($1 \leq k \leq K-1$, $1 \leq n \leq N_{k+1}$) can be a composite function itself as long as they satisfy the assumptions (A1) and (A2).}
\end{Rem}
\begin{Rem}\upshape 
Following the definition in \cite{Boyd2004}, we call a (block-wise) \textcolor{black}{vector} function $f: C \rightarrow \mathbb{R}^{M}$ ($C \subset \mathbb{R}^N$, $M\geq 1$) \textit{non-decreasing} (block-wise) \textcolor{black}{vector} function on $C$ if for any $\mathbf{x} , \mathbf{y} \in C$ satisfying $\mathbf{x} \leq \mathbf{y}$ then $f(\mathbf{x}) \leq f(\mathbf{y})$. 
\end{Rem}
The convexity condition for 2-layered composite functions is presented in \cite{Boyd2004}, and we can easily extend the condition for $K$-layered ones as in the following proposition.
\begin{prop}\label{prop:cvxln}
Assume that all the functions $f^{(k)}: \mathbb{R}^{N_{k}} \rightarrow \mathbb{R}_+^{N_{k+1}}$ ($1 \leq k \leq K$, $\mathbb{R}^{N_{K+1}} = \mathbb{R}_+$) are convex (block-wise vector) functions and $f^{(k)}$ ($2 \leq k \leq K$) are non-decreasing on $\mathbb{R}_+^{N_{k}}$. Then, $f^{(1,K)}$ is a convex function.
\begin{proof}
\begin{align}
&\ f^{(1,K)}(\lambda \mathbf{x} + (1-\lambda) \mathbf{y}) \nonumber\\ 
=&\ f^{(K)}( \cdots ( f^{(2)}(f^{(1)}(\lambda \mathbf{x} + (1-\lambda) \mathbf{y})))\cdots) \nonumber\\
\leq &\ f^{(K)}( \cdots ( f^{(2)}( \lambda  f^{(1)}(\mathbf{x})+ (1-\lambda)  f^{(1)}(\mathbf{y})))\cdots) \nonumber\\
\leq &\  f^{(K)}( \cdots ( \lambda  f^{(1,2)}(\mathbf{x})+ (1-\lambda)  f^{(1,2)}(\mathbf{y})))\cdots)\nonumber\\
\leq &\ \lambda   f^{(1,K)}(\mathbf{x})+ (1-\lambda) f^{(1,K)}(\mathbf{y}),
\end{align}
where the (block-wise) non-decreasing property and the convexity are used from the third line to the last line.
\end{proof}
\end{prop}
According to Proposition \ref{prop:cvxln}, $f^{(1,K)}$ satisfying the assumptions (A1) and (A2), i.e., strictly increasing property, is a convex function.

As important realizations, we can build such functions $f^{(K)}$ and the block functions $f^{(k)}_n$ of $f^{(k)}$ ($2 \leq k \leq K-1$, $1 \leq n \leq N_{k+1}$) based on the $\ell_p$-norm as shown in the following proposition (\textcolor{black}{since the derivation is trivial, the proof is omitted.}). 
\begin{prop}\label{prop:lp}
For $p \in [1,\infty)$, the $\ell_p$-norm and the $\ell_p$-type mixed norm (i.e., all the block functions of all the layers in a composite function are the $\ell_p$-norm) are strictly increasing convex functions on $\mathbb{R}^N_+$.
\end{prop}
Thus, the $\ell_{p,q}$ norm ($p ,q \in [1 , \infty )$)  satisfies \textcolor{black}{the assumptions (A1) and (A2)}.
\subsection{On Solution Conservation by Epigraphical Relaxation for $\ell_p$-type ($p \in [1,\infty)$) Mixed Norms}\label{sec:episplit}
In this section, we consider composite functions satisfying the assumptions (A1) and (A2) that do not have the closed-form proximity operator. First, we reformulate the equation \eqref{eq:genform} by introducing a\textcolor{black}{n} auxiliary variable $\mathbf{z}^{(\textcolor{black}{K})}$ as follows:
\begin{align}
\argmin_{\mathbf{x}, \mathbf{z}^{(\textcolor{black}{K})}}\  f^{(\textcolor{black}{K})}(\mathbf{z}^{(\textcolor{black}{K})}) + g(\mathbf{x}) \ 
 \mathrm{s.t.}\  f^{(\textcolor{black}{1,K-1})}(\mathbf{Ax}) = \mathbf{z}^{(\textcolor{black}{K})}.
\end{align}
Since the constraint is not convex in general, we relax the constraint as
\begin{align}\label{eq:reform2}
&\argmin_{\mathbf{x}, \mathbf{z}^{(\textcolor{black}{K})}} f^{(\textcolor{black}{K})}(\mathbf{z}^{(\textcolor{black}{K})}) + g(\mathbf{x})\  \mathrm{s.t.}\  f^{(\textcolor{black}{1,K-1})}(\mathbf{Ax}) \leq \mathbf{z}^{(\textcolor{black}{K})}.
\end{align}
In the above equation, the block-wise epigraph (see Appendix \ref{app:epi_proj}) constraint appears $(\mathbf{Ax},\mathbf{z}^{(\textcolor{black}{K})}) \in \mathrm{epi}_{f^{(\textcolor{black}{1,K-1})}}$ and this relaxation is termed as \textit{epigraphical relaxation} (ERx). Then, the equation \eqref{eq:reform2} is further manipulated by introducing a new variable $\mathbf{z}^{(\textcolor{black}{K-1})}$, convex relaxation, and the epigraph notation as
\begin{align}\label{eq:reform3}
&\argmin_{\mathbf{x}, \textcolor{black}{\{\mathbf{z}^{(k)}\}_{k=K-1}^{K}}} f^{(\textcolor{black}{K})}(\mathbf{z}^{(\textcolor{black}{K})})+ g(\mathbf{x}) \nonumber \\  
        \mathrm{s.t.}&\ \textcolor{black}{(\mathbf{z}^{(K-1)},\mathbf{z}^{(K)}) \in \mathrm{epi}_{f^{(K-1)}}, f^{(1,K-2)}(\mathbf{Ax}) = \mathbf{z}^{(K-1)},}\nonumber\\
\xrightarrow{\mathrm{ERx}}  &\argmin_{\mathbf{x}, \textcolor{black}{\{\mathbf{z}^{(k)}\}_{k=K-1}^{K}}} f^{(\textcolor{black}{K})}(\mathbf{z}^{(\textcolor{black}{K})})+ g(\mathbf{x}) \nonumber \\  
        \mathrm{s.t.}&\ \textcolor{black}{(\mathbf{z}^{(K-1)},\mathbf{z}^{(K)}) \in \mathrm{epi}_{f^{(K-1)}}, (\mathbf{Ax},\mathbf{z}^{(K-1)}) \in \mathrm{epi}_{f^{(1,K-2)}}}.
\end{align}
In similar fashion, we repeatedly apply ERx to \eqref{eq:reform3} as
\begin{align}\label{eq:reform4}
\widetilde{\mathcal{S}}_{\mathbf{x}} \times \prod_{k=2}^{K}\widetilde{\mathcal{S}}_{\mathbf{z}^{(k)}}  :=&\argmin_{\mathbf{x}, \textcolor{black}{\{\mathbf{z}^{(k)}\}_{k=2}^{K}}} f^{(\textcolor{black}{K})}(\mathbf{z}^{(\textcolor{black}{K})})+ g(\mathbf{x}) \nonumber \\  
        \mathrm{s.t.}&\ \textcolor{black}{ (\mathbf{z}^{(k)},\mathbf{z}^{(k+1)}) \in \mathrm{epi}_{f^{(k)} } \ (2 \leq k \leq K-1),}\nonumber\\  &\ \textcolor{black}{(\mathbf{Ax},\mathbf{z}^{(2)}) \in \mathrm{epi}_{f^{(1)} }.}
\end{align}
Note that the optimization problem \eqref{eq:reform4} is not the same as \eqref{eq:genform} due to ERx. Nevertheless, under the assumptions \textcolor{black}{(A1) and (A2)}, we can derive the equivalence of the solution sets for the original problem and its modified problem by ERx as in the following theorem and corollary.
\begin{theorem}\label{theo:ER}
We assume that functions $f^{(k)}$ ($1\leq k \leq K$) satisfy the assumptions \textcolor{black}{(A1) and (A2)}. Then, the minimizer ${\mathbf{x}}^\star$ and $\{\mathbf{z}^{\star(k)}\}_{k=\textcolor{black}{2}}^{K}$ for \eqref{eq:reform4} satisfies
\begin{align}\label{eq:theorem}
 f^{(k)}(\mathbf{z}^{\star(k)})=&\ \textcolor{black}{\mathbf{z}^{\star(k+1)}\ (2 \leq k \leq K-1),} \nonumber\\ 
        f^{(1)}(\mathbf{A}\mathbf{x}^\star) =&\ \mathbf{z}^{\star(\textcolor{black}{2})}.
\end{align}
Thus, ${\mathbf{x}}^\star$ is also the minimizer of \eqref{eq:genform}, i.e., ${\mathcal{S}}_{\mathbf{x}} =\widetilde{\mathcal{S}}_{\mathbf{x}} $.
\begin{proof}
See Appendix \ref{ap:poof_ER}.
\end{proof}
\end{theorem}
\begin{corol}\label{corol:ER}
Let $f^{(k)}$ (\textcolor{black}{$2\leq k \leq K$}) be a (block-wise \textcolor{black}{vector}) $\ell_p$-type (mixed) norm ($p \in [1,\infty)$) and $f^{(\textcolor{black}{1})}$ be any type of block-wise norm. Then, the set of the minimizers for \eqref{eq:genform} and \eqref{eq:reform4} are identical (${\mathcal{S}}_{\mathbf{x}} =\widetilde{\mathcal{S}}_{\mathbf{x}} $).
\end{corol}
\textcolor{black}{The optimization problem \eqref{eq:reform4} can be solved by the PDS algorithm \eqref{eq:pdsalg}. We convert the constrained optimization problem into the unconstrained one and cast it to the PDS formulation \eqref{eq:pds} as, for example,
\begin{align}\label{eq:erxpds}
&\argmin_{\mathbf{x}, \textcolor{black}{\{\mathbf{z}^{(k)}\}_{k=2}^{K}}} f^{(\textcolor{black}{K})}(\mathbf{z}^{(\textcolor{black}{K})})+ \sum^{M}_{m = 1}g_m (\mathbf{B}_m\mathbf{x}) \nonumber\\
&\qquad + \sum_{k=2}^{K-1} \iota_{\mathrm{epi}_{f^{(k)}}}(\mathbf{z}^{(k)},\mathbf{z}^{(k+1)}) + \iota_{\mathrm{epi}_{f^{(1)}}}(\mathbf{A}\mathbf{x},\mathbf{z}^{(2)}),\nonumber\\
&\mathbf{p} = \begin{bmatrix} \mathbf{x}^\top & \mathbf{z}^{(2)\top} & \cdots & \mathbf{z}^{(K)\top} \end{bmatrix}^\top,\ G(\mathbf{p}) = 0,\nonumber\\
&H(\mathbf{q}) = f^{(K)}(\mathbf{q}_1) + \sum^{M}_{m = 1}g_m (\mathbf{q}_{1+m})  \nonumber\\
&\qquad\quad+ \sum_{k=1}^{K-1} \iota_{\mathrm{epi}_{f^{(k)}}}(\mathbf{q}_{M+1+2k-1},\mathbf{q}_{M+1+2k}), \nonumber\\
&\mathbf{q} = \begin{bmatrix} \mathbf{q}_1^\top & \cdots & \mathbf{q}_{M+1+2(K-1)}^\top \end{bmatrix}^\top = \mathbf{F}\mathbf{p},
\end{align} where the matrix $\mathbf{F}$ consists of $\mathbf{I}$, $\mathbf{O}$, $\mathbf{A}$, $\{\mathbf{B}_m\}_{m=1}^{M}$. As long as $f^{(K)}$ and $\{\iota_{\mathrm{epi}_{f^{(k)}}}\}_{k=1}^{K-1}$ have the closed-form proximity operators, the PDS algorithm can find an optimal solution efficiently. We show practical cases in the following sections.}
\subsubsection{On Computational Complexity for Epigraphical Relaxation}
\textcolor{black}{The main computational bottleneck of the ERx with the PDS algorithm \eqref{eq:erxpds} is broken down into two parts: 1) the matrix multiplication by $\mathbf{F}$ (and $\mathbf{F}^\top$) and 2) computation of proximity operators. The matrix $\mathbf{F}$ typically forms a block-sparse matrix, and its multiplication can be broken down into block-matrix multiplications by $\mathbf{A}$ and $\{\mathbf{B}_m\}_{m=1}^{M}$. Thus, the order of the computational complexity is evaluated as $\mathcal{O}(\max\{N_1 , \{N_m^{(B)}\}_{m=1}^{M}\}N)$ ($N$ denotes the dimension of an input vector). On the other hand, the computational complexity for proximity operators $G$ and $H$ depends on which types of norms are used in a layered mixed norm. While most of them can be calculated with the complexity of $\mathcal{O}(N)$, the projection onto $\ell_1$-norm ball, for example, requires the complexity of $\mathcal{O}(N\log N)$ due to sorting operation \cite{Duchi2008}. In addition, for a matrix $\mathbf{X} \in \mathbb{R}^{M\times N}$, the proximity operator of the nuclear norm and the projection onto the epigraph of the nuclear norm requires the SVD calculation with the order of the complexity bounded by $\mathcal{O}(MN \min\{M,N\})$ \cite{Golub1996}.}

\subsection{On Solution Conservation by Epigraphical Relaxation for $\ell_\infty$-type Mixed Norms}\label{subsec:linf}
This section discusses the case of the $\ell_\infty$-type mixed norms, i.e., composite functions, including the $\ell_\infty$-norms in their layers. Concerning the $\ell_\infty$-norm, the following proposition holds true. \textcolor{black}{Since the derivation is trivial, the proof is omitted.} 
\begin{prop}\label{prop:linf}
The $\ell_\infty$-norm is not a strictly increasing function but a non-decreasing one on $\mathbb{R}_+^N$, i.e., for $\mathbf{x} , \mathbf{y} \in \mathbb{R}_+^N$, if $\mathbf{x} \leq \mathbf{y}$ and $x_{n_0} < y_{n_0}$ $(1\leq {}^\exists {n_0} \leq N)$ then $\|\mathbf{x}\|_\infty \leq \|\mathbf{y}\|_\infty$.
\end{prop}
From the proposition, the $\ell_\infty$-type mixed norms do not satisfy the assumption \textcolor{black}{(A2)} and thus the $\ell_\infty$-type norm cannot keep the minimizers due to the lack of the strictly increasing property. Here, we modify the $\ell_\infty$-type (mixed) norm to make it strictly increasing.
\begin{prop} \label{prop:infeps}
Define the modified $\ell_\infty$-norm as $\|\cdot \|_{{\infty},\epsilon}:= \|\cdot\|_{\infty} + \epsilon \|\cdot\|_2$, where $\epsilon > 0$ is assumed to be a small positive value. Then, the following statements hold true.
\begin{enumerate}
\item $\|\cdot \|_{{\infty},\epsilon}$ is a norm. $\|\cdot\|_\infty = \lim_{\epsilon \rightarrow 0}\|\cdot\|_{\infty,\epsilon}$.
\item $\|\cdot \|_{{\infty},\epsilon}$ is a strictly increasing function on $\mathbb{R}_+^N$. 
\end{enumerate} 
\begin{proof}
The statement 1 can be trivially verified. The statement 2 holds according to Proposition \ref{prop:lp}.
\end{proof}
\end{prop}
From Proposition \ref{prop:infeps}, the following corollary holds.
\begin{corol}\label{corol:ERinf}
Let all the functions $f^{(1)}$ and $f^{(k)}_n$ of $f^{(\textcolor{black}{k})}$ in \eqref{eq:genform} be the $\ell_p$-type ($p \in [1,\infty)$) or the modified $\ell_{\infty}$-type (mixed) norm. Then, the set of the minimizers for \eqref{eq:genform} and \eqref{eq:reform4} are identical (${\mathcal{S}}_{\mathbf{x}} =\widetilde{\mathcal{S}}_{\mathbf{x}} $).
\end{corol}
\subsubsection{Optimization Algorithm Based on Epigraphical Splitting}
This section \textcolor{black}{shows an example of} the modified $\ell_{\infty}$-type mixed norm minimization by the PDS algorithm. For simplicity, we consider the following 2-layered mixed norm:
\begin{align}\label{eq:modlinf}
&\ \mathcal{S}_{\mathbf{x}} = \argmin_{\mathbf{x}\in \mathbb{R}^{N}} f^{(1,2)}(\mathbf{Ax}) + g(\mathbf{x}),
\end{align}
\textcolor{black}{where the first/second layer functions $f^{(1)}$ and $f^{(2)}$ are}
\begin{align}
&\ f^{(1)}:\mathbb{R}^{N_1} \rightarrow \mathbb{R}_{\textcolor{black}{+}}^{N_2}: \nonumber\\
&\quad\begin{bmatrix}
\mathbf{x}_1^\top \  \ldots \  \mathbf{x}_{N_2}^\top 
\end{bmatrix}^\top  \mapsto \begin{bmatrix}
\|\mathbf{x}_1\|_{\infty,\epsilon} \  \ldots \  \|\mathbf{x}_{N_2}\|_{\infty,\epsilon} 
\end{bmatrix}^\top. \nonumber\\
&\ f^{(2)}: \mathbb{R}^{N_2} \rightarrow \mathbb{R}_+: \mathbf{x} \mapsto \|\mathbf{x}\|_{\infty,\epsilon},
\end{align}
and $g \in \Gamma_0(\mathbb{R}^N)$ is a proximable function. By using ERx \textcolor{black}{as in \eqref{eq:reform2}}, the above problem \textcolor{black}{\eqref{eq:modlinf}} can be converted as follows:
\begin{align}
\mathcal{S}_{\mathbf{x}} \times \mathcal{S}_{\mathbf{z}} =&\ \argmin_{\mathbf{x}\in \mathbb{R}^{N},\ \mathbf{z}} \|\mathbf{z}\|_{\infty} + \epsilon\|\mathbf{z}\|_{2} + g(\mathbf{x}) \nonumber \\ 
&\ \mathrm{s.t.}\ \  f^{(\textcolor{black}{1})}(\mathbf{Ax}) \leq \mathbf{z}.
\end{align}
The above constraint can be reformulated by using epigraphical splitting \textcolor{black}{\eqref{eq:afterES} (see Appendix \ref{appsec:ES})} as:
\begin{align}
 f^{(\textcolor{black}{1})}(\mathbf{Ax}) \leq \mathbf{z} \Longleftrightarrow&\  f^{(\textcolor{black}{1})}_{\|\cdot\|_\infty}(\mathbf{Ax}) +  f^{(\textcolor{black}{1})}_{\epsilon\|\cdot\|_2}(\mathbf{Ax})  \leq \mathbf{z}, \nonumber\\
 \Longleftrightarrow&\ f^{(\textcolor{black}{1})}_{\|\cdot\|_\infty}(\mathbf{Ax}) \leq {\bm \eta_1},\  f^{(\textcolor{black}{1})}_{\epsilon \|\cdot\|_2}(\mathbf{Ax}) \leq {\bm \eta_2},\nonumber\\
&\ {\bm \eta_1} + {\bm \eta_2} - \mathbf{z} \leq \mathbf{0},\nonumber\\
f^{(\textcolor{black}{1})}_{\|\cdot\|_{\infty}}:\mathbb{R}^{N_1} \rightarrow \mathbb{R}_{\textcolor{black}{+}}^{N_2}:& \nonumber\\
\quad\begin{bmatrix}
\mathbf{x}_1^\top \  \ldots \  \mathbf{x}_{N_2}^\top 
\end{bmatrix}^\top & \mapsto \begin{bmatrix}
\|\mathbf{x}_1\|_{\infty} \  \ldots \  \|\mathbf{x}_{N_2}\|_{\infty} 
\end{bmatrix}^\top, \nonumber\\
f^{(\textcolor{black}{1})}_{\epsilon \|\cdot\|_2}:\mathbb{R}^{N_1} \rightarrow \mathbb{R}_{\textcolor{black}{+}}^{N_2}:&  \nonumber\\
\quad\begin{bmatrix}
\mathbf{x}_1^\top \  \ldots \  \mathbf{x}_{N_2}^\top 
\end{bmatrix}^\top & \mapsto \begin{bmatrix}
\epsilon \|\mathbf{x}_1\|_{2} \  \ldots \ \epsilon  \|\mathbf{x}_{N_2}\|_{2} 
\end{bmatrix}^\top.
\end{align}
The minimization problem can be solve by PDS as
\begin{align} 
\mathbf{p} =&\  \begin{bmatrix}
\mathbf{x}^\top &  \mathbf{z}^\top  & {\bm \eta_1}^\top & {\bm \eta_2}^\top
\end{bmatrix}^\top,\ G(\mathbf{p}) = \mathbf{0}, \nonumber\\
	H(\mathbf{q}) =&\ \|\mathbf{q}_1\|_{\infty} + \epsilon\|\mathbf{q}_2\|_{2}+ g(\mathbf{q}_3)  +\iota_{\mathrm{epi}_{\|\cdot\|_\infty}}(\mathbf{q}_4,\mathbf{q}_5) \nonumber\\ 
	&\ + \iota_{\mathrm{epi}_{\epsilon\|\cdot\|_2}}(\mathbf{q}_6,\mathbf{q}_7) + \iota_{(-\infty,0]^{N_2}}(\mathbf{q}_8),\nonumber\\
	\mathbf{q} =&\  \begin{bmatrix} \mathbf{q}_1^\top & \cdots & \mathbf{q}_{8}^\top \end{bmatrix}^\top =\mathbf{F}\mathbf{p},
\nonumber\\
	\mathbf{F}=&\ \textcolor{black}{\begin{bmatrix}
		\mathbf{O} & \mathbf{O}  & \mathbf{I} & \mathbf{A}^\top & \mathbf{O} & \mathbf{A}^\top & \mathbf{O} & \mathbf{O}\\
		\mathbf{I} & \mathbf{I} & \mathbf{O} & \mathbf{O} & \mathbf{O} & \mathbf{O} & \mathbf{O} & -\mathbf{I} \\
		\mathbf{O} & \mathbf{O} & \mathbf{O} & \mathbf{O} & \mathbf{I} & \mathbf{O} & \mathbf{O} & \mathbf{I} \\
		\mathbf{O} & \mathbf{O} & \mathbf{O} & \mathbf{O} & \mathbf{O} & \mathbf{O} & \mathbf{I} & \mathbf{I} 
\end{bmatrix}^\top},
\end{align}
\textcolor{black}{where the proximity operators of $\|\cdot\|_{2}$, $\|\cdot\|_{\infty}$, and $\iota_{(-\infty,0]^{N_2}}$ are shown in Table \ref{tab:complexity}. The proximity operator of $\iota_{\mathrm{epi}_{\epsilon\|\cdot\|_2}}$ and that of $\iota_{\mathrm{epi}_{\|\cdot\|_\infty}}$ are calculated by \eqref{eq:epil2} and \eqref{eq:epiinf}.}

\textcolor{black}{The order of the computational complexity for the above algorithm depends on the matrix multiplication by $\mathbf{A}$ $(\mathcal{O}(N_1 N))$, the proximity operator of $\iota_{\mathrm{epi}_{\|\cdot\|_\infty}}$ $(\mathcal{O}(N_1 \log N_1))$ (see Table \ref{tab:complexity}), and the proximable function $g$.}
\subsection{Epigraphical Relaxation for Mixed Matrix Norms}\label{subsec:MMN}
This section discusses ERx on mixed matrix norms. To apply the formulation based on vectors in \eqref{eq:genform} for matrices, we use the (block-wise) vectorized notation as $\mathrm{bvec}_{(M_{b,1},N_{b,1})}(\mathbf{X}) \in \mathbb{R}^{MN}$, instead of $\mathbf{X} \in \mathbb{R}^{M\times N}$, and seek the minimizer $\mathbf{x}^\star$, then reshape it to the matrix ${\mathbf{X}}^\star = \mathrm{bmat}_{(M_{b,1},N_{b,1})}(\mathbf{x}^\star)$  (see the definitions of $\mathrm{bvec}$ and $\mathrm{bmat}$ in Table \ref{tab:Notations}).
\begin{align}\label{eq:genformmat}
\mathbf{x}^\star \in&\ \argmin_{\mathbf{x} = \mathrm{bvec}_{(M_{b,1},N_{b,1})}(\mathbf{X})\in \mathbb{R}^{MN}} f^{(1,K)}(\mathbf{Ax}) + g(\mathbf{x}),\nonumber\\ 
\mathbf{X}^\star =&\ \mathrm{bmat}_{(M_{b,1},N_{b,1})}(\mathbf{x}^\star),
\end{align}
where the specifications for the $k$-th layer function are
\begin{itemize}
\item $f^{(K)}:\mathbb{R}^{N_K} \rightarrow \mathbb{R}_+$, $f^{(1)}:\mathbb{R}^{N_1} \rightarrow \mathbb{R}_+^{N_2}$,
\item for an intermediate output vector $\mathrm{bvec}_{(M_{b,k},N_{b,k})}(\mathrm{mat}(\mathbf{x})) = \begin{bmatrix}
\mathbf{x}_1^\top & \ldots & \mathbf{x}_{{N_{k+1}}}^\top
\end{bmatrix}^\top \in \mathbb{R}^{N_k}$ ($\mathbf{x}_n \in \mathbb{R}^{M_{b,k}N_{b,k}}$), \textcolor{black}{$f^{(k)}:\mathbb{R}^{N_k} \rightarrow \mathbb{R}_+^{N_{k+1}}$} $(2 \leq k \leq K-1)$ are
\begin{align}
 &\ f^{(k)}(\mathrm{bvec}_{(M_{b,k},N_{b,k})}(\mathrm{mat}(\mathbf{x}))) = f^{(k)}(\mathbf{P}_k\mathbf{x})   \nonumber\\
=&\ \begin{bmatrix}
f_1^{(k)}(\mathbf{x}_1) & \ldots & f_{N_{k+1}}^{(k)}(\mathbf{x}_{N_{k+1}})
\end{bmatrix}^\top,
\end{align}
where $\mathbf{P}_k$ is a certain permutation matrix corresponding to the reordering operation of $\mathrm{bvec}_{(M_{b,k},N_{b,k})}(\mathrm{mat}(\cdot))$. An example of a composite function $f^{(1,3)}$ for matrices is illustrated in Fig. \ref{fig:matrixnorm}.
\begin{figure}[t]
\centering
			\begin{minipage}[b]{1\linewidth}
			\centering
		\scalebox{0.5}{\includegraphics[keepaspectratio=true]{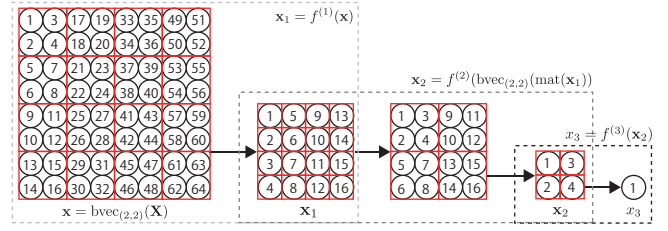}}
	\end{minipage}
	\caption{Example of a composite function for matrices.}\label{fig:matrixnorm}
\end{figure}
\end{itemize}
\subsubsection{$\ell_p$-type Mixed Matrix Norm}
From the definition, the $\ell_p$-matrix norm is equivalent to the $\ell_p$-norm for their vectorized versions $\|\mathbf{X}\|_p = \|\mathrm{vec}(\mathbf{X})\|_p = \|\mathrm{bvec}(\mathbf{X})\|_p$. Thus, the $\ell_p$-matrix norm ($p \in [1,\infty)$) is a strictly increasing function on $\mathbb{R}^{M \times N}$ and keeps the minimizers. Similarly, if the modified $\ell_\infty$-norm $\|\cdot\|_{\infty,\epsilon}$, instead of the $\ell_\infty$-norm, is used for layered mixed norms, the minimizers of the original problem are kept. 
\subsubsection{Schatten-$p$-type Mixed Matrix Norm}
Besides the $\ell_p$-matrix norm, the Schatten-$p$-matrix norms are widely used in many research fields of signal processing. This section discusses the important cases $p = 1, 2, \infty$.
\begin{prop}\label{prop:S2Sinf}
The Schatten-2 norm  $\|\cdot\|_{S_2}$ is a strictly increasing function and the Schatten-$\infty$ norm $\|\cdot\|_{S_\infty}$ is a non-decreasing function on $\mathbb{R}_+^{M \times N}$.
\begin{proof}
See Appendix \ref{ap:S2Sinf}.
\end{proof}
\end{prop}
From Proposition \ref{prop:S2Sinf}, the Schatten-$\infty$ norm cannot be introduced into layered mixed norms for keeping the minimizer. 

Inspired from the modified $\ell_\infty$-norm in Sec. \ref{subsec:linf}, we define the modified Schatten-$\infty,\epsilon$ norm $\|\cdot\|_{S_\infty,\epsilon} = \|\cdot\|_{S_\infty} + \epsilon \|\cdot\|_F$. In a similar discussion in Proposition \ref{prop:infeps}, we can verify that Schatten-$\infty,\epsilon$ norm is a strictly increasing function.

Unfortunately, the nuclear norm (the Schatten-1 norm) is neither a strictly increasing function nor a non-decreasing one, which can be confirmed by a simple example:
\begin{align}
\mathbf{A} = \begin{bmatrix} 1 & 1 \\ 1 & 0.9 \end{bmatrix}, \ \mathbf{B} = \begin{bmatrix} 1 & 1 \\ 1 & 1 \end{bmatrix} \Longrightarrow \|\mathbf{A}\|_\ast > \|\mathbf{B}\|_\ast.
\end{align}
Thus, \textcolor{black}{$K$-layered mixed norms having the nuclear norm in from the second to $K$-th layer} do not satisfy the assumption (\textcolor{black}{A2}) and are not guaranteed to be convex.
\begin{Rem}\upshape 
\textcolor{black}{The assumptions (A1) and (A2) do not require the strictly increasing property for the first layer function to satisfy Proposition \ref{prop:cvxln} and Theorem \ref{theo:ER}. Thus, we can use an arbitrary norm, including the nuclear norm, for the first layer (innermost) function without violating the assumptions for Proposition \ref{prop:cvxln} and Theorem \ref{theo:ER} (as long as the other layer functions satisfy the assumptions (A1) and (A2)).}
\end{Rem}
\section{Application of Epigraphical Relaxation in Signal Recovery}\label{sec:AppER}
This section introduces two practical examples, the DSTV and the ASNN regularizers, that do not have the closed-form proximity operators and cannot be tractable without using ERx.
\subsection{Decorrelated Structure-tensor Total Variation}\label{sec:EpiDSTV}
This section introduces a new regularization function by extending the decorrelated vectorial variation (DVTV) proposed in \cite{Ono2014DVTV}. VTV \cite{Bresson2008} evaluates the variation of the RGB channels for color images (see Appendix \ref{ap:VTVPDS}) and DVTV is the extended version of VTV that calculates VTV on the luma-chroma color space. For an $N$-sample color image $\mathbf{x} \in \mathbb{R}^{3N}$, DVTV is defined as
\begin{align}\label{eq:DVTVdef}
\|\mathbf{x}\|_{\mathrm{DVTV}} :=&  \sum^{N}_{n=1} \left\{w\sqrt{\sum^{2}_{j=1}d_{\mathrm{y},n,j}^2} + \sqrt{\sum^{2}_{c=1}\sum^{2}_{j=1}d_{c,n,j}^2}\right\},\nonumber\\
 =&\  \|\mathbf{P}^{(1)}\mathbf{D}(\mathbf{C}_0 \otimes \mathbf{I} )\mathbf{x}\|_{\textcolor{black}{2,1}}^{(w,2,4)}, \nonumber\\
\|\mathbf{x}\|_{\textcolor{black}{2,1}}^{(w,2,4)} :=&\  w\|\mathbf{x}_1\|_{\textcolor{black}{2,1}} + \|\mathbf{x}_2\|_{\textcolor{black}{2,1}},\nonumber\\
\mathbf{x} =&\ \begin{bmatrix} \mathbf{x}_1^\top & \mathbf{x}_2^\top \end{bmatrix}^\top,\  (\mathbf{x}_1 \in \mathbb{R}^{2N},\ \mathbf{x}_2 \in \mathbb{R}^{4N}),\nonumber\\
\mathbf{D} =&\ \mathrm{diag}(\mathbf{D}_0,\mathbf{D}_0,\mathbf{D}_0),\nonumber\\ \mathbf{D}_0 =&\ \begin{bmatrix}
\mathbf{D}_v^\top & \mathbf{D}_h^\top 
\end{bmatrix}^\top,
\end{align}
where $d_{\mathrm{y},n,1}$ and $d_{\mathrm{y},n,2}$ are the $n$-th luma vertical and horizontal differences, $d_{c,n,1}$ and $d_{c,n,2}$ are the $n$-th chroma ones. $\mathbf{D}_v,\ \mathbf{D}_h \in \mathbb{R}^{N \times N}$ are the vertical and horizontal difference matrices, respectively, and $\mathbf{C}_0 \in \mathbb{R}^{3\times 3}$ the DCT matrix which acts as the RGB-luma/chroma color space transformation. $\mathbf{P}^{(1)}$ permutes the vertical and horizontal differences of the luma and chroma channels of each sample as 
\begin{align}
\mathbf{P}^{(1)}\mathbf{D}\mathbf{x} =&\ \begin{bmatrix} \mathbf{d}_{\mathrm{y}}^\top & \mathbf{d}_{\mathrm{c}}^\top \end{bmatrix}^\top,\nonumber\\ 
\mathbf{d}_{\mathrm{y}} =&\ \begin{bmatrix} \mathbf{d}_{\mathrm{y},1}^\top & \ldots & \mathbf{d}_{\mathrm{y},N}^\top\end{bmatrix}^\top,\nonumber\\
 \mathbf{d}_{\mathrm{y},n} =&\ \begin{bmatrix} d_{\mathrm{y},n,1} & d_{\mathrm{y},n,2} \end{bmatrix}^\top,\nonumber\\
\mathbf{d}_{\mathrm{c}} =&\ \begin{bmatrix} \mathbf{d}_{\mathrm{c},1}^\top & \ldots & \mathbf{d}_{\mathrm{c},N}^\top\end{bmatrix}^\top,\nonumber\\
 \mathbf{d}_{\mathrm{c},n} =&\ \begin{bmatrix} d_{1,n,1} & d_{1,n,2} & d_{2,n,1} & d_{2,n,2} \end{bmatrix}^\top.
\end{align}

Now we generalize DVTV to DSTV which is defined as
\begin{align}
\|\mathbf{x}\|_{\mathrm{DSTV}} :=&\  \sum^{N}_{n=1} w\|\mathbf{X}_{\mathrm{y},n}\|_\ast + \sum^{N}_{n=1}\sqrt{\sum^{2}_{c=1}\|\mathbf{X}_{c,n}\|_\ast^2},\nonumber\\
\mathbf{X}_{\mathrm{y},n} =&\ \begin{bmatrix}
\mathbf{d}^{(\mathbf{X})}_{\mathrm{y},n,1} & \mathbf{d}^{(\mathbf{X})}_{\mathrm{y},n,2}
\end{bmatrix},\ \mathbf{X}_{c,n} = \begin{bmatrix}
\mathbf{d}^{(\mathbf{X})}_{c,n,1} & \mathbf{d}^{(\mathbf{X})}_{c,n,2}
\end{bmatrix},
\end{align}
where $\mathbf{d}^{(\mathbf{X})}_{\mathrm{y},n,1} \in \mathbb{R}^{W^2}$ and $\mathbf{d}^{(\mathbf{X})}_{\mathrm{y},n,2} \in \mathbb{R}^{W^2}$ are the vertical and horizontal differences in the $W \times W$ local patch centered at the $n$-th luma sample, and $\mathbf{d}_{c,n,1}$ and $\mathbf{d}_{c,n,2}$ are those of the chroma sample. DSTV is reduced to DVTV when $W = 1$. Since the structure tensor captures first-order information around a local region, it carries more flexible and robust measures of image variation than DVTV. Moreover, DSTV evaluates the mixed norm of chroma STVs that promotes the group sparsity of the STVs. This formulation is natural because if a region is very smooth, both chroma STVs are expected to be zero simultaneously.

DSTV can be expressed by using a mixed norm as
\begin{align}\label{eq:DSTVdef}
\|\mathbf{x}\|_{\mathrm{DSTV}}=&\ \|\mathbf{E}\mathbf{P}^{(1)}\mathbf{D}(\mathbf{C}_0 \otimes \mathbf{I} )\mathbf{x}\|_{\textcolor{black}{\ast,2,1}}^{(w)}\nonumber\\ 
=&\  f^{(1,2)}(\mathbf{E}\mathbf{P}^{(1)}\mathbf{D}(\mathbf{C}_0 \otimes \mathbf{I} )\mathbf{x}),\nonumber\\
\|\mathbf{x}\|_{\textcolor{black}{\ast,2,1}}^{(w)}:=&\  \sum^{N}_{n=1}w \| \mathbf{x}_n \|_\ast + \sum^{N}_{n=1} \sqrt{\sum^{2}_{c=1}\|\mathbf{x}_{c,n}\|_\ast^2},
\end{align}
where $\mathbf{E} \in \mathbb{R}^{6W^2N \times 6N}$ is an expansion operator that duplicates all the gradients $\mathbf{P}^{(1)}\mathbf{D}(\mathbf{C}_0 \otimes \mathbf{I} )\mathbf{x}$ in all the patches. Since $f^{(1,2)}=f^{(2)} \circ f^{(1)}$, where $f^{(1)}$ is a block-wise nuclear norm and $f^{(2)} = \|\cdot\|_{\textcolor{black}{2,1}}^{(w,2,4)}$ is the $\ell_p$-type mixed norm, $f^{(1,2)}$ satisfies the assumptions (A1) and (A2). 

Here, let us consider to recover the original image $\widehat{\mathbf{x}} \in [0,1]^{3N}$ from the incomplete measurement $\mathbf{y} \in \mathbb{R}^M$. We assume that the observation is contaminated by additive Gaussian noise $\mathbf{n} \in \mathbb{R}^M$ after some degradation ${\mathbf{\Phi}} \in \mathbb{R}^{M\times 3N}$ as $\mathbf{y}= {\mathbf{\Phi}}\widehat{\mathbf{x}} + \mathbf{n}$. Image recovery by the DSTV minimization with the $\ell_2$-ball data fidelity constraint is formulated as
\begin{align}\label{eq:DSTVmin}
{\mathcal{S}}_{\mathbf{x}} =&\ \argmin_{\mathbf{x} \in [0,1]^{3N}} \|\mathbf{E}\mathbf{P}^{(1)}\mathbf{D}(\mathbf{C}_0 \otimes \mathbf{I} )\mathbf{x}\|_{\textcolor{black}{\ast,2,1}}^{(w)}\nonumber\\  &\mathrm{s.t.}\ \ \mathbf{\Phi}\mathbf{x} \in \mathcal{B}_2(\mathbf{y}, \epsilon).
\end{align} 
Since $\|\cdot\|_{\textcolor{black}{\ast,2,1}}$ does not have a closed-form expression of the proximity operator, we apply ERx for \eqref{eq:DSTVmin} as
\begin{align}\label{eq:DSTVminepi}
{\mathcal{S}}_{\mathbf{x}} \times {\mathcal{S}}_{\mathbf{z}}  
 =&\ \argmin_{\mathbf{x}\in\mathbb{R}^{3N}, \mathbf{z}}  \|\mathbf{z}\|_{\textcolor{black}{2,1}}^{(w,2,4)}+ \iota_{[0,1]^{3N}}(\mathbf{x})+\iota_{\mathcal{B}_2(\mathbf{y},\epsilon)}(\mathbf{\Phi}\mathbf{x})\nonumber\\
 & + \iota_{\mathrm{epi}_{\|\cdot\|_\ast}}(\mathbf{E}\mathbf{P}^{(1)}\mathbf{D}(\mathbf{C}_0 \otimes \mathbf{I} )\mathbf{x},\mathbf{z}).
\end{align}
In this paper, we construct the solver for \eqref{eq:DSTVminepi} by PDS. The detail algorithm is summarized in Algorithm \ref{alg:DSTVminepi}, where
\begin{align} \label{eq:DSTVpds}
\mathbf{p} =&\  \begin{bmatrix}
\mathbf{x}^\top & \mathbf{z}^\top 
\end{bmatrix}^\top, G(\mathbf{p}) = \iota_{[0, 1]^{3N}}(\mathbf{x}), \nonumber\\
	H(\mathbf{q}) =&\  \|\mathbf{q}_1\|\textcolor{black}{_{2,1}^{(w,2,4)}} + \iota_{\mathcal{B}_2(\mathbf{y},\epsilon)}(\mathbf{q}_2) + \iota_{\mathrm{epi}_{\|\cdot\|_\ast}}(\mathbf{q}_3,\mathbf{q}_4),\nonumber\\ \mathbf{q} =&\ \begin{bmatrix}
\mathbf{q}_1^\top & \mathbf{q}_2^\top & \mathbf{q}_3^\top & \mathbf{q}_4^\top
\end{bmatrix}^\top = \mathbf{F}\mathbf{p}, \nonumber\\
	\mathbf{F}=&\  
	\begin{bmatrix}
		\mathbf{O} & \mathbf{\Phi}^\top & (\mathbf{E}\mathbf{P}^{(1)}\mathbf{D}(\mathbf{C}_0 \otimes \mathbf{I}))^\top & \mathbf{O}\\
		\mathbf{I} & \mathbf{O} & \mathbf{O} & \mathbf{I}\\
	\end{bmatrix}^\top.
\end{align}

\begin{algorithm}[t]
	\caption{Solver for \eqref{eq:DSTVminepi} by PDS}
	\label{alg:DSTVminepi}
	\begin{algorithmic}[1]
		{\scriptsize
			\STATE set $n=0$ and choose initial parameters for $\mathbf{p}^{(0)}$, $\mathbf{q}^{(0)}$.
			\WHILE{$\|{\mathbf{p}}^{(n)} - {\mathbf{p}}^{(n-1)}\|_2 > \epsilon_{\mathrm{stop}}$}
			\STATE $\widetilde{\mathbf{p}}^{(n)} :=  \begin{bmatrix}
\widetilde{\mathbf{x}}^{(n)\top} & \widetilde{\mathbf{z}}^{(n)\top}
\end{bmatrix}^\top = \mathbf{p}^{(n)} - \gamma_1 \mathbf{F}^{\top}\mathbf{q}^{(n)}$
			\STATE $\mathbf{p}^{(n+1)} = \begin{bmatrix}
\mathrm{prox}_{\iota_{[0, 1]^N}}(\widetilde{\mathbf{x}}^{(n)})^\top & \widetilde{\mathbf{z}}^{(n)\top}
\end{bmatrix}^\top$
			\STATE $\mathbf{t}^{(n)} = \mathbf{q}^{(n)}+\gamma_2 \mathbf{F}(2\mathbf{p}^{(n+1)}-\mathbf{p}^{(n)})$
			\STATE $\mathbf{q}_1^{(n+1)}=\mathbf{t}_1^{(n)}-\gamma_2\mathrm{prox}_{\frac{1}{\gamma_2}\|\cdot\|_{\textcolor{black}{2,1}}^{(w,2,4)} }\left(\frac{1}{\gamma_2}\mathbf{t}_{1}^{(n)}\right)$.
			\STATE $\mathbf{q}_2^{(n+1)}=\mathbf{t}_2^{(n)}-\gamma_2\mathrm{prox}_{\frac{1}{\gamma_2}\iota_{\mathcal{B}_2(\mathbf{y}, \epsilon)}}\left(\frac{1}{\gamma_2}\mathbf{t}_{2}^{(n)}\right)$.
						\STATE $(\mathbf{q}_3^{(n+1)}, \mathbf{q}_4^{(n+1)}) = (\mathbf{t}_3^{(n)},\mathbf{t}_4^{(n)})-\gamma_2\mathrm{prox}_{\frac{1}{\gamma_2}\mathrm{epi}_{\|\cdot\|_\ast}}\left(\frac{1}{\gamma_2}(\mathbf{t}_3^{(n)},\mathbf{t}_4^{(n)})\right)$.
			\STATE $n=n+1$.
			\ENDWHILE
			\STATE Output $\mathbf{x}^{(n)}$.}
	\end{algorithmic}
\end{algorithm}
The proximity operator of $\|\cdot\|_{\textcolor{black}{2,1}}^{(w,2,4)}$ is a group thresholding operator given in \cite{Ono2014DVTV} and that of $\iota_{\mathcal{B}_2(\mathbf{y}, \epsilon)}$ is in Table \ref{tab:complexity}.
 The epigraphical projection of the nuclear norm ($\mathcal{P}_{\mathrm{epi}_{\|\cdot\|_{S_1}}}$ in \eqref{eq:Sp}) requires the computation of the epigraphical projection of the $\ell_1$-norm ($\mathcal{P}_{\mathrm{epi}_{\|\cdot\|_{1}}}$). Although the characterization of the epigraphical projection of the $\ell_1$-norm by a certain thresholding parameter $\lambda^\star$, as in \eqref{eq:epil1}, is shown in \cite{Beck2017}, the explicit form of $\lambda^\star$ is not. The closed-form solution $\lambda^\star$ is given in the following proposition.
\begin{prop}\label{prop:lambda_solution}
Let $\lambda^\star$ be a positive root of the function \eqref{eq:phi_lambda}, and $\rho: \{1,\ldots,N\} \rightarrow \{1,\ldots,N\}$ be a mapping to permute the elements of $\mathbf{x}\in \mathbb{R}^N$ in descending order in terms of the absolute value ($|x_{\rho(1)}| \geq \ldots \geq |x_{\rho(N)}|$). $\lambda^\star$ can be represented as
\begin{align}\label{eq:lambda_solution}
\lambda^\star =&\  \begin{cases}
-\textcolor{black}{\xi} & (\textcolor{black}{\xi} < -|x_{\rho(1)}|) \\
\frac{S_{N_0}-\textcolor{black}{\xi}}{N_0+1} & (\widehat{S}_{N_0,0} \leq  \textcolor{black}{\xi} < \widehat{S}_{N_0,1})
\end{cases},\nonumber\\
\widehat{S}_{N_0,m} =&\  S_{N_0}-(N_0+1)|x_{\rho(N_0 + m)}|\ (m =0,1),
\end{align}
where $1\leq N_0 \leq N$ and $S_{N_0} = \sum^{N_0}_{n=1}|x_{\rho(n)}|$. We assume that $x_{\rho(N+1)}=0$.
\begin{proof}
See Appendix \ref{ap:lambda_solution}.
\end{proof}
\end{prop}
\textcolor{black}{Regarding the computational complexity of the DSTV minimization by the PDS algorithm \eqref{eq:DSTVpds}, the orders for the matrix multiplication by $\mathbf{F}$ and the proximity operators are $\mathcal{O}(MN)$ and $\mathcal{O}(W^2N)$, respectively.}
\subsection{Amplitude Spectrum Nuclear Norm Promoting and Its Application to Frequency-domain RPCA}\label{subsec:FRPCA}
\begin{figure}[t]
	\centering
		\begin{minipage}{0.3\linewidth}
		\centering
		\scalebox{0.17}{\includegraphics[keepaspectratio=true]{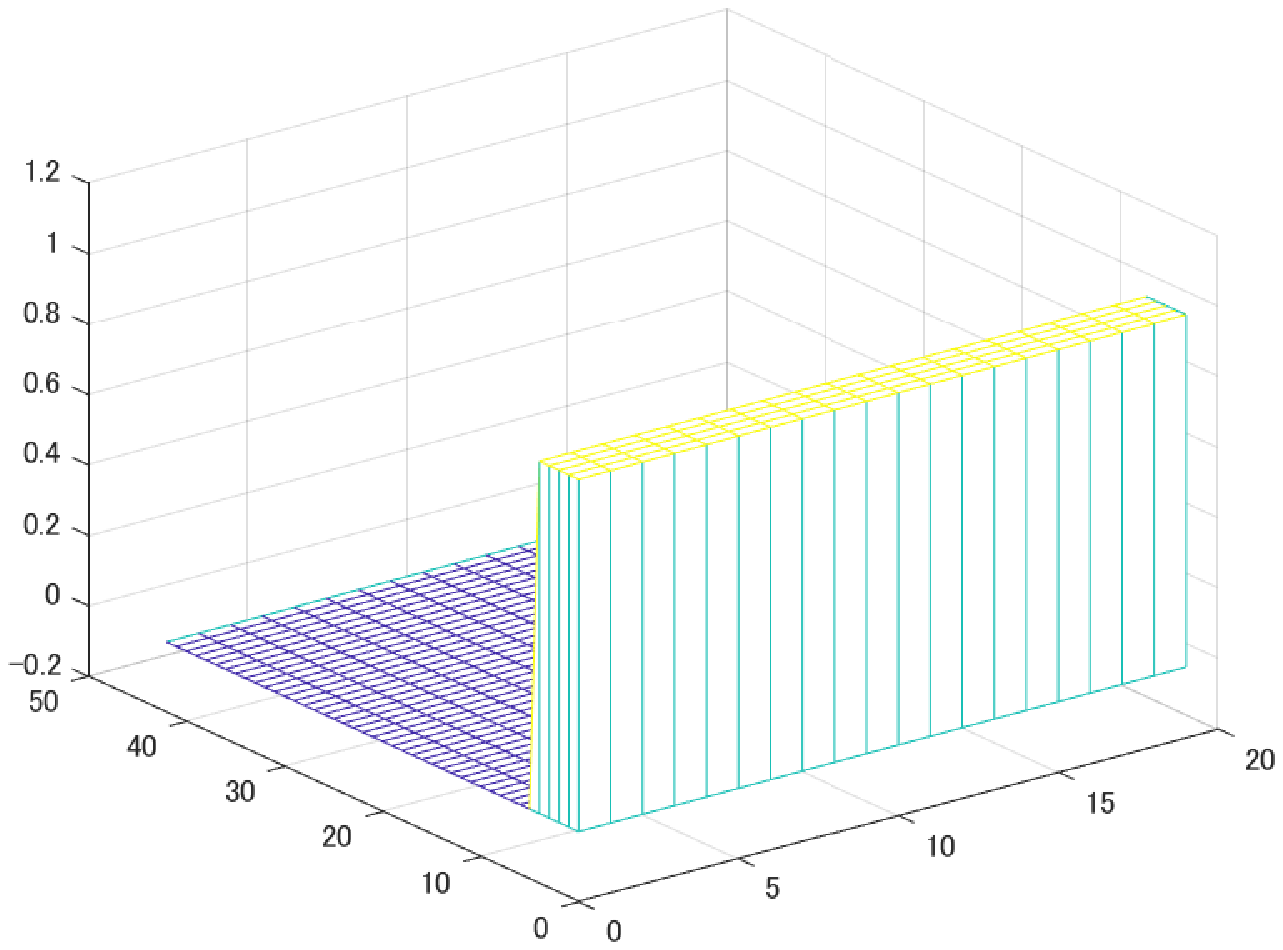}}
				\centerline{(a) Shift:0, $\widehat{\mathbf{L}}_0$}
	\end{minipage}
			\begin{minipage}{0.3\linewidth}
		\centering
		\scalebox{0.17}{\includegraphics[keepaspectratio=true]{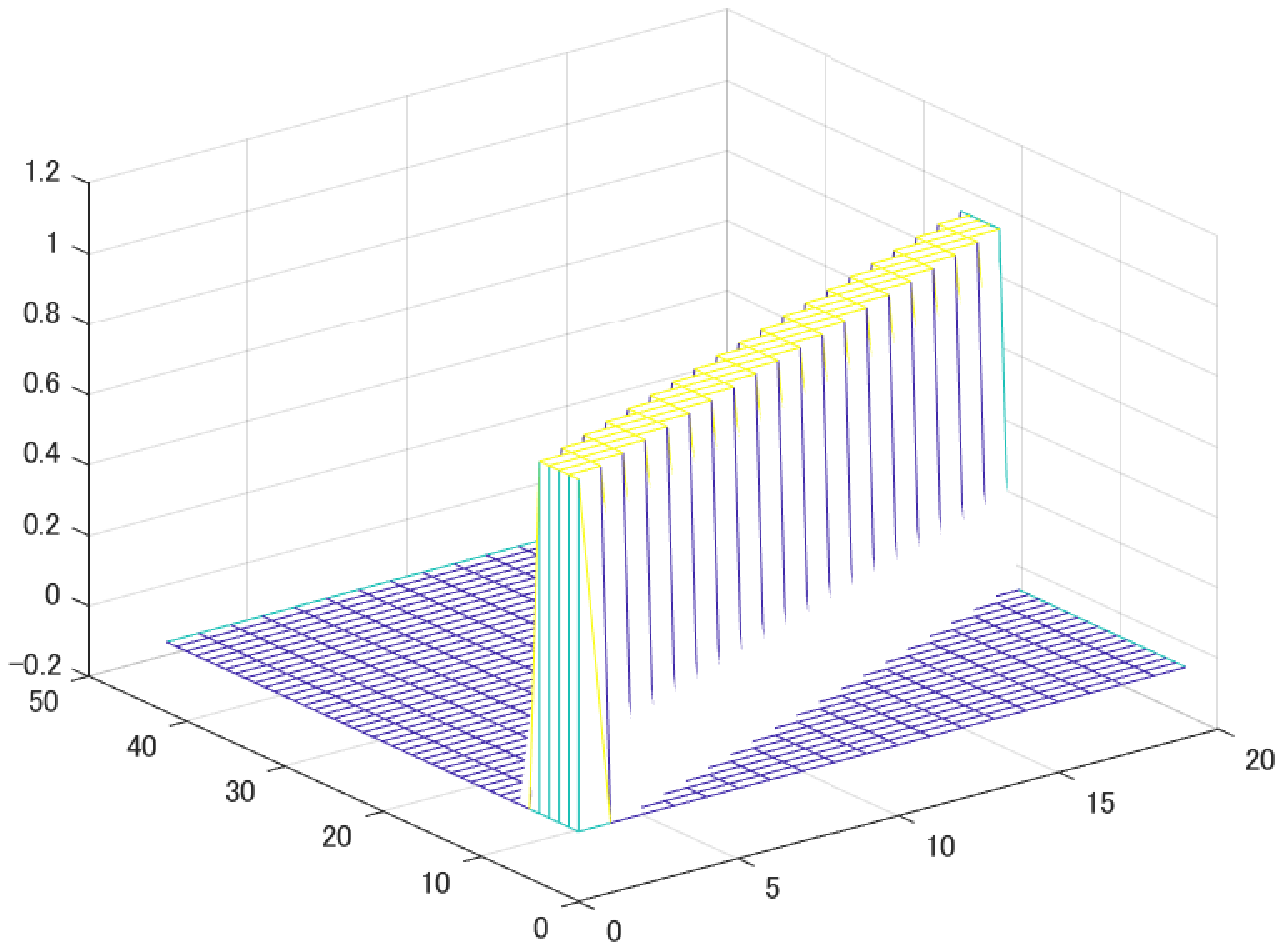}}
				\centerline{(b) Shift:1, $\widehat{\mathbf{L}}_1$}
	\end{minipage}
	\begin{minipage}{0.3\linewidth}
		\centering
		\scalebox{0.17}{\includegraphics[keepaspectratio=true]{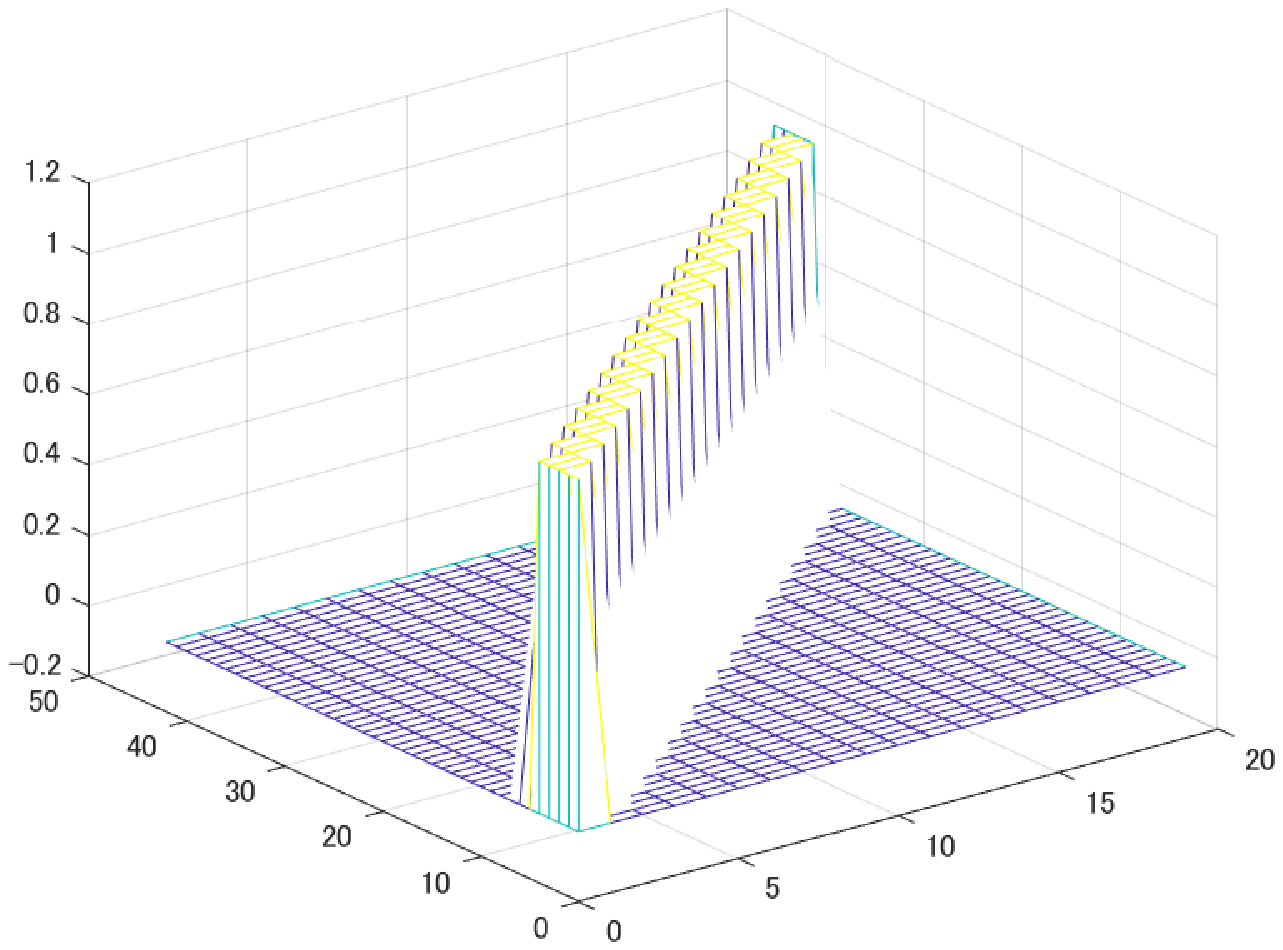}}
				\centerline{(c) Shift:2, $\widehat{\mathbf{L}}_2$}
	\end{minipage}
	\caption{Examples of target signals.}\label{fig:shiftsignal}
\end{figure}

As well as sparse modeling, low-rank modeling is also one of the most effective approaches that can analyze and extract \textcolor{black}{isomorphic components}\footnote{\textcolor{black}{In this paper, we denote isomorphic components that one component corresponds to the others through the shift deformation.}} from observations \cite{Wright2009,Candes2011,Liu2013,Ono2014BNN, Ono2016LCNN,Ono2016, Song2017, Yin2018,Tivive2019}. For example, robust PCA (RPCA) \cite{Wright2009,Candes2011,Song2017,Tivive2019} can extract \textcolor{black}{isomorphic components} lying a low-dimensional subspace from an observed matrix in the presence of outliers. An example of \textcolor{black}{1-dimensional (1D) isomorphic components} $\widehat{\mathbf{L}}_0 \in \mathbb{R}^{43\times 20}$ is shown in Fig. \ref{fig:shiftsignal}(a), where $m$-th column of $\widehat{\mathbf{L}}_s$ is set as
\begin{align}
[\widehat{\mathbf{L}}_s]_{m,n} = \begin{cases}
1 & (s(n-1) + 1\leq m \leq s(n-1) + 5) \\
0 & (\mathrm{otherwise})
\end{cases}.
\end{align} 
The problem formulation for RPCA is given as follows \cite{Candes2011}:
\begin{align}
\{\mathbf{L}^\star,\ \mathbf{S}^\star\} \in \argmin_{\mathbf{L},\ \mathbf{S}} \|\mathbf{L}\|_\ast + \phi(\mathbf{S}) \ \ \mathrm{s.t.} \ \ \mathbf{X} = \mathbf{L} + \mathbf{S},
\end{align}
where $\phi$ is a sparsity-promoting function, e.g., the $\ell_1$-norm and the $\ell_1$-norm ball constraint\footnote{Although the original RPCA \cite{Candes2011} employs the $\ell_1$-norm as $\phi$, we use the $\ell$-norm ball constraint for experiments in Sec. \ref{subsec:expFRPCA} for fair performance comparison between the nuclear norm and the ASNN.}.

One of the drawback of RPCA is the sensitivity against the shift or the misalignment of target components. Fig. \ref{fig:shiftsignal}(b) and (c) show target components shifted every one and two sample $\widehat{\mathbf{L}}_1,\ \widehat{\mathbf{L}}_2 \in \mathbb{R}^{43\times 20}$, respectively.
When target components are not aligned exactly due the shift (Fig. \ref{fig:shiftsignal}(b) and (c)), they are not lying low-dimensional subspace any more. 
Consequently, RPCA cannot extract the \textcolor{black}{isomorphic components} precisely. 

To resolve this drawback, we utilize the fundamental signal processing theory: the amplitude spectra of an original and its (circularly) shifted signals are the same. For example, for given $\widehat{\mathbf{L}}_s = \begin{bmatrix} \widehat{{\bm \ell}}_{s,1} & \cdots & \widehat{{\bm \ell}}_{s,N} \end{bmatrix} \in \mathbb{R}^{M \times N}$, 
$
|\mathbf{W}\widehat{{\bm \ell}}_{s,n_1}| = |\mathbf{W}\widehat{{\bm \ell}}_{s,n_2}|,
$
where $\mathbf{W}$ denotes the \textcolor{black}{nomalized} FFT matrix ($\textcolor{black}{\tfrac{1}{\sqrt{M}}}[\mathbf{W}]_{m,n} = \textcolor{black}{\tfrac{1}{\sqrt{M}}}\exp\left( -\mathrm{j} \frac{2\pi}{M}mn\right)$) and the definition $|\cdot|$ for vectors and matrices is given in Table \ref{tab:Notations}.
According to this nature, we design ASNN $\|\cdot\|_{\mathrm{ASNN}} : \mathbb{R}^{M\times N} \rightarrow \mathbb{R}$ \textcolor{black}{for a set of 1D signals $\mathbf{X}=\begin{bmatrix} \mathbf{x}_1 & \cdots & \mathbf{x}_{N} \end{bmatrix}$} as:
\begin{align}
\|\mathbf{X}\|_{\mathrm{ASNN}} := \||\mathbf{W}\mathbf{X}|\|_{\ast},
\end{align}
\textcolor{black}{where $|\mathbf{W}\mathbf{X}|$ denotes the amplitude spectra of input 1D signals $\mathbf{X}$}. The ASNN takes small value when columns in $\mathbf{X}$ have an isomorphic structure even in the presence of shift. 
\begin{figure}[t]
	\centering
		\begin{minipage}{1\linewidth}
				\centering
		\scalebox{0.5}{\includegraphics[keepaspectratio=true]{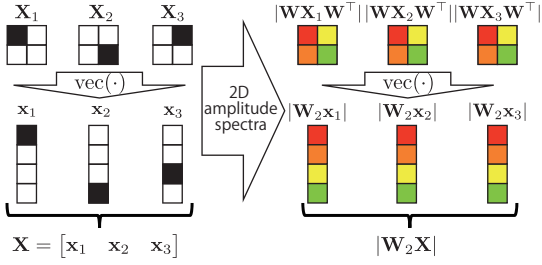}}
	\end{minipage}	
	\caption{\textcolor{black}{(Vectorized) amplitude spectra for 2D signals.}}\label{fig:nd_AS}
	\vspace{-3mm}
\end{figure}

Now we modify the cost function of RPCA by using the ASNN, which is termed as frequency-domain RPCA ($\mathcal{F}$-RPCA), as:
\begin{align}\label{eq:FRPCA}
\{\mathbf{L}^\star,\ \mathbf{S}^\star\} \in&\ \argmin_{\mathbf{L},\ \mathbf{S}} \|\mathbf{X}\|_{\mathrm{ASNN}}  + \phi(\mathbf{S}) \ \ \mathrm{s.t.} \ \ \mathbf{X} = \mathbf{L} + \mathbf{S} \nonumber\\
 = &\ \argmin_{\mathbf{L},\ \mathbf{S}} \||\mathbf{W}\mathbf{L}|\|_\ast + \phi(\mathbf{S}) +\iota_{\{\mathbf{X}\}}(\mathbf{L} + \mathbf{S})\nonumber\\
  = &\ \argmin_{\mathbf{L},\ \mathbf{S}} \||\mathbf{W}\mathbf{L}|\|_\ast + \widetilde{\phi}(\mathbf{L},\mathbf{S},\mathbf{X}),
\end{align}
where $\widetilde{\phi}(\mathbf{L},\mathbf{S},\mathbf{X}) = \phi(\mathbf{S}) +\iota_{\{\mathbf{X}\}}(\mathbf{L} + \mathbf{S})$. Note that the absolute value of a complex number can be considered as the $\ell_2$ norm of the real and imaginary parts:
$
|a + \mathrm{j}b| = \sqrt{a^2 + b^2} = \left\| \begin{bmatrix} a & b \end{bmatrix}^\top \right\|_2
$. With this in mind, $|\mathbf{W}\mathbf{L}|$ can be reformulated as:
\begin{align}\label{eq:probFRPCA}
\||\mathbf{W}\mathbf{L}|\|_{\ast} =&\ \||(\mathbf{W}^{[c]}-\mathrm{j}\mathbf{W}^{[s]})\mathbf{L}|\|_{\ast} \nonumber\\ 
=&\ \left\| \mathbf{P}^{(2)}  \begin{bmatrix} \mathbf{W}^{[c]\top} & -\mathbf{W}^{[s]\top} \end{bmatrix}^\top \mathbf{L}  \right\|_{\textcolor{black}{2,\ast}}= \left\| \mathbf{T} \mathbf{L}  \right\|_{\textcolor{black}{2,\ast}},
\end{align}
where $[\mathbf{W}^{[c]}]_{m,n} = \tfrac{1}{\sqrt{M}}\cos\left( \frac{2\pi}{M}mn\right)$, $[\mathbf{W}^{[s]}]_{m,n} = \tfrac{1}{\sqrt{M}}\sin\left( \frac{2\pi}{M}mn\right)$, $\mathbf{P}^{(2)}$ is a permutation matrix, and $\|\cdot\|_{\textcolor{black}{2,\ast}} : \mathbb{R}^{2M\times N} \rightarrow \mathbb{R}$ is defined as:
\begin{align}
\mathbf{X}:=&\
\begin{bmatrix} 
\mathbf{x}_{1,1} & \cdots & \mathbf{x}_{1,N} \\
\vdots  & \ddots & \vdots   \\
\mathbf{x}_{M,1} & \cdots & \mathbf{x}_{M,N}
\end{bmatrix},\ (\mathbf{x}_{m,n} \in \mathbb{R}^2)\nonumber\\
\|\mathbf{X}\|_{\textcolor{black}{2,\ast}} :=&\ \|g_{\|\cdot\|_2}(\mathbf{X})\|_\ast,\nonumber\\
g_{\|\cdot\|_2}(\mathbf{X}):=&\
\begin{bmatrix} 
\| \mathbf{x}_{1,1}\|_2 & \cdots & \|\mathbf{x}_{1,N}\|_2  \\
\vdots  & \ddots & \vdots   \\
\| \mathbf{x}_{M,1}\|_2 & \cdots & \|\mathbf{x}_{M,N}\|_2  
\end{bmatrix}.
\end{align}
As shown in Sec. \ref{subsec:MMN}, the nuclear norm is not a non-decreasing function on $\mathbb{R}_+^{M\times N}$. Thus, $\left\| \cdot  \right\|_{\textcolor{black}{2,\ast}}$ is not even a convex function.

To find an approximated solution for the problem \eqref{eq:FRPCA}, we introduce ERx as follows:
\begin{align}\label{eq:FRPCAERx}
 &\ \argmin_{\mathbf{L},\ \mathbf{S}} \left\| \mathbf{T} \mathbf{L}  \right\|_{\textcolor{black}{2,\ast}} + \widetilde{\phi}(\mathbf{L},\mathbf{S},\mathbf{X}) \nonumber\\
= &\ \argmin_{\mathbf{L},\ \mathbf{S}} \left\| \mathbf{Z}  \right\|_{\ast} + \widetilde{\phi}(\mathbf{L},\mathbf{S},\mathbf{X}) \ \mathrm{s.t.}\ g_{\|\cdot\|_2}\left(\mathbf{T} \mathbf{L}\right) = \mathbf{Z}, \nonumber\\
\xrightarrow{\mathrm{ERx}} &\ \argmin_{\mathbf{L},\ \mathbf{S}} \left\| \mathbf{Z}  \right\|_{\ast} + \widetilde{\phi}(\mathbf{L},\mathbf{S},\mathbf{X})+\iota_{\mathrm{epi}_{\|\cdot\|_2}}\left(\mathbf{T}  \mathbf{L}, \mathbf{Z}\right),\nonumber\\
\{\textcolor{black}{\widetilde{\bm \ell}},\ \widetilde{\mathbf{s}}\} \in &\ \argmin_{\textcolor{black}{\bm \ell},\ \mathbf{s}} \left\| \mathbf{z}  \right\|_{\ast} + \widetilde{\phi}(\textcolor{black}{\bm \ell},\mathbf{s}, \mathbf{x}) +\iota_{\mathrm{epi}_{\|\cdot\|_2}}\left(\widehat{\mathbf{T}} \textcolor{black}{\bm \ell}, \mathbf{z}\right),
\end{align}
where $\mathbf{x} = \mathrm{vec}(\mathbf{X})$, ${\bm \ell} = \mathrm{vec}(\mathbf{L})$, $\mathbf{s} = \mathrm{vec}(\mathbf{S})$, and $\widehat{\mathbf{T}} = \mathbf{I} \otimes \mathbf{T}$. Since the objective function in the last line of \eqref{eq:FRPCAERx} is convex, the global minimizer can be found by the PDS algorithm. Note that the problem \eqref{eq:FRPCAERx} is obtained by epigraphical relaxation, the minimizer for \eqref{eq:FRPCAERx} is not the one for \eqref{eq:FRPCA}.
The minimization problem can be \textcolor{black}{solved} by the PDS algorithm as
\begin{align} 
\mathbf{p} =&\  \begin{bmatrix}
 {\bm \ell}^\top &  \mathbf{s}^\top & \mathbf{z}^\top 
\end{bmatrix}^\top,\ G(\mathbf{p}) = \mathbf{0}, \nonumber\\
	H(\mathbf{q}) =&\ \|\mathbf{q}_1\|_\ast + \phi(\mathbf{q}_2) +\iota_{\{\mathbf{x}\}}(\mathbf{q}_3) + \iota_{\mathrm{epi}_{\|\cdot\|_2}}(\mathbf{q}_4,\mathbf{q}_5), \nonumber\\
	\mathbf{q} =&\  \mathbf{F}\mathbf{p},\  \mathbf{F}=  
	\begin{bmatrix}
		\mathbf{O} & \mathbf{O} & \mathbf{I} & \widehat{\mathbf{T}}^\top & \mathbf{O} \\
		\mathbf{O} & \mathbf{I} & \mathbf{I} & \mathbf{O} & \mathbf{O}\\
		\mathbf{I}  & \mathbf{O} & \mathbf{O} & \mathbf{O} & \mathbf{I} \\
\end{bmatrix}^\top.
\end{align}
The detailed algorithm is summarized in Algorithm \ref{alg:FRPCA}, \textcolor{black}{where the proximity operator of the nuclear norm $\|\cdot\|_{\ast}$ is in Table \ref{tab:complexity} and that of $\iota_{\mathrm{epi}_{\|\cdot\|_2}}$ is computed by \eqref{eq:epil2}}.

\textcolor{black}{The computation for the $\mathcal{F}$-RPCA is dominated by the SVD calculation for the proximity operator of the nuclear norm. The order of the computational complexity is bounded by $\mathcal{O}(MN\min\{M,N\})$.}
\begin{Rem}\upshape 
\textcolor{black}{Since using the nuclear norm as the second-layer function violates the assumption (A2), the ASNN does not satisfy Theorem \ref{theo:ER}. Thus, ERx act as the convex relaxation on \eqref{eq:FRPCA} and the minimizer of \eqref{eq:FRPCAERx} is an estimate for \eqref{eq:FRPCA}}.
\end{Rem}
\begin{Rem}\upshape 
We can easily extend the ASNN in the 1D signal case to the multidimensional (MD) signal case. We define ASNN for given $D$-dimensional array tensors $\{\mathcal{X}_n\}_{n=1}^{N} \subset \mathbb{R}^{N_1 \times N_2 \times \cdots \times N_{D}}$ as
\begin{align}
\|\mathcal{X}\|_{\mathrm{ASNN}} :=&\  \||\mathbf{W}_D\mathbf{X}|\|_{\ast}= \left\| \mathbf{T} \mathbf{X}  \right\|_{\textcolor{black}{2,\ast}}, \nonumber\\
\mathbf{X} =&\  \begin{bmatrix} \mathrm{vec}(\mathcal{X}_1) &\cdots& \mathrm{vec}(\mathcal{X}_{N})\end{bmatrix} \in \mathbb{R}^{\prod_{d=1}^{D} N_d \times \textcolor{black}{N}}, \nonumber\\
\mathbf{W}_D =&\  \mathbf{W} \otimes \cdots \otimes \mathbf{W}, \ 
\mathbf{T} =  \mathbf{P}^{(3)}  \begin{bmatrix} \mathbf{T}_r^\top & \mathbf{T}_i^\top \end{bmatrix}^\top,
\end{align}
where $\mathbf{T}_r$ and $\mathbf{T}_i$ are the combination of the Kronecker products of $\mathbf{W}^{[c]}$ and $\mathbf{W}^{[s]}$, and $\mathbf{P}^{(3)}$ is a permutation matrix (we assume $N_1 = \cdots = N_D = M$). For example $D=2$, \textcolor{black}{the construction of $|\mathbf{W}_2\mathbf{X}|$ is illustrated in Fig. \ref{fig:nd_AS}, where}
\begin{align}
\mathbf{W}_2 =&\  \mathbf{W} \otimes \mathbf{W} = (\mathbf{W}^{[c]}-j\mathbf{W}^{[s]}) \otimes (\mathbf{W}^{[c]}-j\mathbf{W}^{[s]}), \nonumber\\
  \mathbf{T}_r=&\ (\mathbf{W}^{[c]} \otimes \mathbf{W}^{[c]}) - (\mathbf{W}^{[s]} \otimes \mathbf{W}^{[s]}),\nonumber\\ 
  \mathbf{T}_i=&\ - ((\mathbf{W}^{[c]} \otimes \mathbf{W}^{[s]}) + (\mathbf{W}^{[s]} \otimes \mathbf{W}^{[c]})).
\end{align}
\end{Rem}
\begin{Rem}\upshape 
\textcolor{black}{Tensor nuclear norm (TNN) based on t-SVD has been proposed for evaluating low-rankness of tensors \cite{Semerci2014}. The TNN is analogous to the ASNN because DFT is employed in its calculation. However, the usages of DFT in the TNN and the ASNN are different, and also, both norms evaluate different perspectives in low-rankness as in the following.
\begin{enumerate}
\item DFT is employed in the TNN for fast calculation. Since the mode-3 DFT results in the block-wise diagonalization of a block-circulant matrix, we can circumvent the large-scale SVD calculation. In addition, the TNN evaluates similarity of vectors lying along every mode.  
\item DFT is employed in the ASNN for amplitude spectra calculation. Since MD isomorphic components share an identical amplitude spectrum, the ASNN evaluates the structural similarity of input MD signals.
\end{enumerate}}
\end{Rem}

\begin{algorithm}[t]
	\caption{Solver for \eqref{eq:FRPCAERx}}
	\label{alg:FRPCA}
	\begin{algorithmic}[1]
		{\scriptsize
			\STATE set $n=0$ and choose initial parameters for $\mathbf{p}^{(0)}$, $\mathbf{q}^{(0)}$.
			\WHILE{$\|{\mathbf{p}}^{(n)} - {\mathbf{p}}^{(n-1)}\|_2 > \epsilon_{\mathrm{stop}}$}
			\STATE $\mathbf{p}^{(n+1)} = \mathbf{p}^{(n)} - \gamma_1 \mathbf{F}^{\top}\mathbf{q}^{(n)}$
			\STATE $\mathbf{t}^{(n)} = \mathbf{q}^{(n)}+\gamma_2 \mathbf{F}(2\mathbf{p}^{(n+1)}-\mathbf{p}^{(n)})$
			\STATE $\mathbf{q}_1^{(n+1)}=\mathbf{t}_1^{(n)}-\gamma_2\mathrm{prox}_{\frac{1}{\gamma_2}\|\cdot\|_{\ast} }\left(\frac{1}{\gamma_2}\mathbf{t}_{1}^{(n)}\right)$.
			\STATE $\mathbf{q}_2^{(n+1)}=\mathbf{t}_2^{(n)}-\gamma_2\mathrm{prox}_{\frac{1}{\gamma_2}\phi}\left(\frac{1}{\gamma_2}\mathbf{t}_{2}^{(n)}\right)$.
						\STATE $\mathbf{q}_3^{(n+1)}=\mathbf{t}_3^{(n)}-\gamma_2\mathrm{prox}_{\frac{1}{\gamma_2}\iota_{\{\mathbf{x}\}}}\left(\frac{1}{\gamma_2}\mathbf{t}_{3}^{(n)}\right)$.
						\STATE $(\mathbf{q}_4^{(n+1)}, \mathbf{q}_5^{(n+1)}) = (\mathbf{t}_4^{(n)},\mathbf{t}_5^{(n)})-\gamma_2\mathrm{prox}_{\frac{1}{\gamma_2}\mathrm{epi}_{\|\cdot\|_2}}\left(\frac{1}{\gamma_2}(\mathbf{t}_4^{(n)},\mathbf{t}_5^{(n)})\right)$.
			\STATE $n=n+1$.
			\ENDWHILE
			\STATE Output $\mathbf{x}^{(n)}$.}
	\end{algorithmic}
\end{algorithm}
\section{Experimental Results}
\label{sec:Experimental}
This section evaluates the proposed regularizers, DSTV and the ASNN, in practical applications. DSTV is applied to image recovery in Sec. \ref{subsec:expDSTV} and the ASNN signal decomposition in Sec. \ref{subsec:expFRPCA}. The following experiments were performed using MATLAB (R2019b, 64bit) on Mac OS X (Version 10.15.4) with an Intel Core i7 2.8 GHz quad-core processor and 16 GB LPDDR3 memory. \textcolor{black}{From Sec. \ref{subsec:pds}, it is sufficient for convergence to set the parameters $\gamma_1$ and $\gamma_2$ in the PDS algorithm as small values. In both experiments, we heuristically chose the parameters $\gamma_1 = 0.01$ and $\gamma_2 = \frac{1}{12\gamma_1}$.}
\subsection{Image Recovery by DSTV}\label{subsec:expDSTV}
\begin{figure}[t]
	\centering
			\begin{minipage}{0.23\linewidth}
		\centering
		\scalebox{0.9}{\includegraphics[keepaspectratio=true]{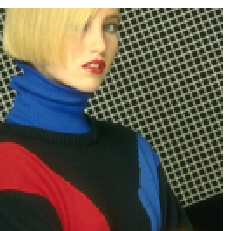}}
				\centerline{(a) Image 1}
	\end{minipage}
			\begin{minipage}{0.23\linewidth}
		\centering
		\scalebox{0.9}{\includegraphics[keepaspectratio=true]{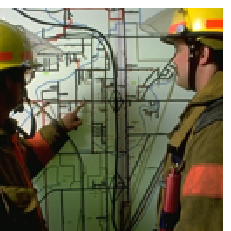}}
				\centerline{(a) Image 2}
	\end{minipage}
			\begin{minipage}{0.23\linewidth}
		\centering
		\scalebox{0.9}{\includegraphics[keepaspectratio=true]{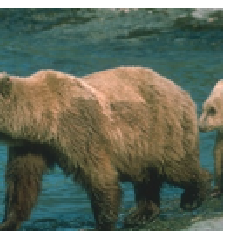}}
				\centerline{(a) Image 3}
	\end{minipage}
			\begin{minipage}{0.23\linewidth}
		\centering
		\scalebox{0.9}{\includegraphics[keepaspectratio=true]{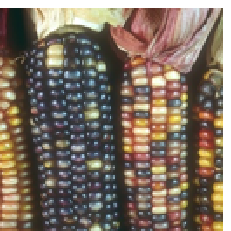}}
				\centerline{(a) Image 4}
	\end{minipage}
\caption{Original images}
	\label{fig:oriim}
\end{figure}
\begin{figure*}[t]
	\centering
			\begin{minipage}{0.13\linewidth}
		\centering
		\scalebox{0.5}{\includegraphics[keepaspectratio=true]{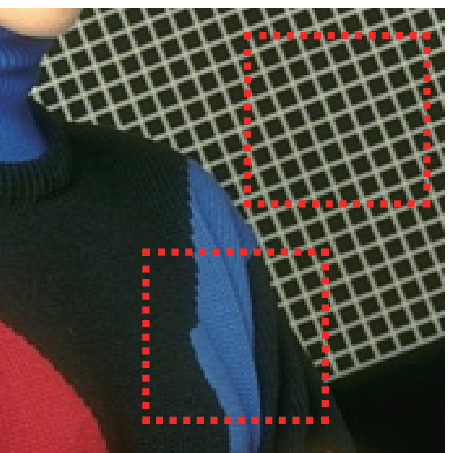}}
				\centerline{(a) Image 1}
	\end{minipage}
			\begin{minipage}{0.13\linewidth}
		\centering
		\scalebox{0.5}{\includegraphics[keepaspectratio=true]{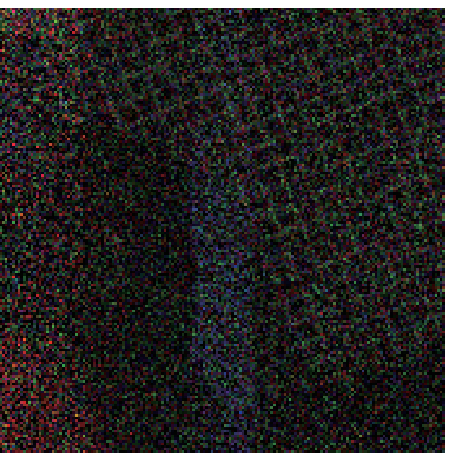}}
				\centerline{(b) Observation}
	\end{minipage}
		\begin{minipage}{0.13\linewidth}
		\centering
		\scalebox{0.5}{\includegraphics[keepaspectratio=true]{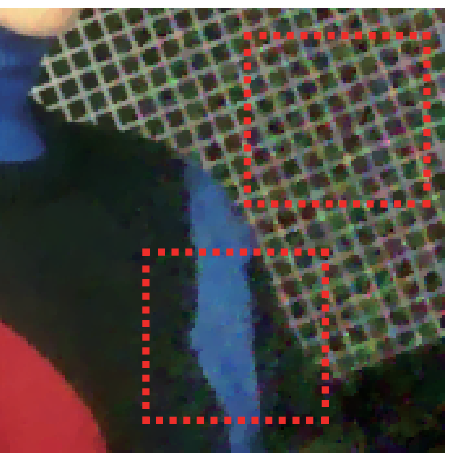}}
				\centerline{(c) VTV}
	\end{minipage}
			\begin{minipage}{0.13\linewidth}
			\centering
		\scalebox{0.5}{\includegraphics[keepaspectratio=true]{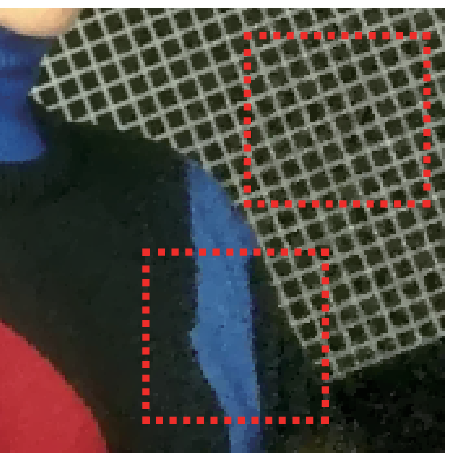}}
				\centerline{(d) DVTV}
	\end{minipage}
			\begin{minipage}{0.13\linewidth}
		\centering
		\scalebox{0.5}{\includegraphics[keepaspectratio=true]{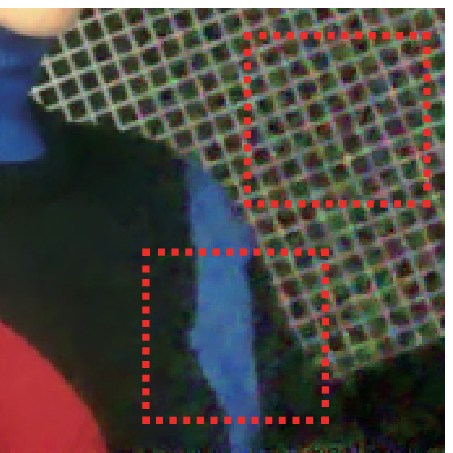}}
				\centerline{(e) STV}
	\end{minipage}
		\begin{minipage}{0.13\linewidth}
		\centering
		\scalebox{0.5}{\includegraphics[keepaspectratio=true]{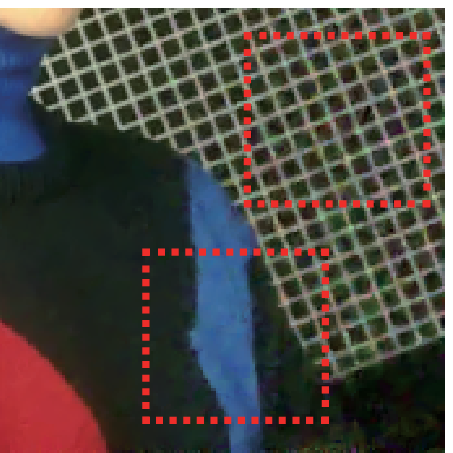}}
				\centerline{(f) ASTV}
	\end{minipage}
			\begin{minipage}{0.13\linewidth}
			\centering
		\scalebox{0.5}{\includegraphics[keepaspectratio=true]{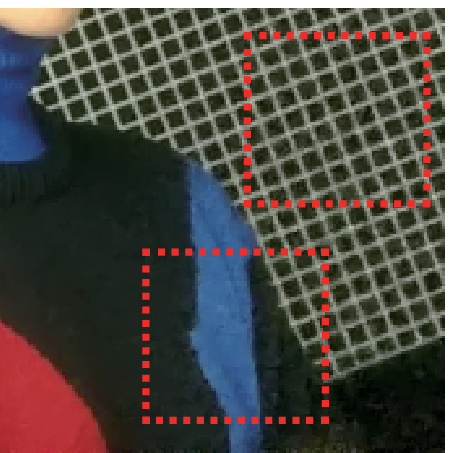}}
				\centerline{(g) DSTV}
	\end{minipage}
		\\
				\begin{minipage}{0.13\linewidth}
		\centering
		\scalebox{0.5}{\includegraphics[keepaspectratio=true]{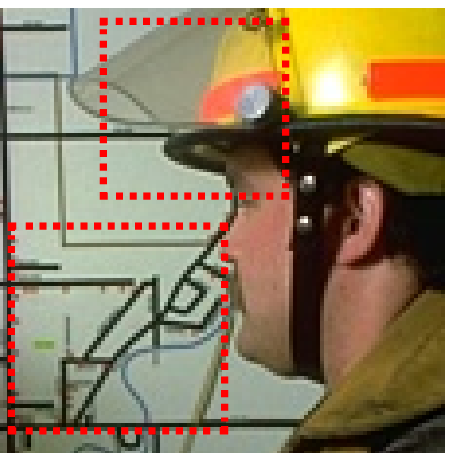}}
				\centerline{(h) Image 2}
	\end{minipage}
			\begin{minipage}{0.13\linewidth}
		\centering
		\scalebox{0.5}{\includegraphics[keepaspectratio=true]{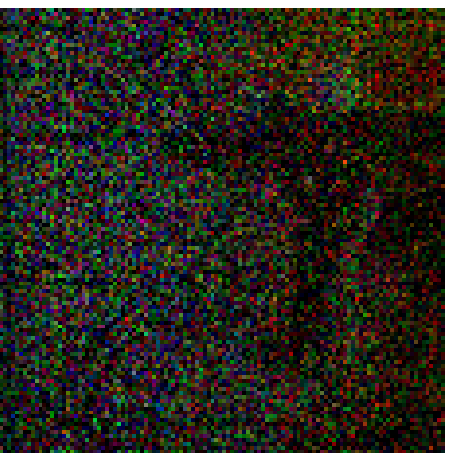}}
				\centerline{(i) Observation}
	\end{minipage}
		\begin{minipage}{0.13\linewidth}
		\centering
		\scalebox{0.5}{\includegraphics[keepaspectratio=true]{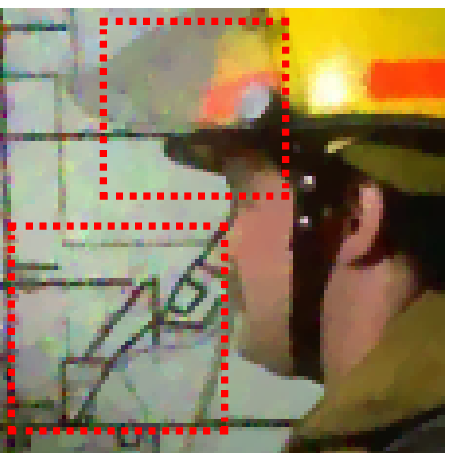}}
				\centerline{(j) VTV}
	\end{minipage}
			\begin{minipage}{0.13\linewidth}
			\centering
		\scalebox{0.5}{\includegraphics[keepaspectratio=true]{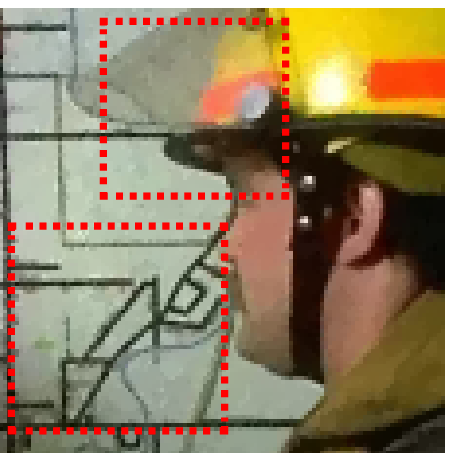}}
				\centerline{(k) DVTV}
	\end{minipage}
			\begin{minipage}{0.13\linewidth}
		\centering
		\scalebox{0.5}{\includegraphics[keepaspectratio=true]{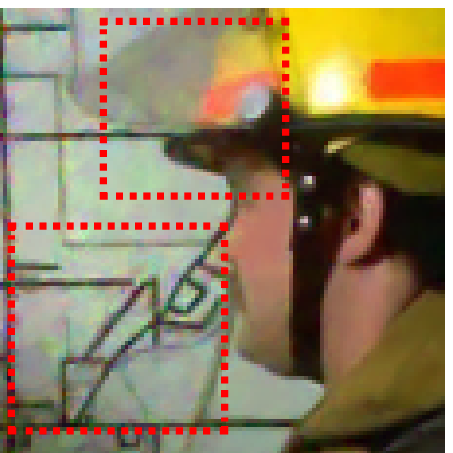}}
				\centerline{(l) STV}
	\end{minipage}
		\begin{minipage}{0.13\linewidth}
		\centering
		\scalebox{0.5}{\includegraphics[keepaspectratio=true]{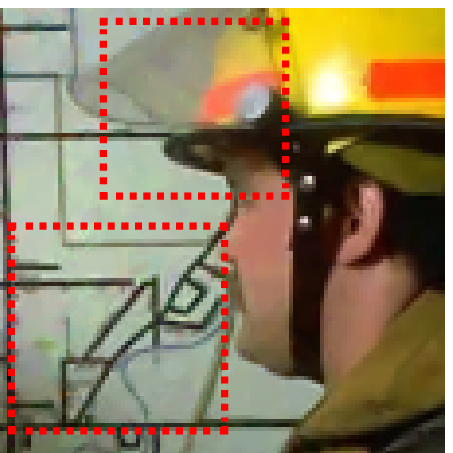}}
				\centerline{(m) ASTV}
	\end{minipage}
			\begin{minipage}{0.13\linewidth}
			\centering
		\scalebox{0.5}{\includegraphics[keepaspectratio=true]{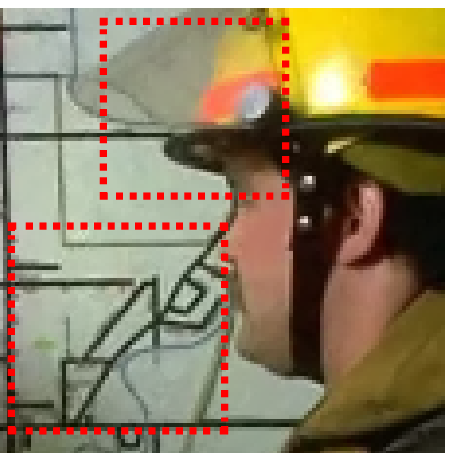}}
				\centerline{(n) DSTV}
	\end{minipage}
			\\
				\begin{minipage}{0.13\linewidth}
		\centering
		\scalebox{0.5}{\includegraphics[keepaspectratio=true]{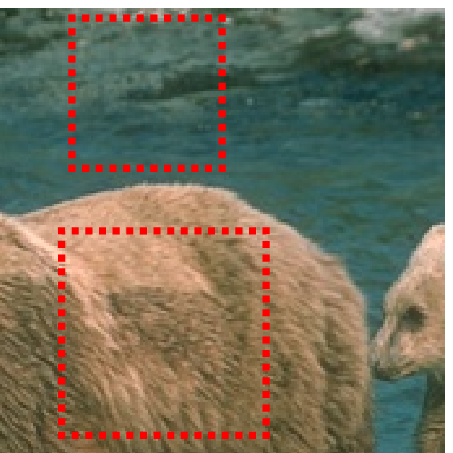}}
				\centerline{(o) Image 3}
	\end{minipage}
			\begin{minipage}{0.13\linewidth}
		\centering
		\scalebox{0.5}{\includegraphics[keepaspectratio=true]{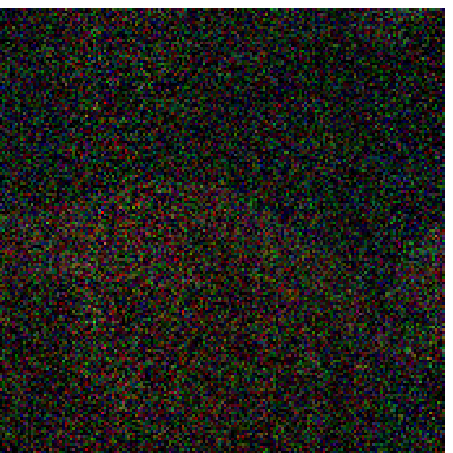}}
				\centerline{(p) Observation}
	\end{minipage}
		\begin{minipage}{0.13\linewidth}
		\centering
		\scalebox{0.5}{\includegraphics[keepaspectratio=true]{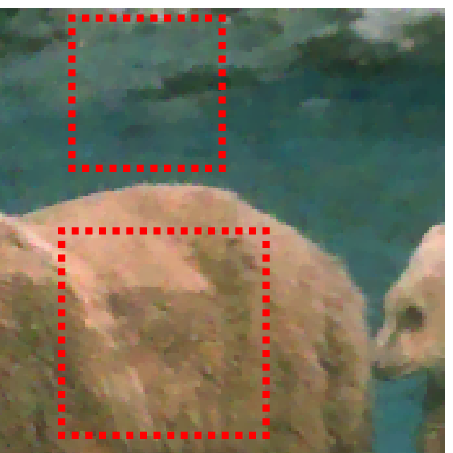}}
				\centerline{(q) VTV}
	\end{minipage}
			\begin{minipage}{0.13\linewidth}
			\centering
		\scalebox{0.5}{\includegraphics[keepaspectratio=true]{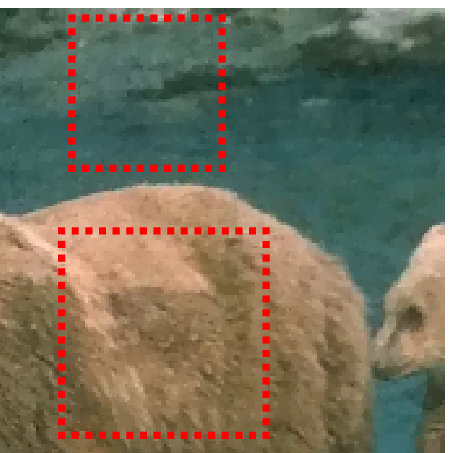}}
				\centerline{(r) DVTV}
	\end{minipage}
	\begin{minipage}{0.13\linewidth}
		\centering
		\scalebox{0.5}{\includegraphics[keepaspectratio=true]{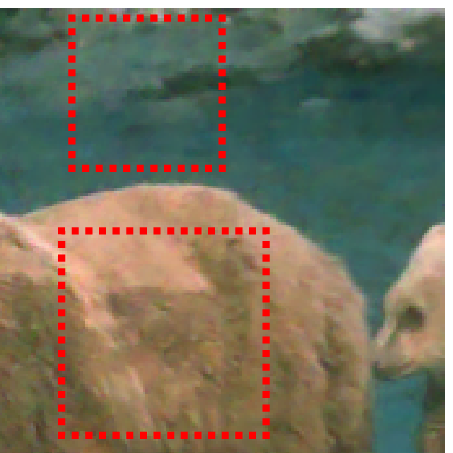}}
				\centerline{(s) STV}
	\end{minipage}
		\begin{minipage}{0.13\linewidth}
		\centering
		\scalebox{0.5}{\includegraphics[keepaspectratio=true]{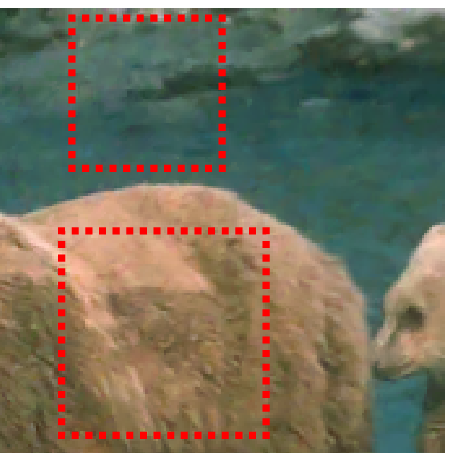}}
				\centerline{(t) ASTV}
	\end{minipage}
			\begin{minipage}{0.13\linewidth}
			\centering
		\scalebox{0.5}{\includegraphics[keepaspectratio=true]{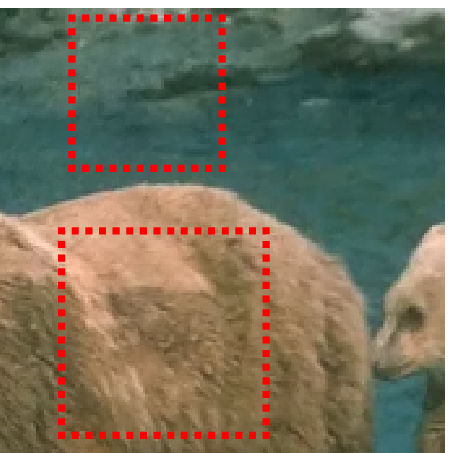}}
				\centerline{(u) DSTV}
	\end{minipage}
				\begin{minipage}{0.13\linewidth}
		\centering
		\scalebox{0.5}{\includegraphics[keepaspectratio=true]{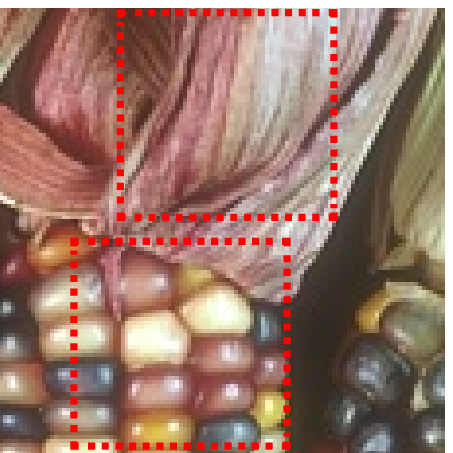}}
				\centerline{(v) Image 4}
	\end{minipage}
			\begin{minipage}{0.13\linewidth}
		\centering
		\scalebox{0.5}{\includegraphics[keepaspectratio=true]{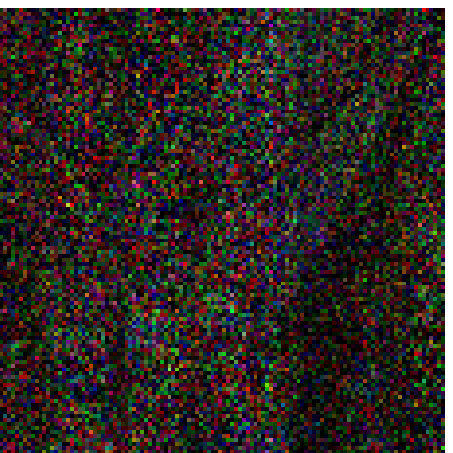}}
				\centerline{(w) Observation}
	\end{minipage}
		\begin{minipage}{0.13\linewidth}
		\centering
		\scalebox{0.5}{\includegraphics[keepaspectratio=true]{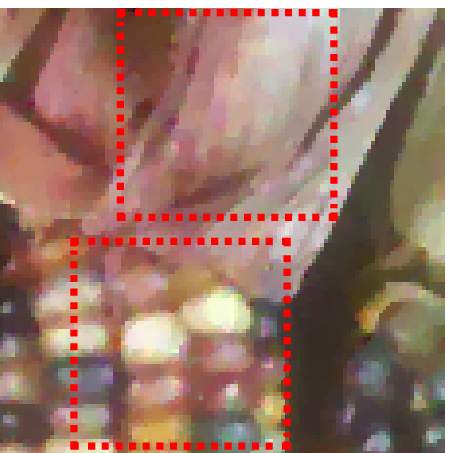}}
				\centerline{(x) VTV}
	\end{minipage}
			\begin{minipage}{0.13\linewidth}
			\centering
		\scalebox{0.5}{\includegraphics[keepaspectratio=true]{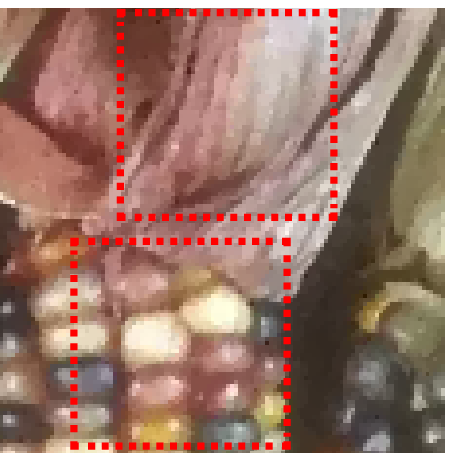}}
				\centerline{(y) DVTV}
	\end{minipage}
	\begin{minipage}{0.13\linewidth}
		\centering
		\scalebox{0.5}{\includegraphics[keepaspectratio=true]{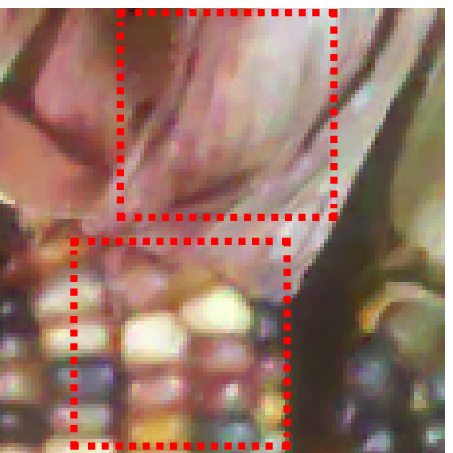}}
				\centerline{(z) STV}
	\end{minipage}
		\begin{minipage}{0.13\linewidth}
		\centering
		\scalebox{0.5}{\includegraphics[keepaspectratio=true]{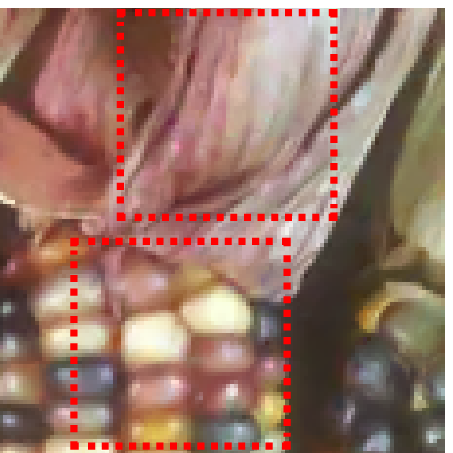}}
				\centerline{(aa) ASTV}
	\end{minipage}
			\begin{minipage}{0.13\linewidth}
			\centering
		\scalebox{0.5}{\includegraphics[keepaspectratio=true]{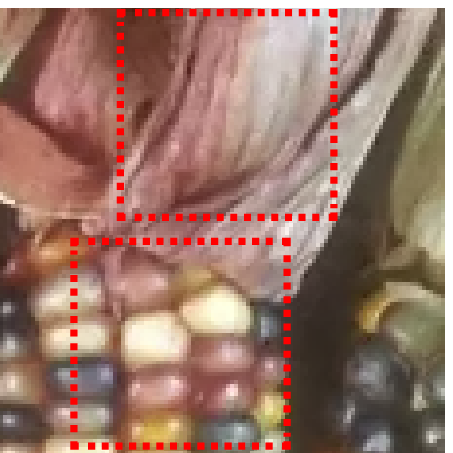}}
				\centerline{(ab) DSTV}
	\end{minipage}
\caption{\textcolor{black}{Reconstructed images of compressed image sensing.}}
	\label{fig:recim}
\end{figure*}
\begin{figure}[t]
\centering
			\begin{minipage}[b]{1\linewidth}
			\centering
		\scalebox{0.5}{\includegraphics[keepaspectratio=true]{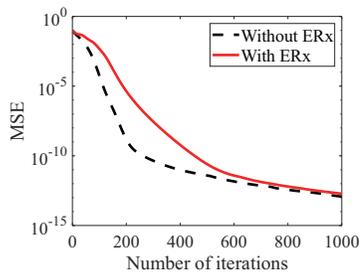}}
	\end{minipage}
	\caption{Profile of MSE: (red) the minimizer of VTVwoERx/the updated solution of VTVwERx, (dashed black) the minimizer and the updated solution of VTVwoERx.}\label{fig:profile}
	\vspace{-3mm}
\end{figure}
We evaluated the performance of the proposed DSTV in compressed sensing reconstructions by using the minimization problem in \eqref{eq:DSTVmin}. We also evaluated VTV \cite{Bresson2008}, DVTV \cite{Ono2014DVTV}, STV \cite{Lefkimmiatis2015}, and ASTV \cite{Ono2016} as conventional methods by replacing DSTV to them in \eqref{eq:DSTVmin}. For STV regularizers, the patch size $W \times W$ is $3 \times 3$. For DVTV and DSTV, we set the weighting parameter $w = 0.5$ in \eqref{eq:DVTVdef} and \eqref{eq:DSTVdef}. Four test images from \textit{Berkeley Segmentation Database} (BSDS300) \cite{Martin2001} shown in Fig. \ref{fig:oriim} were used. The size of the cropped images was set to $256\times 256$. 
In compressed image sensing, each incomplete observation $\mathbf{y}= {\mathbf{\Phi}}\widehat{\mathbf{x}} + \mathbf{n}$ $({\mathbf{\Phi}} := \mathbf{S}\widetilde{\mathbf{\Phi}})$ was obtained by the Noiselet transform \cite{COIFMAN2001} $\widetilde{\mathbf{\Phi}}$ followed by random downsampling $\mathbf{S} \in \mathbb{R}^{L\times 3N}$ $(L=0.2 \times 3N )$ in the presence of additive white Gaussian noise $\mathbf{n}$ with standard derivation $\sigma = 0.1$. The radius $\epsilon$ was set to the oracle value, i.e., $\epsilon = \|\mathbf{\Phi}\widehat{\mathbf{x}} - \mathbf{y}\|_2$.

First, we demonstrate the identity of the minimizers between the VTV minimization problems without and with ERx (VTVwERx and VTVwoERx) by using the test image \textit{Image} 1. Appendix \ref{ap:VTVPDS} describes the PDS algorithms for VTVwERx and VTVwoERx. In Fig. \ref{fig:profile}, the red curve shows the profile of the MSE between the minimizer of VTVwoERx ($\mathbf{x}_{\mathrm{woERx}}^\star$) and the updated solution of VTVwERx ($\mathbf{x}_{\mathrm{wERx}}^{(n)}$) in each iteration ($\frac{1}{3N}\|\mathbf{x}_{\mathrm{wERx}}^{(n)}-\mathbf{x}_{\mathrm{woERx}}^\star\|_2$) and the dashed black curve shows the one between $\mathbf{x}_{\mathrm{woERx}}^\star$ and the updated solution of VTVwoERx ($\mathbf{x}_{\mathrm{woERx}}^{(n)}$) in each iteration ($\frac{1}{3N}\|\mathbf{x}_{\mathrm{woERx}}^{(n)}-\mathbf{x}_{\mathrm{woERx}}^\star\|_2$). The stop criteria $\epsilon_{\mathrm{stop}}$ (as used in Algorithm \ref{alg:DSTVminepi} and \ref{alg:FRPCA}) was set to $\epsilon_{\mathrm{stop}} = 10^{-7}$. As the red curve indicates, the updated solution to VTVwERx is monotonically approaching the minimizer to VTVwoERx. In addition, although the convergence rate of VTVwERx is slightly slower than that of VTVwoERx, the difference is not remarkable as the two curves show. Average times for one iteration for VTVwoERx and VTVwERx are 0.013 [sec] and 0.016 [sec], respectively.

\textcolor{black}{Regarding the computational complexity, all the methods share the same order for the matrix multiplication by $\mathbf{F}$ with $\mathcal{O}(\sqrt{N}\log \sqrt{N})$ derived from the Noiselet transform. On the other hand, the orders for the proximity operator of the methods are summarized in Table \ref{tab:CCprox_ex}. The orders of VTV (wERx/woERx) and DVTV depend on the group-thresholding operation, while the ones of STV, ASTV, DSTV depend on the singular-value thresholding for each patch.}

Next, we show the reconstructed images in Fig. \ref{fig:recim}. DSTV can reduce the false-color and stair-casing artifacts better due to the benefits from decorrelation and STV \textcolor{black}{(particularly in the red-dashed regions)}. As well as subjective quality, DSTV achieves the best reconstruction performance in terms of the reconstruction error (PSNR [dB]).
\begin{table}[t]
\textcolor{black}{
\caption{Numerical erros (PSNR [dB]) in compressive sensing experiments}
	\label{tab:exFRPCA}
	\begin{center}
		\scalebox{1}{
			\begin{tabular}{ccccccc}
				\thline
				\multicolumn{1}{c|}{} &\multicolumn{1}{c|}{VTV}          & \multicolumn{1}{c|}{DVTV}        & \multicolumn{1}{c|}{STV}  & \multicolumn{1}{c|}{ASTV} & \multicolumn{1}{c}{DSTV} \\
				\hline
				\multicolumn{1}{c|}{Image 1} &\multicolumn{1}{c|}{21.44} &\multicolumn{1}{c|}{26.22} & \multicolumn{1}{c|}{21.42} & \multicolumn{1}{c|}{25.49} & \multicolumn{1}{c}{\textbf{27.24}}\\  
				\multicolumn{1}{c|}{Image 2} &\multicolumn{1}{c|}{23.20} & \multicolumn{1}{c|}{27.59} & \multicolumn{1}{c|}{23.34} & \multicolumn{1}{c|}{27.64} & \multicolumn{1}{c}{\textbf{29.14}} \\ 
				\multicolumn{1}{c|}{Image 3} &\multicolumn{1}{c|}{29.14} &\multicolumn{1}{c|}{31.24} & \multicolumn{1}{c|}{29.19} & \multicolumn{1}{c|}{30.98} & \multicolumn{1}{c}{\textbf{32.09}} \\  
				\multicolumn{1}{c|}{Image 4} &\multicolumn{1}{c|}{24.89} &\multicolumn{1}{c|}{27.83} &\multicolumn{1}{c|}{24.98} & \multicolumn{1}{c|}{28.72} & \multicolumn{1}{c}{\textbf{29.17}} \\ 
				\thline
			\end{tabular}
		}
	\end{center}
	\caption{Computational complexity for proximity operator}
	\label{tab:CCprox_ex}
	\begin{center}
		\scalebox{1}{
			\begin{tabular}{cccccc}
				\thline
				\multicolumn{1}{c|}{VTV (wo/w)ERx}          & \multicolumn{1}{c|}{DVTV}        & \multicolumn{1}{c|}{STV}  & \multicolumn{1}{c|}{ASTV} & \multicolumn{1}{c}{DSTV} \\
				\hline
				\multicolumn{1}{c|}{$\mathcal{O}(N)$} &\multicolumn{1}{c|}{$\mathcal{O}(N)$} & \multicolumn{1}{c|}{$\mathcal{O}(W^2N)$} & \multicolumn{1}{c|}{$\mathcal{O}(W^2N)$} & \multicolumn{1}{c}{$\mathcal{O}(W^2N)$}\\  
				\thline
			\end{tabular}
		}
	\end{center}
}
	\vspace{-0.6cm}
\end{table}
\subsection{Frequency-domain Robust PCA}\label{subsec:expFRPCA}
This section demonstrates $\mathcal{F}$-RPCA by the ASNN presented in Sec. \ref{subsec:FRPCA}. In this experiment, we extract target components with an isomorphic structure $\widehat{\mathbf{L}}_s$ from an observed signal corrupted with sparse noise as $\mathbf{X} = \widehat{\mathbf{L}}_s + \widehat{\mathbf{S}}_s$. The target components used in this experiment are $\widehat{\mathbf{L}}_0$, $\widehat{\mathbf{L}}_1$, and $\widehat{\mathbf{L}}_2$ given in Fig. \ref{fig:shiftsignal}. The sparse noise $\widehat{\mathbf{S}}_s \in \{0,1\}^{43 \times 20}$ contains 1s in the region of $\mathcal{N}_s = \{(m,n) \in \{1,2,\ldots,43\}\times\{1,2,\ldots,20\} \ |\  [\mathbf{L}_s]_{m,n}=0\}$ with the probability of $p = 0.025, 0.05, 0.1$. We apply RPCA and $\mathcal{F}$-PCA to extract the target component $\widehat{\mathbf{L}}_s$. $\phi$ is set to the $\ell_1$-ball constraint as $\phi(\mathbf{s}) = \mathcal{B}_1(\mathbf{s}, \epsilon)$ where $\epsilon$ is set as the oracle value, i.e., $\epsilon = \|\widehat{\mathbf{S}}_s\|_1$. The stop criteria in Algorithm \ref{alg:FRPCA} is $\epsilon_{\mathrm{stop}} = 10^{-5}$.

\begin{figure*}[t]
	\centering
		\begin{minipage}{0.16\linewidth}
		\centering
		\scalebox{0.15}{\includegraphics[keepaspectratio=true]{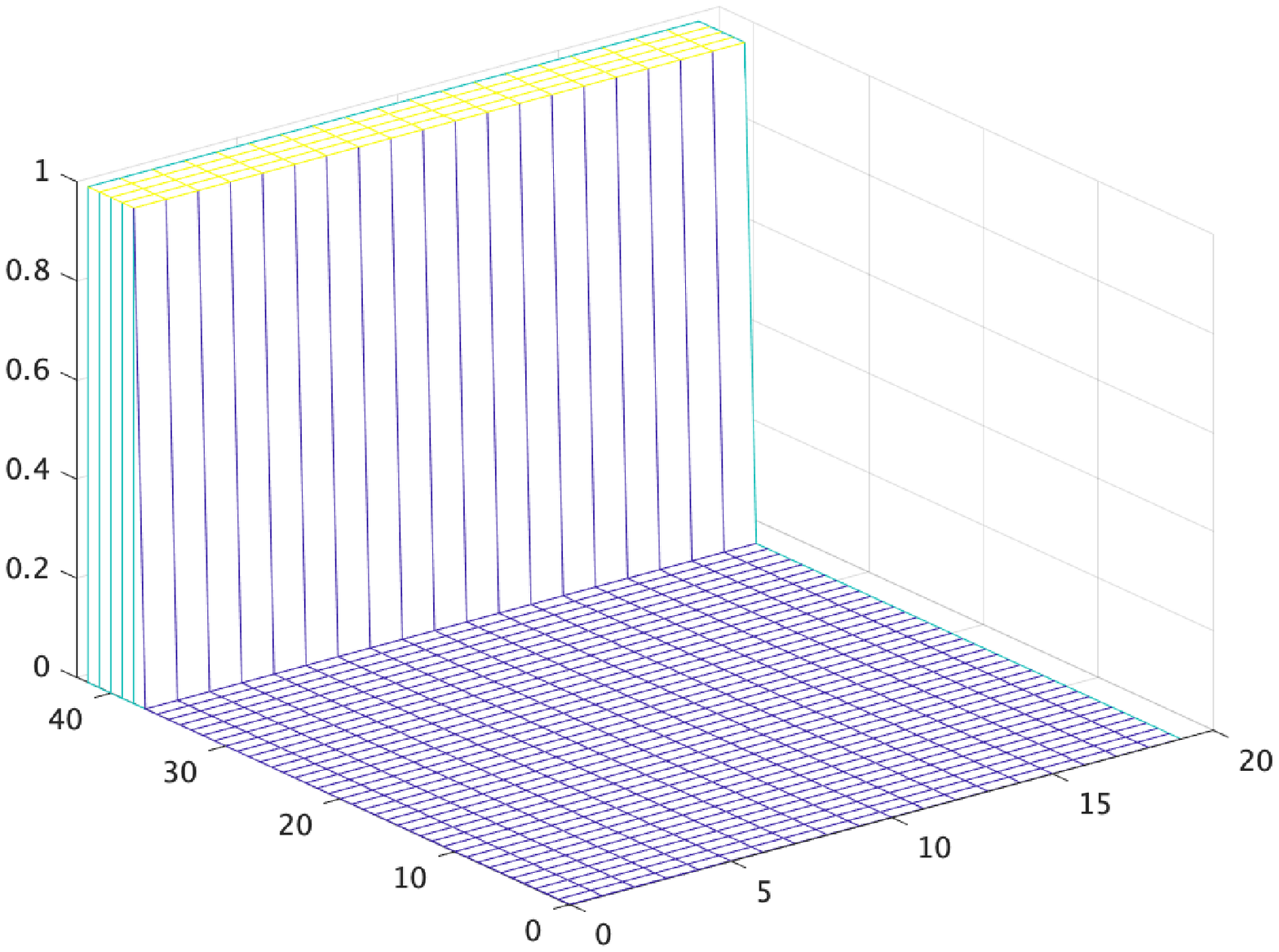}}
				\centerline{(a) Original $\mathbf{J}\widehat{\mathbf{L}}_0$}
	\end{minipage}
			\begin{minipage}{0.16\linewidth}
			\centering
		\scalebox{0.15}{\includegraphics[keepaspectratio=true]{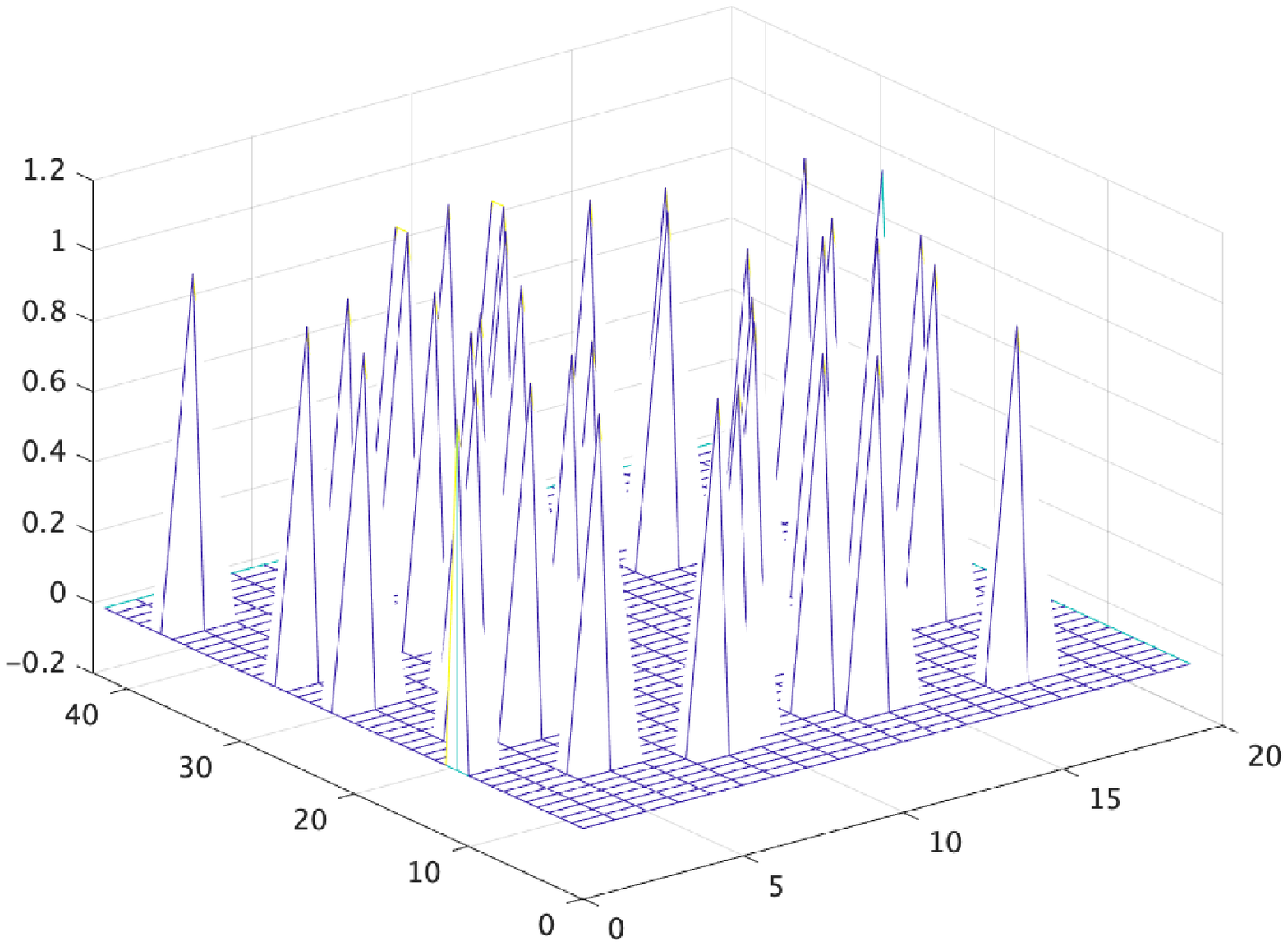}}
				\centerline{(b) Noise $\mathbf{J}\widehat{\mathbf{S}}_0$}
	\end{minipage}
			\begin{minipage}{0.16\linewidth}
		\centering
		\scalebox{0.15}{\includegraphics[keepaspectratio=true]{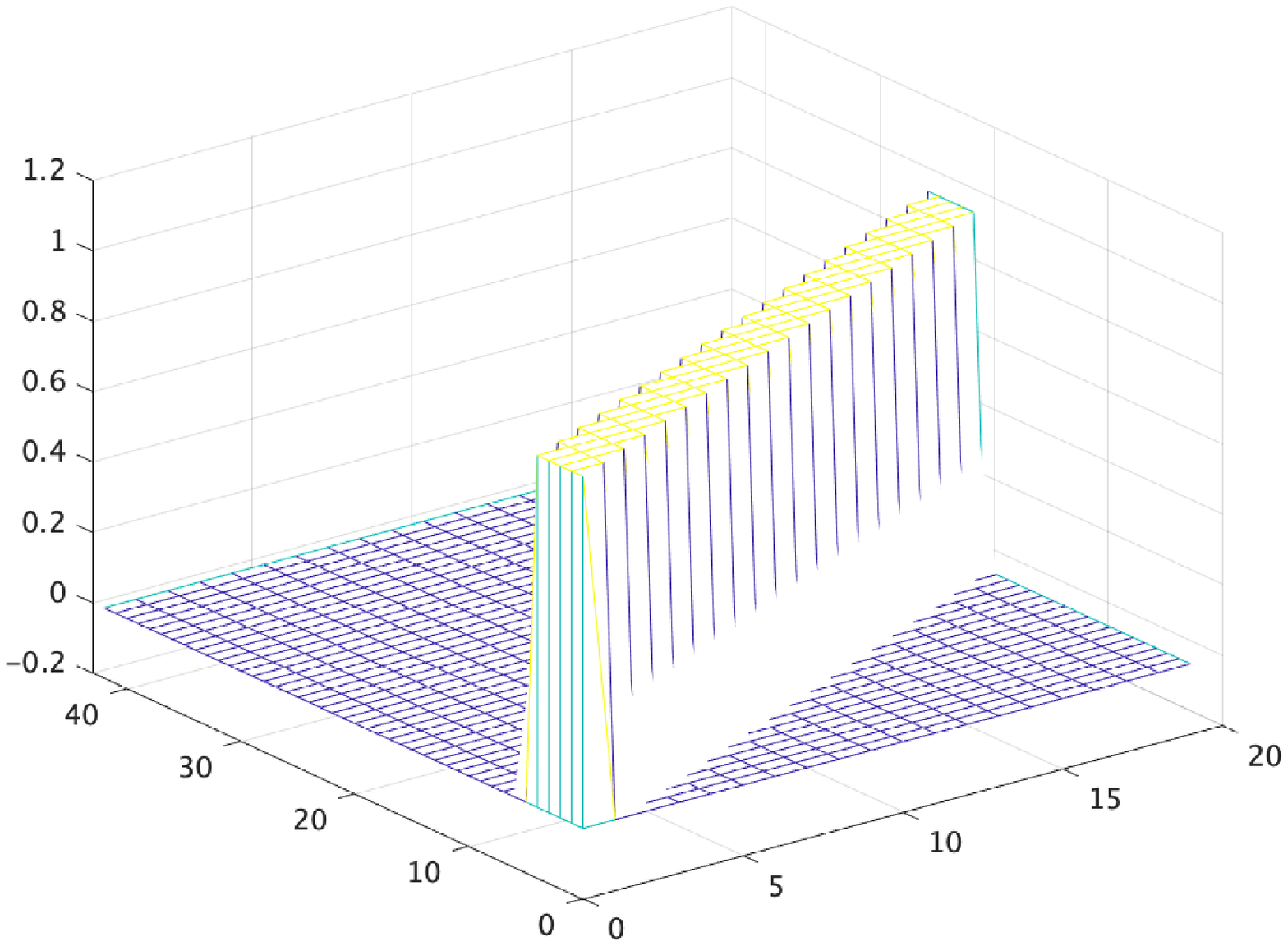}}
				\centerline{(c) Original $\widehat{\mathbf{L}}_1$}
	\end{minipage}
			\begin{minipage}{0.16\linewidth}
			\centering
		\scalebox{0.15}{\includegraphics[keepaspectratio=true]{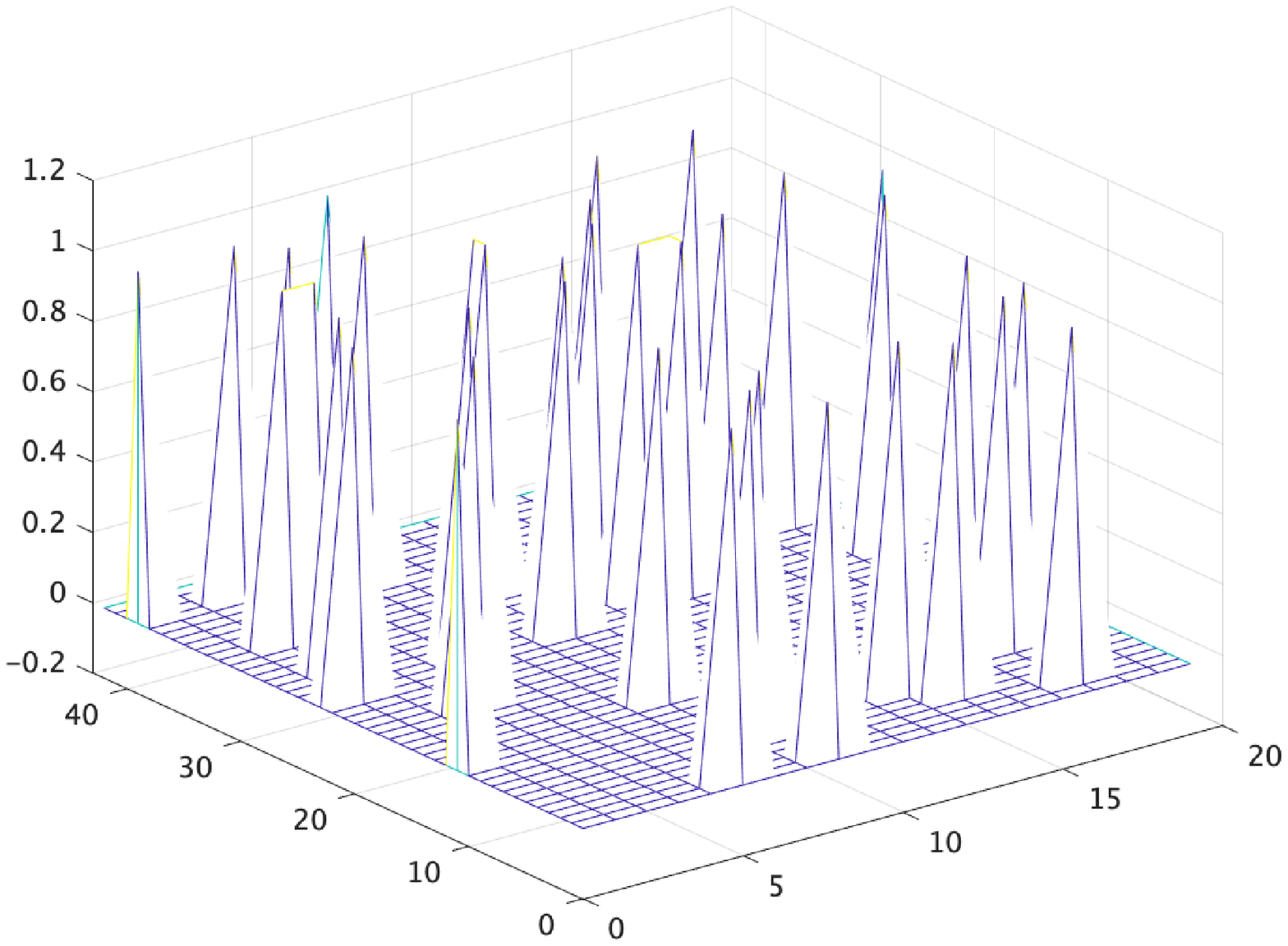}}
				\centerline{(d) Noise $\widehat{\mathbf{S}}_1$}
	\end{minipage}
			\begin{minipage}{0.16\linewidth}
		\centering
		\scalebox{0.15}{\includegraphics[keepaspectratio=true]{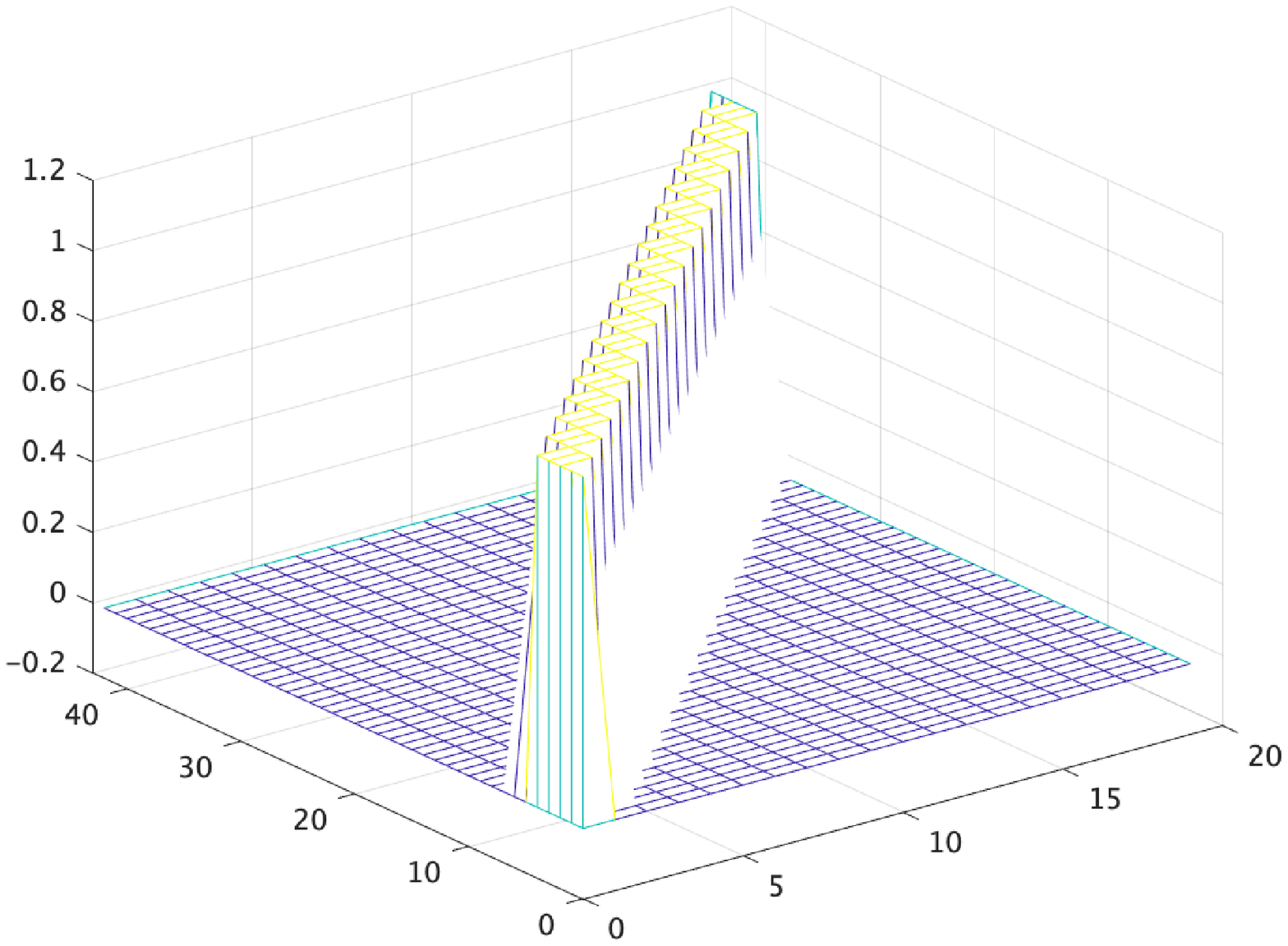}}
				\centerline{(e) Original $\widehat{\mathbf{L}}_2$}
	\end{minipage}
			\begin{minipage}{0.16\linewidth}
			\centering
		\scalebox{0.15}{\includegraphics[keepaspectratio=true]{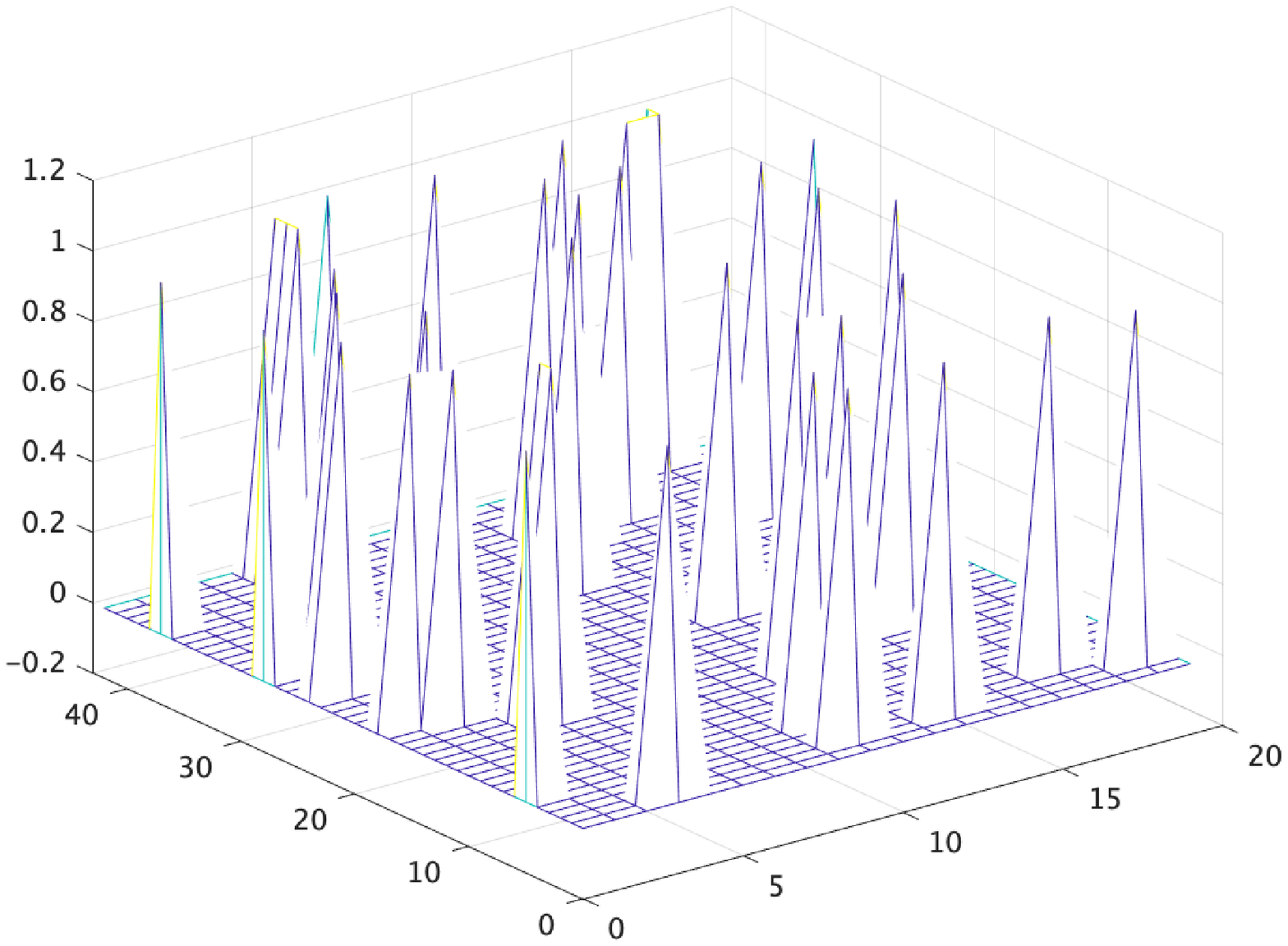}}
				\centerline{(f) Noise $\widehat{\mathbf{S}}_2$}
	\end{minipage}
	\\
		\begin{minipage}{0.16\linewidth}
		\centering
		\scalebox{0.15}{\includegraphics[keepaspectratio=true]{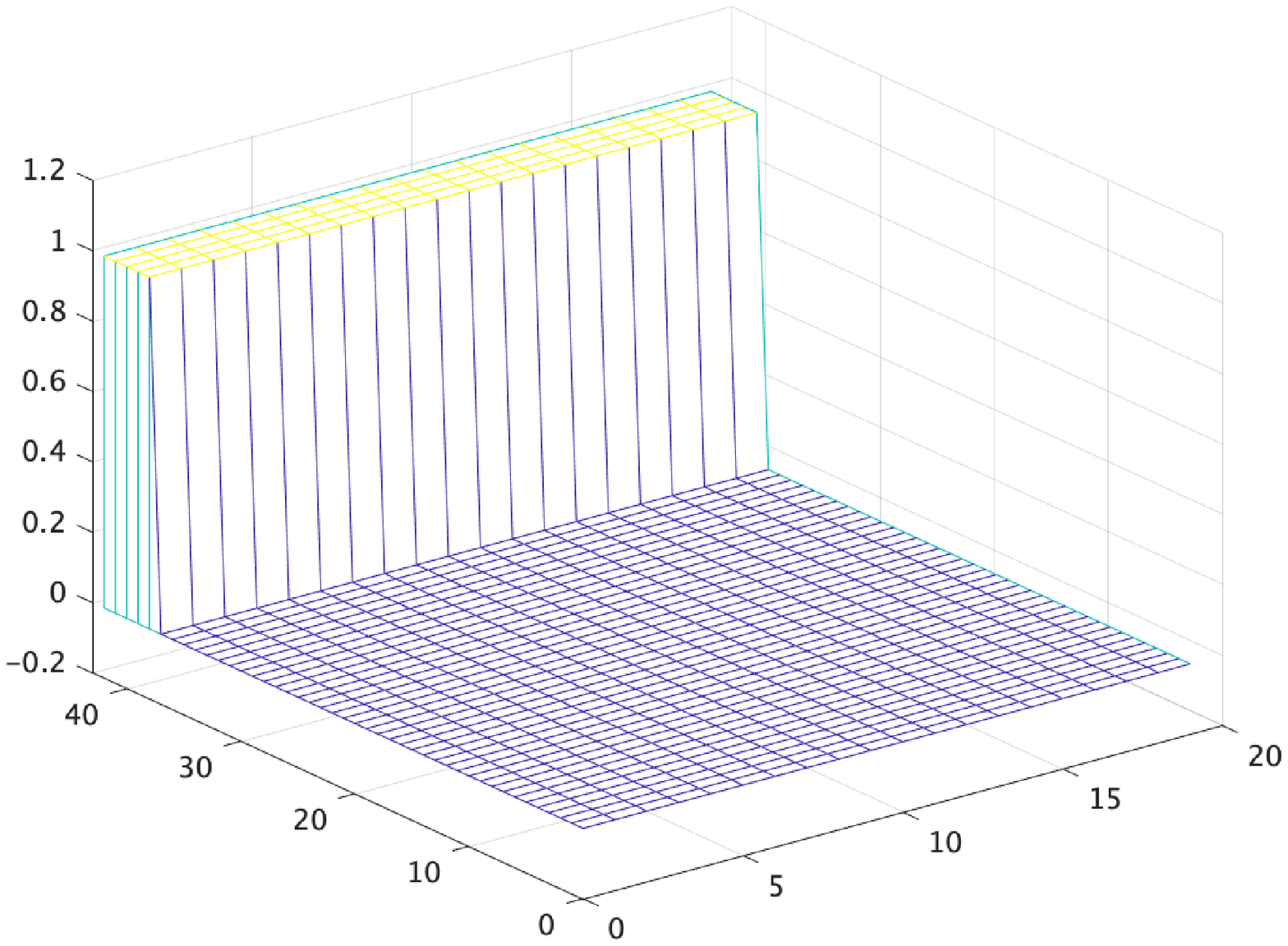}}
				\centerline{(g) RPCA $\mathbf{J}\mathbf{L}_0^\star$}
	\end{minipage}
			\begin{minipage}{0.16\linewidth}
			\centering
		\scalebox{0.15}{\includegraphics[keepaspectratio=true]{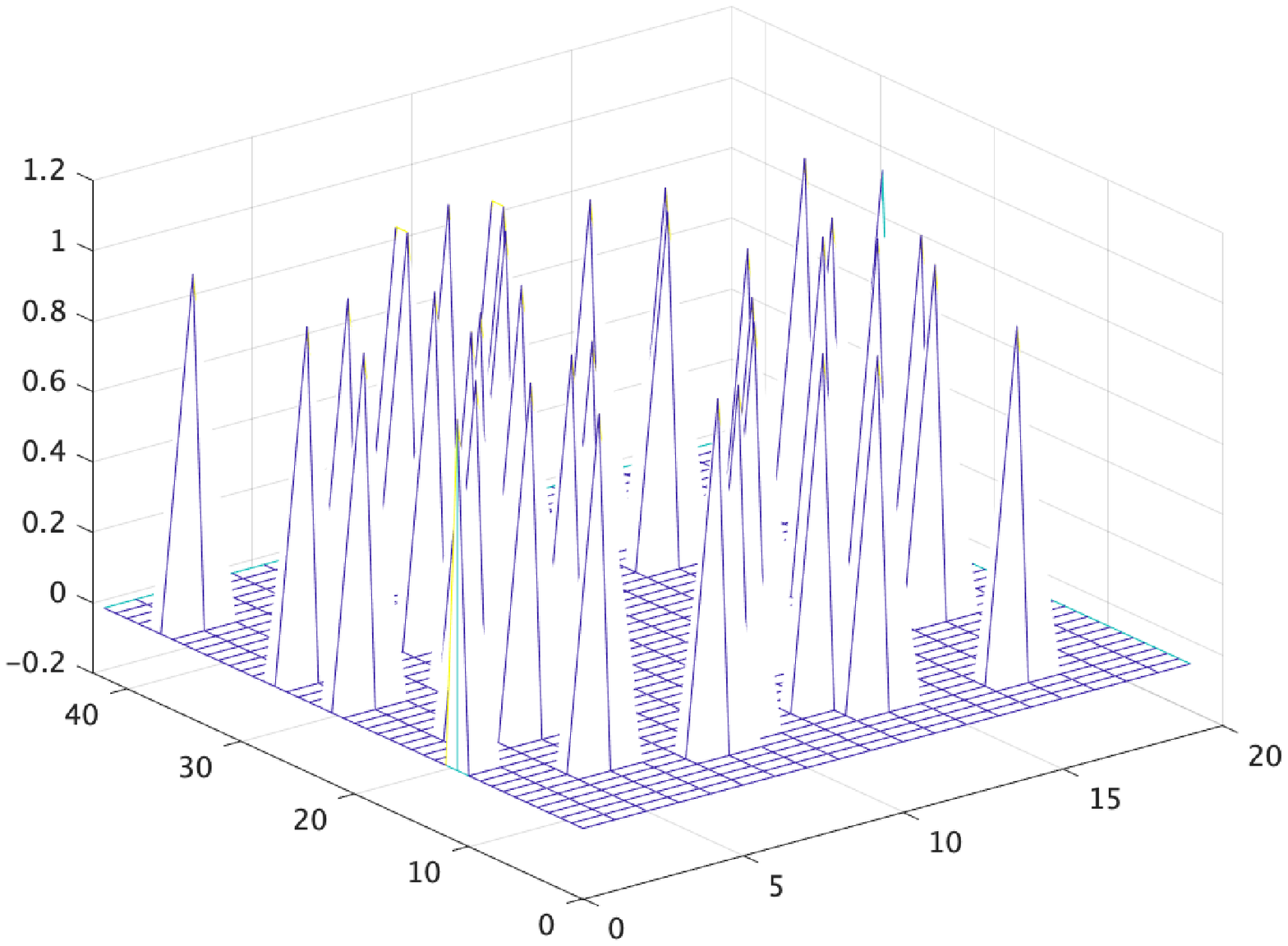}}
				\centerline{(h) RPCA $\mathbf{J}\mathbf{S}_0^\star$}
	\end{minipage}
			\begin{minipage}{0.16\linewidth}
		\centering
		\scalebox{0.15}{\includegraphics[keepaspectratio=true]{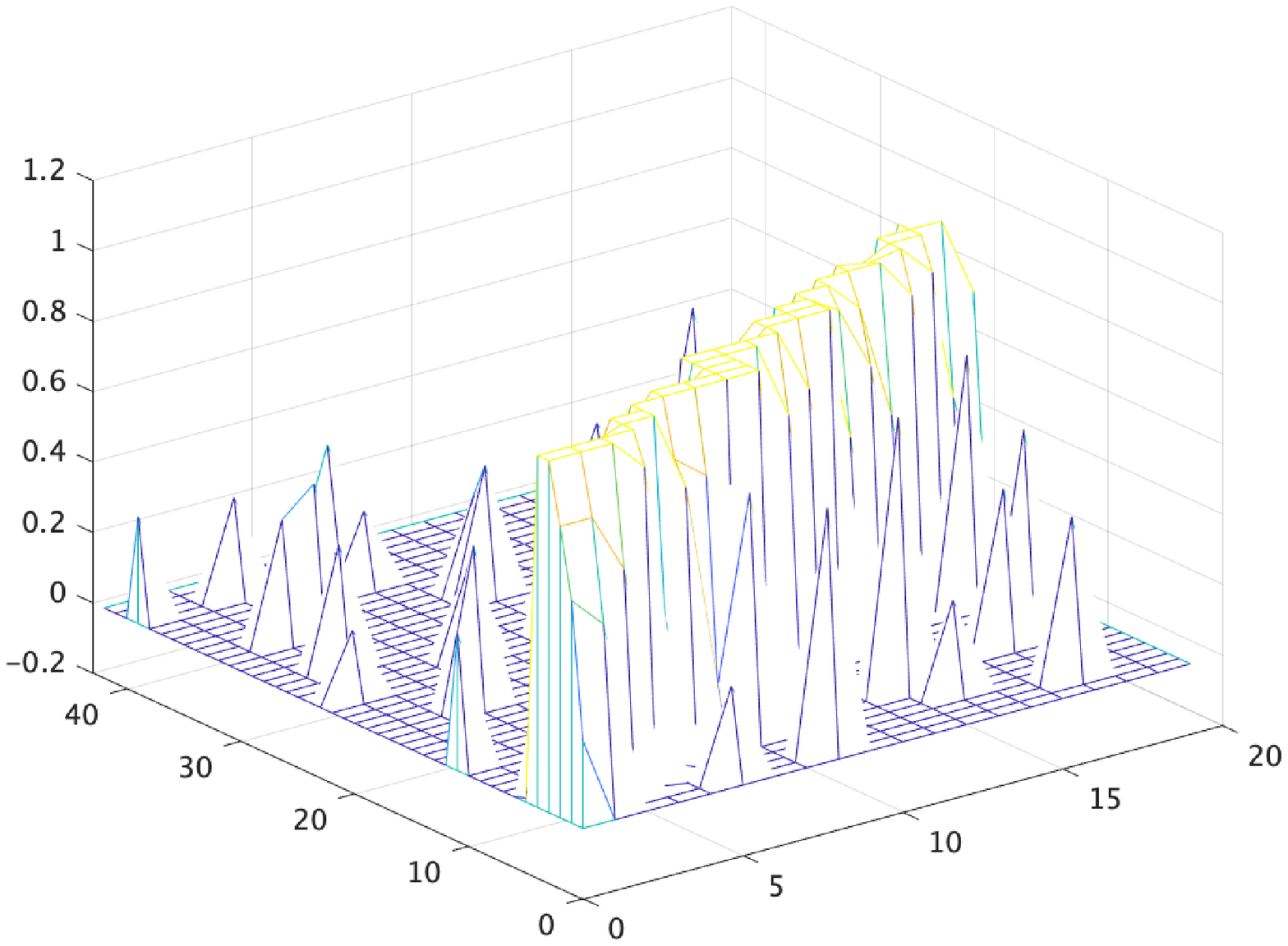}}
				\centerline{(i) RPCA $\mathbf{L}_1^\star$}
	\end{minipage}
			\begin{minipage}{0.16\linewidth}
			\centering
		\scalebox{0.15}{\includegraphics[keepaspectratio=true]{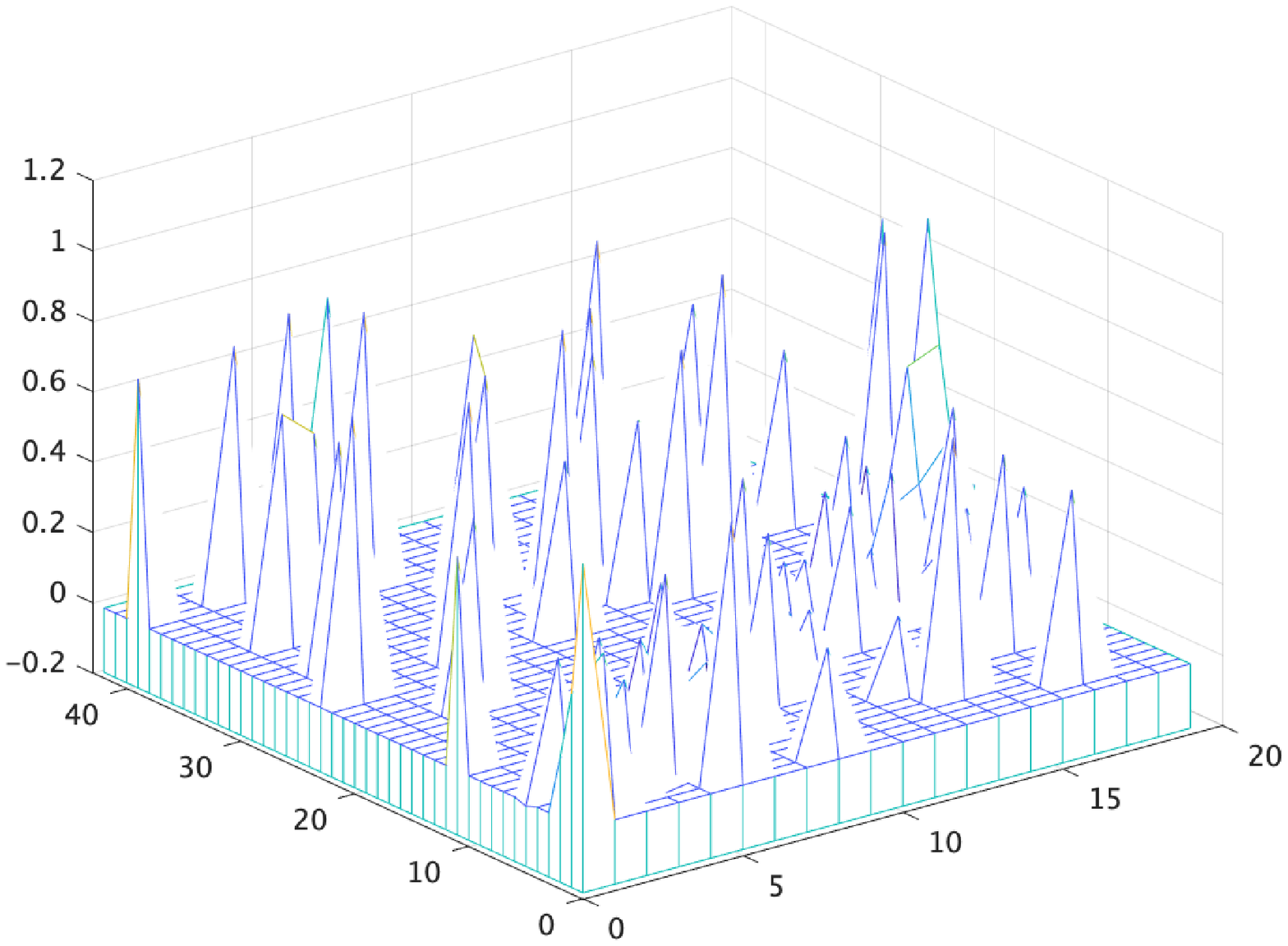}}
				\centerline{(j) RPCA $\mathbf{S}_1^\star$}
	\end{minipage}		
	\begin{minipage}{0.16\linewidth}
		\centering
		\scalebox{0.15}{\includegraphics[keepaspectratio=true]{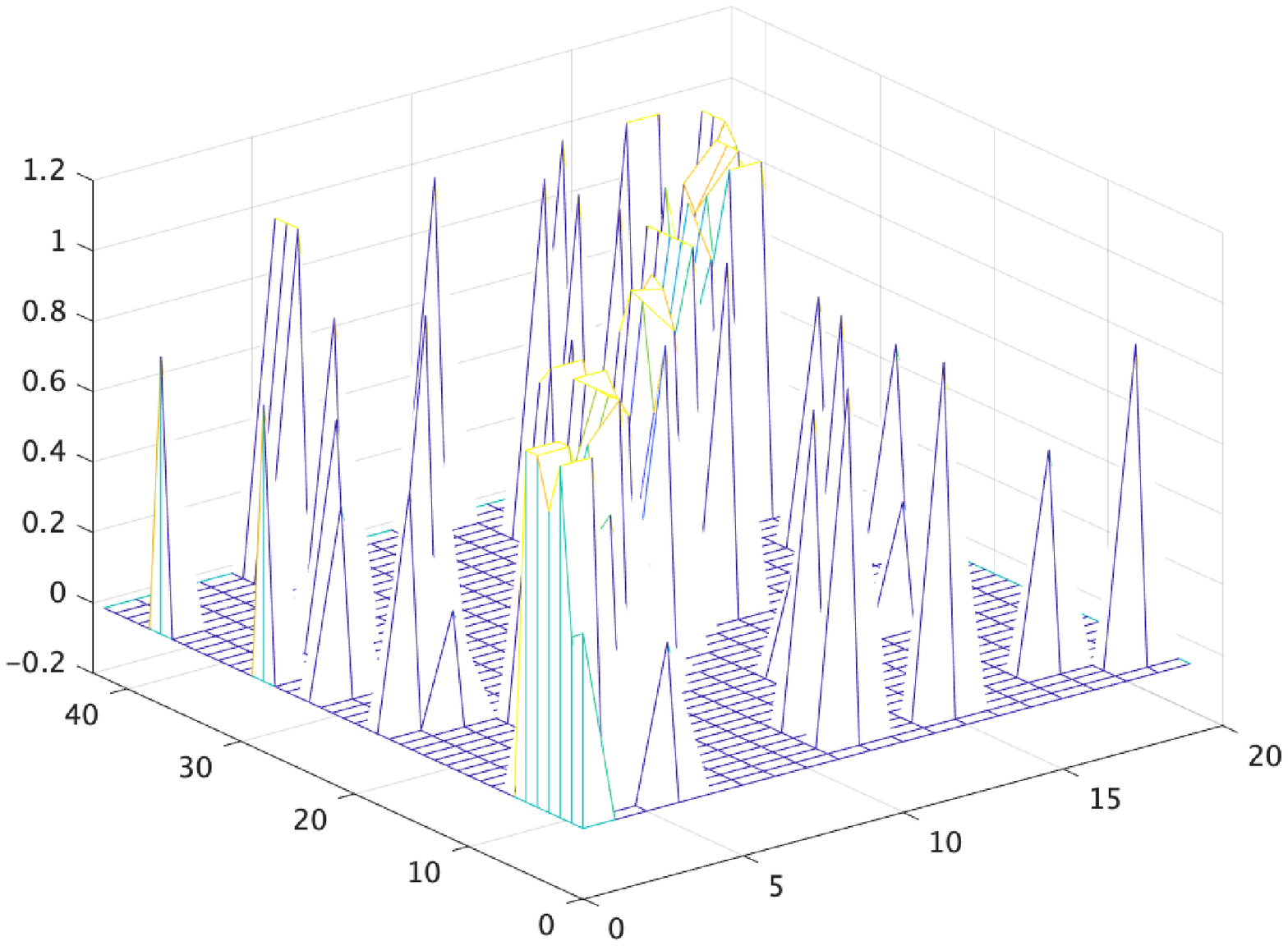}}
				\centerline{(k) RPCA $\mathbf{L}_2^\star$}
	\end{minipage}
			\begin{minipage}{0.16\linewidth}
			\centering
		\scalebox{0.15}{\includegraphics[keepaspectratio=true]{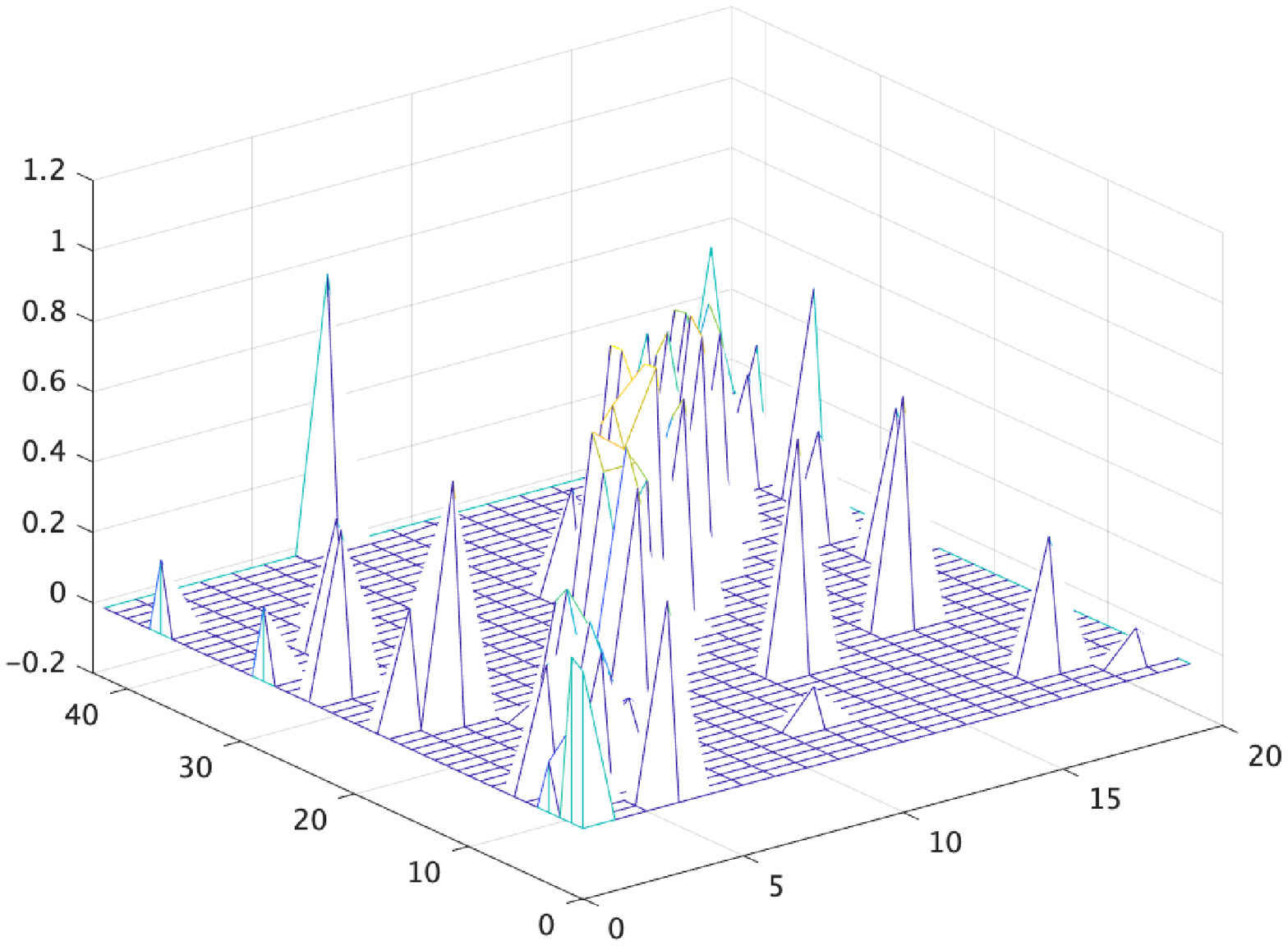}}
				\centerline{(l) RPCA $\mathbf{S}_2^\star$}
	\end{minipage}
	\\
			\begin{minipage}{0.16\linewidth}
		\centering
		\scalebox{0.15}{\includegraphics[keepaspectratio=true]{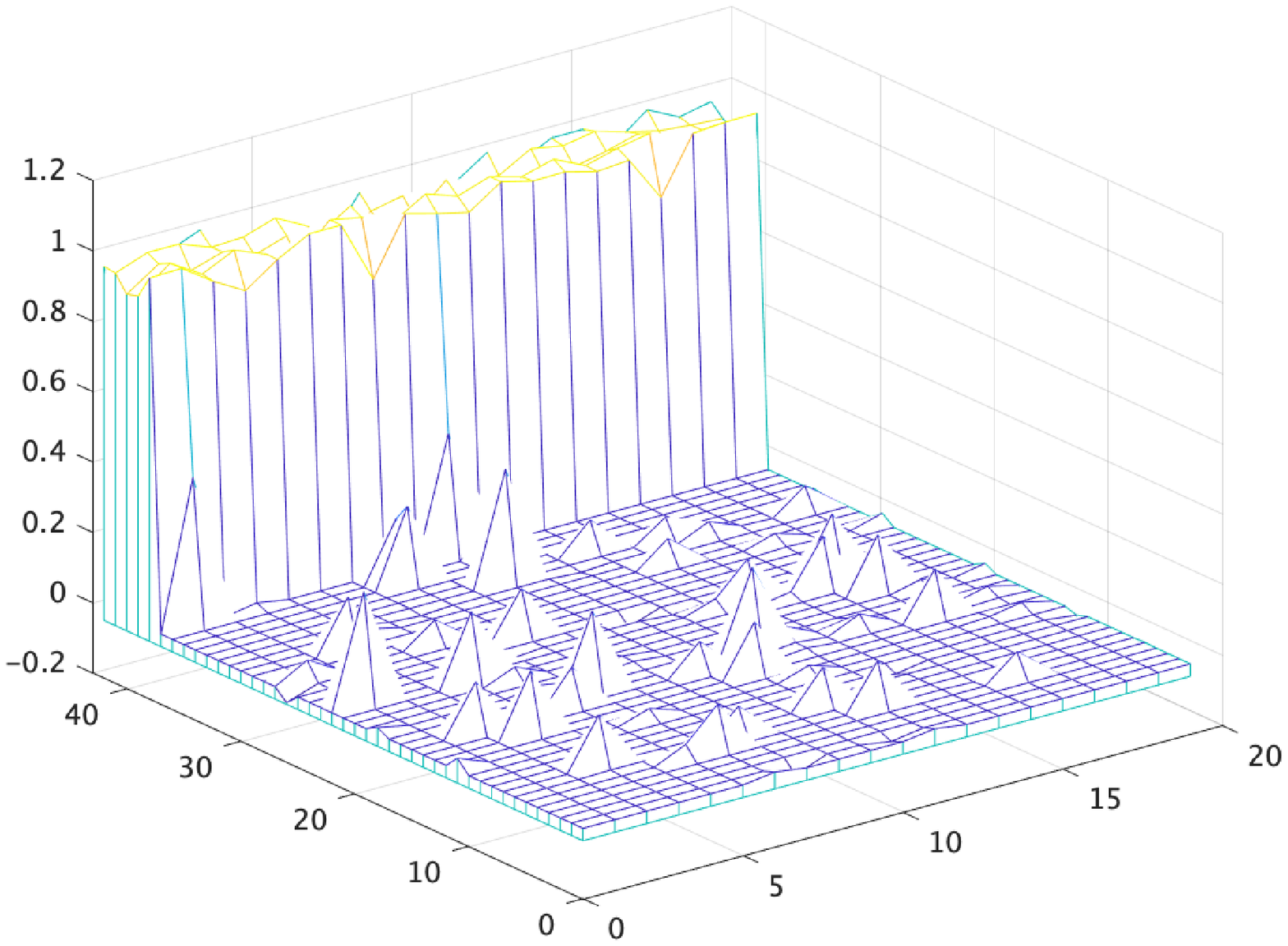}}
				\centerline{(m) $\mathcal{F}-$RPCA $\mathbf{J}\mathbf{L}_0^\star$}
	\end{minipage}
			\begin{minipage}{0.16\linewidth}
			\centering
		\scalebox{0.15}{\includegraphics[keepaspectratio=true]{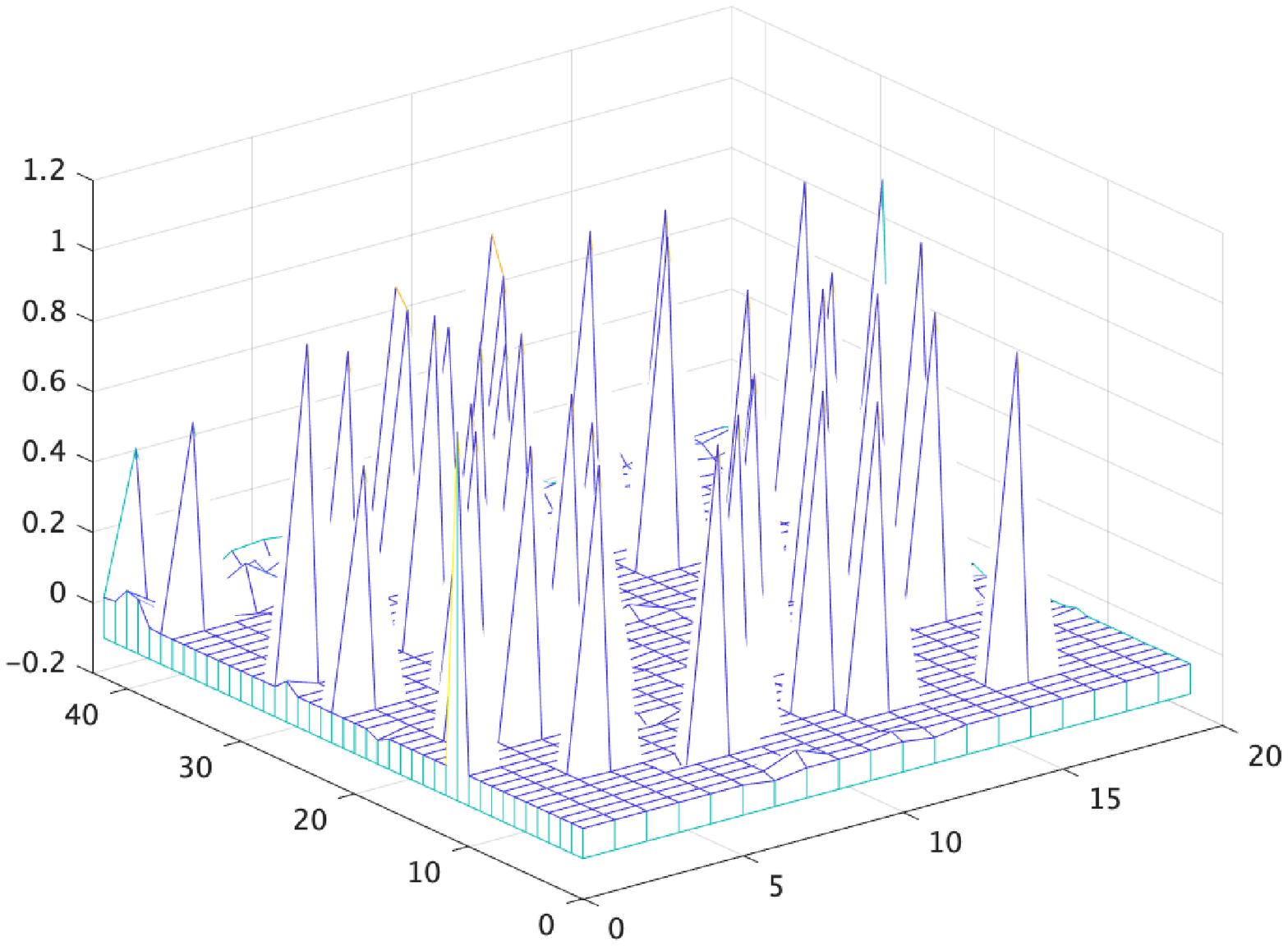}}
				\centerline{(n) $\mathcal{F}-$RPCA $\mathbf{J}\mathbf{S}_0^\star$}
	\end{minipage}
				\begin{minipage}{0.16\linewidth}
		\centering
		\scalebox{0.15}{\includegraphics[keepaspectratio=true]{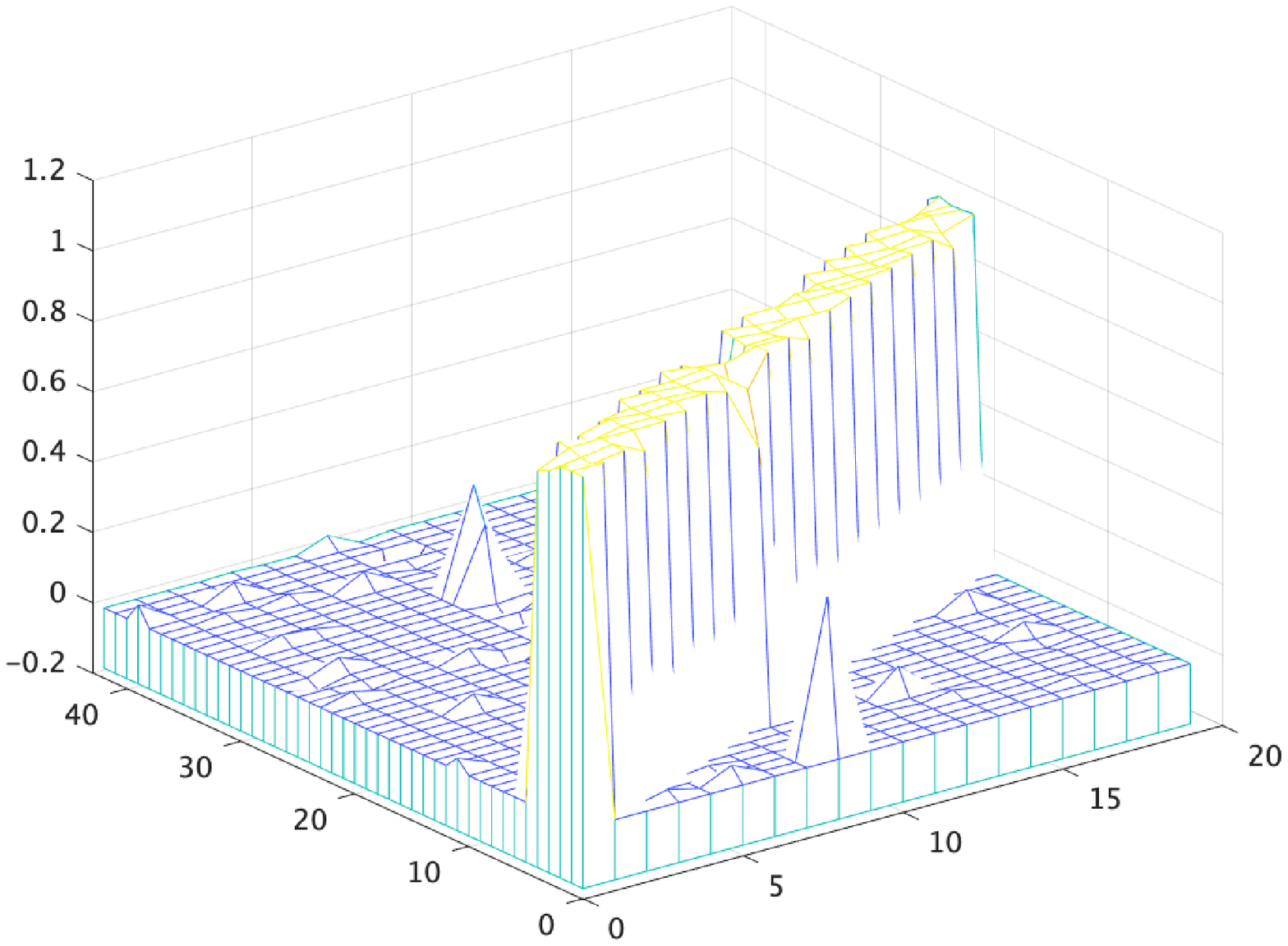}}
				\centerline{(o) $\mathcal{F}-$RPCA $\mathbf{L}_1^\star$}
	\end{minipage}
			\begin{minipage}{0.16\linewidth}
			\centering
		\scalebox{0.15}{\includegraphics[keepaspectratio=true]{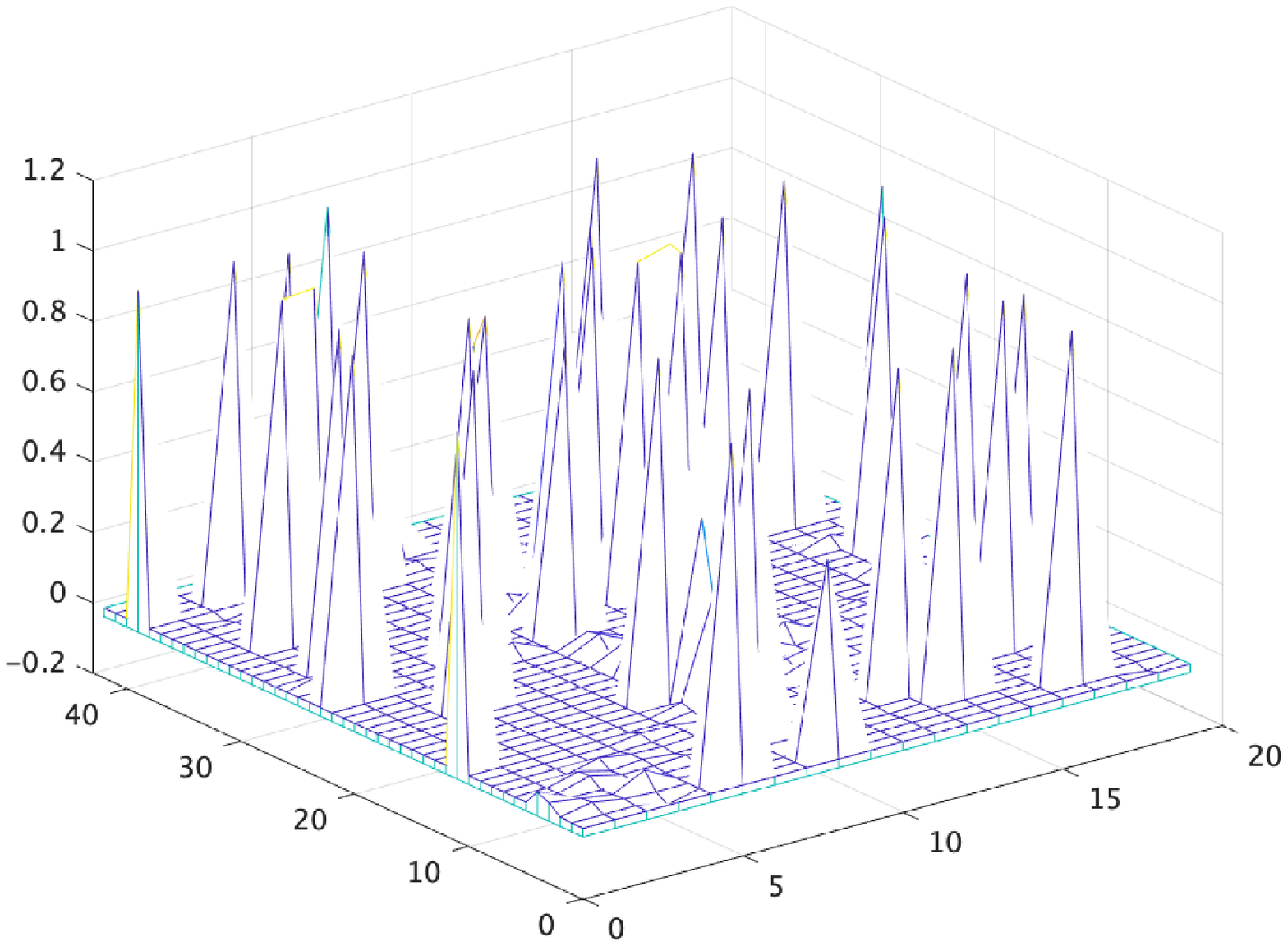}}
				\centerline{(p) $\mathcal{F}-$RPCA $\mathbf{S}_1^\star$}
	\end{minipage}
				\begin{minipage}{0.16\linewidth}
		\centering
		\scalebox{0.15}{\includegraphics[keepaspectratio=true]{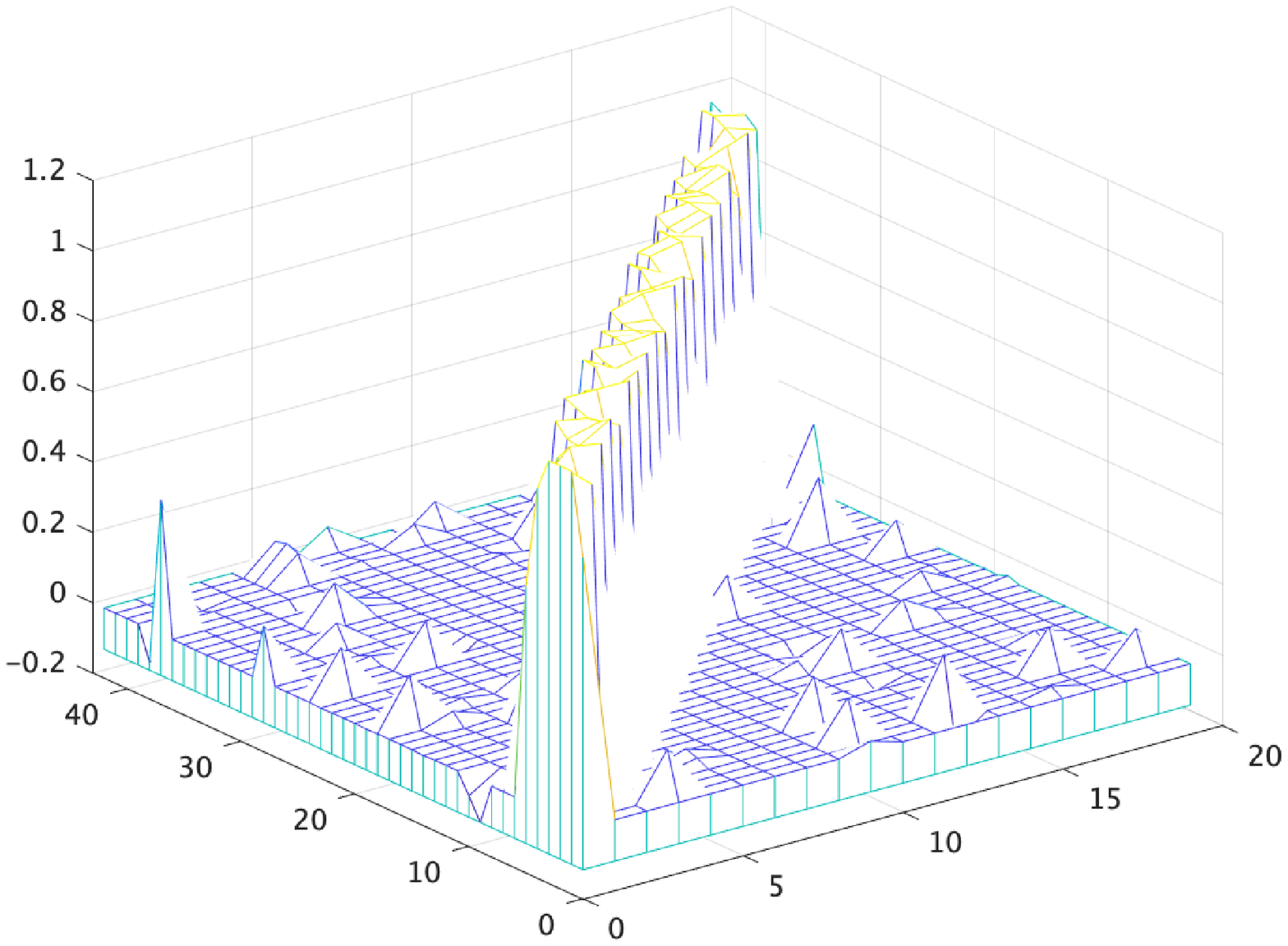}}
				\centerline{(q) $\mathcal{F}-$RPCA $\mathbf{L}_2^\star$}
	\end{minipage}
			\begin{minipage}{0.16\linewidth}
			\centering
		\scalebox{0.15}{\includegraphics[keepaspectratio=true]{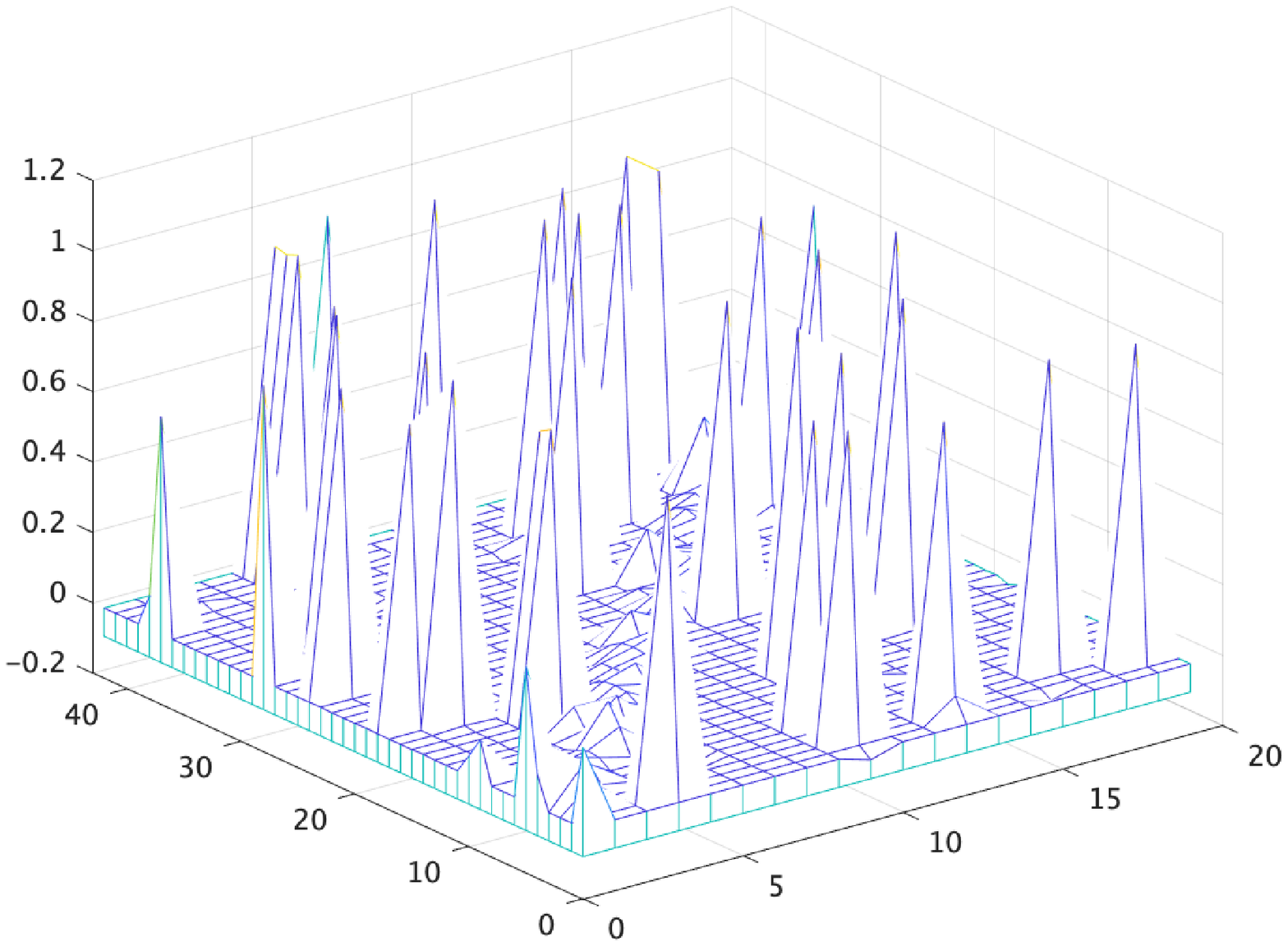}}
				\centerline{(r) $\mathcal{F}-$RPCA $\mathbf{S}_2^\star$}
	\end{minipage}
	\caption{Original and estimated components obtained by RPCA and $\mathcal{F}$-RPCA. Numerical errors (PSNR [dB]) between the original target components $\widehat{\mathbf{L}}_s$ and their estimates $\mathbf{L}_s^\star$ are (Shift:0) RPCA: 102.23, $\mathcal{F}$-PCA: 26.93, (Shift:1) RPCA: 17.9, $\mathcal{F}$-RPCA: 27.57, (Shift:2) RPCA: 12.98, $\mathcal{F}$-RPCA: 27.68.}\label{fig:exFRPCA}
\end{figure*}

Fig. \ref{fig:exFRPCA} show the estimated components ${\mathbf{L}}_s^\star$ and ${\mathbf{S}}_s^\star$ obtained from RPCA and $\mathcal{F}$-RPCA, respectively. Note that, in Fig. \ref{fig:exFRPCA}, the estimates are flipped as $\mathbf{J}\widehat{\mathbf{L}}_0$ and $\mathbf{J}\widehat{\mathbf{S}}_0$ for clearer visualization. In the case of the shift of 0, the target components $\widehat{\mathbf{L}}_0$ are lying in 1-dimensional subspace. In this case, RPCA decomposes the target components and the noise components almost perfectly by minimizing the nuclear norm. On the other hand, the accuracy for decomposition by $\mathcal{F}$-RPCA is worse than RPCA because the ASNN cannot distinguish the similar components with an identical frequency structure. However, in the case of the shift of 1 and 2, $\mathcal{F}$-RPCA can estimate the original component $\widehat{\mathbf{L}}_s$ much better than RPCA due to low-rankness promoting for amplitude spectrum. As well as subjective quality, $\mathcal{F}$-PCA achieved higher PSNR than RPCA as shown in Table \ref{tab:exFRPCA}.

\begin{figure}[t]
\centering
		\begin{minipage}{0.15\linewidth}
		\centering
		\scalebox{1.2}{\includegraphics[keepaspectratio=true]{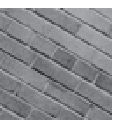}}
				\centerline{(a) $\widehat{\mathbf{L}}$}
		\end{minipage}
		\begin{minipage}{0.15\linewidth}
		\centering
		\scalebox{1.2}{\includegraphics[keepaspectratio=true]{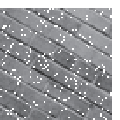}}
				\centerline{(b) $\mathbf{X}$}
		\end{minipage}
		\begin{minipage}{0.15\linewidth}
		\centering
		\scalebox{1.2}{\includegraphics[keepaspectratio=true]{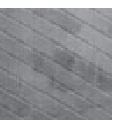}}
		\centerline{(c) $\mathbf{L}^\star$}
		\end{minipage}
		\begin{minipage}{0.15\linewidth}
		\centering
		\scalebox{1.2}{\includegraphics[keepaspectratio=true]{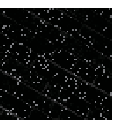}}
		\centerline{(d) $\mathbf{S}^\star$}
		\end{minipage}
		\begin{minipage}{0.15\linewidth}
		\centering
		\scalebox{1.2}{\includegraphics[keepaspectratio=true]{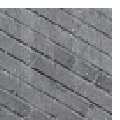}}
				\centerline{(e) $\mathbf{L}^\star$}
		\end{minipage}
		\begin{minipage}{0.15\linewidth}
		\centering
			\scalebox{1.2}{\includegraphics[keepaspectratio=true]{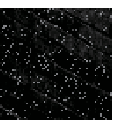}}
		\centerline{(f) $\mathbf{S}^\star$}
		\end{minipage}
\caption{(a) Original image (b) observation, (c) and (d) decomposed components by RPCA, (e) and (f) decomposed components by $\mathcal{F}$-RPCA. Numerical errors (PSNR [dB]) between the original image $\widehat{\mathbf{L}}$ and the estimate $\mathbf{L}^\star$ are RPCA: 24.40 and $\mathcal{F}$-RPCA: 24.21.}
	\label{fig:exprealdata}
	\vspace{-5mm}
\end{figure}

\textcolor{black}{In addition to the experiment with synthetic data shown above, as an example of real-world signal processing, here we conducted another demonstration of RPCA/$\mathcal{F}$-RPCA using an observation $\mathbf{X}$ in Fig. \ref{fig:exprealdata}(a) which was produced from an original image $\widehat{\mathbf{L}} \in \mathbb{R}^{64\times 64}$ (obtained from the \textit{USC-SIPI Image Database}) by adding the sparse noise with the probability of $p = 0.1$. The original image $\widehat{\mathbf{L}}$ was estimated by the component $\mathbf{L}^\star$ in the RPCA/$\mathcal{F}$-RPCA decomposition. The decomposed components and the estimation error are shown in Fig. \ref{fig:exprealdata}. Although the numerical error of $\mathcal{F}$-RPCA is worse than that of RPCA, the edges are estimated more precisely by $\mathcal{F}$-RPCA, whereas RPCA significantly smoothes the edges. Since the columns of the original image can be characterized as (approximately) circularly-shifted versions of each other, the ASNN is more suitable for regularization.}
\begin{table}[t]
\caption{Numerical errors (PSNR [dB]) in target extraction experiments}
	\label{tab:exFRPCA}
	\begin{center}
		\scalebox{1}{
			\begin{tabular}{ccccc}
				\thline
				\multicolumn{1}{c|}{} &\multicolumn{1}{c|}{}          & \multicolumn{1}{c|}{$p=0.025$}        & \multicolumn{1}{c|}{$p=0.05$}  & \multicolumn{1}{c}{$p=0.1$}  \\
				\hline
				\multicolumn{1}{c|}{\multirow{2}{*}{Shift:0}} &\multicolumn{1}{c|}{RPCA} &\multicolumn{1}{c|}{103.57} & \multicolumn{1}{c|}{103.95} & \multicolumn{1}{c}{102.63} \\  
				\multicolumn{1}{c|}{} &\multicolumn{1}{c|}{$\mathcal{F}$-RPCA} & \multicolumn{1}{c|}{42.56} & \multicolumn{1}{c|}{26.68} & \multicolumn{1}{c}{16.47} \\ 
				\thline
				\multicolumn{1}{c|}{\multirow{2}{*}{Shift:1}} &\multicolumn{1}{c|}{RPCA} &\multicolumn{1}{c|}{18.92} & \multicolumn{1}{c|}{16.75} & \multicolumn{1}{c}{13.69}  \\  
				\multicolumn{1}{c|}{} &\multicolumn{1}{c|}{$\mathcal{F}$-RPCA} &\multicolumn{1}{c|}{43.32} &\multicolumn{1}{c|}{27.41} & \multicolumn{1}{c}{16.55} \\ 
				\thline
				\multicolumn{1}{c|}{\multirow{2}{*}{Shift:2}} &\multicolumn{1}{c|}{RPCA} &\multicolumn{1}{c|}{16.19} &\multicolumn{1}{c|}{13.02} & \multicolumn{1}{c}{10.79} \\  
				\multicolumn{1}{c|}{} &\multicolumn{1}{c|}{$\mathcal{F}$-RPCA} &\multicolumn{1}{c|}{45.32} &\multicolumn{1}{c|}{27.39} & \multicolumn{1}{c}{16.54} \\
				\thline
			\end{tabular}
		}
	\end{center}
	\vspace{-0.3cm}
\end{table}

\subsubsection{Promising Applications of ASNN}
Besides the demonstrations shown above, the ASNN for promoting low-rankness of the amplitude spectrum is expected to be effective in many research fields of signal processing.

In speech and audio denoising, enhancement, and separation, many methods \cite{Ozerov2010,Wilson2008} typically attempt to find low-rank components from the amplitude spectrogram obtained by the short-time Fourier transform (STFT) of an input signal, because the targeted amplitude spectrogram of interest often tends to be low-rank. However, since the amplitude spectrogram-based approaches do not pay attention to the phase spectra estimation, the estimated signal is not necessarily proper. In order to find a more reasonable estimate, recent studies have suggested phase-aware approaches by using complex-valued spectrograms \cite{Kameoka2009,Krawczyk2014,Mowlaee2016,Emiya2018,Masuyama2019}. From the viewpoint of low-rank modeling approach, Emiya \textit{et al.} have clarified the fact that the rank of the complex spectrogram is the number of sinusoids forming an input signal \cite{Emiya2018}, \textcolor{black}{then Masuyama \textit{et al.} have proposed an effective way of promoting low-rankness of the complex spectrograms \cite{Masuyama2019} by instantaneous phase correction. Here, we should note that in the method of instantaneous phase correction, the instantaneous frequency should be accurately estimated from the observed signal. Thus, the situation would be severe if the observation suffers from some interference, e.g., a high level of noise. Since the proposed ASNN can evaluate the low-rankness of amplitude spectra directly without prior information, a more reasonable estimate could be obtained robustly (even if the observation suffers from severe degradation) by supporting the phase-aware approaches with the ASNN}. 

Low-rankness-aware approaches also contribute to many image processing tasks, e.g., foreground/background decomposition, illumination and occlusion correction, and reflection removal from multiple images \cite{Wright2009}. In those tasks, image misalignment due to the shift during observation, e.g., handshaking and camera pan, causes severe degradation to accuracy. 

\textcolor{black}{
This problem is recognized by many researchers and tremendous efforts have been paid (a comprehensive survey has been presented in \cite{Bouwmans2018}).
}
\textcolor{black}{
One of the efficient approaches in conventional works is RPCA with deformation-operator. Problem formulations for these approaches are roughly summarized as follows:
\begin{align}
&\argmin_{\mathbf{L},\mathbf{S},\mathcal{D}} \|\mathbf{L}\|_{\ast} + \lambda\|\mathbf{S}\|_1 \ \ \mathrm{s.t.}\ \ \mathcal{D}(\mathbf{X}) = \mathbf{L} + \mathbf{S}, \label{eq:RPCA1}\\
&\argmin_{\mathbf{L},\mathbf{S},\mathcal{D}} \|\mathcal{D}(\mathbf{L})\|_{\ast} + \lambda\|\mathbf{S}\|_1 \ \ \mathrm{s.t.}\ \ \mathbf{X} = \mathbf{L} + \mathbf{S}. \label{eq:RPCA2}
\end{align}
The equation \eqref{eq:RPCA1} finds the best deformation operator on images for searching the isomorphic components (low-rank components) \cite{Peng2012}, while the equation \eqref{eq:RPCA2} the best one on the isomorphic components \cite{Rodriguez2015,Silva2015}. Through the optimization with \eqref{eq:RPCA1} and \eqref{eq:RPCA2}, an observation is decomposed into the isomorphic components and the sparse noise. These approaches tend to suffer from some difficulties, for example, additional computational cost due to iterative updates for finding the best deformation operator parameters, and local minima due to highly non-convex nature of the problems \eqref{eq:RPCA1} and \eqref{eq:RPCA2}, and so on. On the other hand, our approach, i.e., ASNN with ERx, does not require any deformation operator to be optimized and can find a global optimal solution of the relaxed problem.
}
\section{Concluding Remarks}
\label{sec:Conc}
In this paper, we proposed a generic framework of ERx for non-proximable mixed norms. As long as the closed-form proximity operators of a decoupled norm and epigraphical constraints can be computed in closed-form, we can efficiently handle multi-layered non-proximable mixed norms by proximal splitting algorithms. We theoretically proved that if each function, except for the innermost function, was a strictly increasing function, we could guarantee the equivalence of the minimizers between the original problem and the modified problem by ERx. Then, we thoroughly investigated the strictly increasing property for the $\ell_p$ vector/matrix norms ($p \in [1,\infty)$) and the Schatten-$p$ norms ($p = 1, 2$). Moreover, for the $\ell_\infty$-vector/matrix norms and the Schatten-$\infty$ norm, we modified them strictly increasing and presented optimization procedure by using epigraphical splitting. Then, we introduced two non-proximable regularization functions. The first is DSTV. For DSTV, the epigraphical projection of the $\ell_1$-norm required for the DSTV minimization was shown. Finally, the DSTV-based image recovery achieved better performance than the conventional TV and STV regularizers. The second is the ASNN that can robustly evaluate the similarity of the structure among signals in the presence of the misalignment. In the experiments, $\mathcal{F}$-RPCA with the ASNN can extract intrinsic components from an observed matrix with sparse noise, even if the target components are misaligned due to the shift.

\appendices
\section{Epigraphical Projection}\label{app:epi_proj}
Here, several examples of the epigraphical projection $\mathcal{P}_{\mathrm{epi}_{f}}$ are given in the following.
\subsubsection{$\ell_p$-norm ($p =1,2,\infty$)}
\begin{figure}[t]
	\centering
		\begin{minipage}[b]{\linewidth}
		\centering
		\scalebox{0.4}{\includegraphics[keepaspectratio=true]{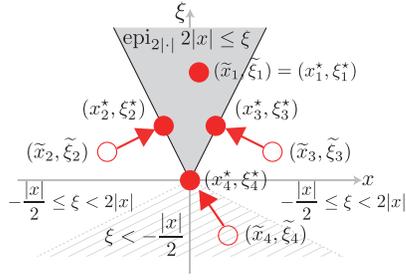}}
					\end{minipage}	
	\caption{\textcolor{black}{Epigraphical projection of the (scaled) $\ell_2$-norm $\tau\|\cdot\|_2:\mathbb{R}^N\rightarrow \mathbb{R}_+$ ($N=1$, $\tau = 2$). The red-open circles $\{(\widetilde{x}_i,\widetilde{\xi}_i)\}_{i=1}^4$ and the red-filled circles $\{({x}^\star_i,{\xi}^\star_i)\}_{i=1}^4$ denote the initial points and the projected points, respectively.} 
	}\label{fig:epi_abs}
\end{figure}

For $^\forall (\mathbf{x},\xi) \in \mathbb{R}^N \times \mathbb{R}$, the projection onto the epigraph of the \textcolor{black}{(scaled) $\ell_2$-norm ($\tau\|\cdot\|_2$, $\tau > 0$)} is expressed as\textcolor{black}{\cite{Chierchia2015}}:
\begin{align}\label{eq:epil2}
\mathcal{P}_{\mathrm{epi}_{\tau\|\cdot\|_2}}(\mathbf{x},\xi)
=&\ \begin{cases}
(\mathbf{x},\xi) & (\textcolor{black}{\tau\|\mathbf{x}\|_2 \leq \xi})\\
(\mathbf{0},0) & (\|\mathbf{x}\|_2 < -\tau\xi)\\
\alpha\left(\mathbf{x},\tau\|\mathbf{x}\|_2\right) & (\mathrm{otherwise})
\end{cases},
\end{align}
where $\alpha = \frac{1}{1+\tau^2}\left(1+\frac{\tau \xi}{\|\mathbf{x}\|_2}\right)$ (an example is illustrated in Fig. \ref{fig:epi_abs}).
The projection onto the epigraph of the $\ell_1$-norm is expressed as \cite{Beck2017}:
\begin{align}\label{eq:epil1}
\mathcal{P}_{\mathrm{epi}_{\|\cdot\|_1}}(\mathbf{x},\xi) =&\ \begin{cases}
(\mathbf{x},\xi) & (\|\mathbf{x}\|_1 \leq \xi) \\
(\mathcal{T}_{\lambda^\star}(\mathbf{x}),\xi+\lambda^\star) & (\|\mathbf{x}\|_1 > \xi) 
\end{cases},
\end{align}
where $\lambda^\star$ is any positive root of the non-increasing function:
\begin{align}\label{eq:phi_lambda}
\varphi(\lambda) =\|\mathcal{T}_{\lambda}(\mathbf{x})\|_1 - \lambda - \xi.
\end{align}
$\mathcal{T}_{\lambda}:\mathbb{R}^{N} \rightarrow \mathbb{R}^{N}$ is the soft-thresholding operator (see Table \ref{tab:complexity}). The projection onto the epigraph of the $\ell_\infty$-norm\textcolor{black}{\cite{Chierchia2015}} is expressed as $\mathcal{P}_{\mathrm{epi}_{\|\cdot\|_\infty}}(\mathbf{x},\xi) = (\mathbf{x}^\star, \mathbf{\xi}^\star)$, where
\begin{align}\label{eq:epiinf}
x_n^\star =&\ \mathrm{sign}(x_n)\min\{|x_n|, \xi^\star \},\nonumber\\
\xi^\star =&\ \max \left\{ \frac{\xi + \sum_{n = \overline{n}}^{N} v_n}{N-\overline{n}+2 },0\right\}.
\end{align}
$\{v_n\}_{n=1}^{N} \subset \mathbb{R}_+ $ is a sequence obtained by sorting the sequence $\{|x_n|\}_{n=1}^{N}$ in ascending order and set $v_{-1}=-\infty$, $v_{N} = \infty$. $\overline{n} \in \{ 1, \ldots, N+1\}$ is the unique integer such that
\begin{align}
v_{\overline{n}-1} < \frac{\xi + \sum_{n = \overline{n}}^{N} v_n}{N-\overline{n}+2 } \leq v_{\overline{n}}.
\end{align}
\subsubsection{Schatten $p$-norm ($S_p$-norm) \cite{Chierchia2013}}
For $^\forall (\mathbf{X},\xi) \in \mathbb{R}^{M\times N} \times \mathbb{R}$ and the singular value decomposition  $\mathbf{X} = \mathbf{U}\mathbf{\Lambda}\mathbf{V}^\top$, $\mathcal{P}_{\mathrm{epi}_{\|\cdot\|_{S_p}}}$ is represented as
\begin{align}\label{eq:Sp}
(\mathbf{U}\mathbf{\Lambda}^\star\mathbf{V}^\top , \xi^\star)
=& \mathcal{P}_{\mathrm{epi}_{\|\cdot\|_{S_p}}}(\mathbf{X},\xi),
\end{align}
where $\mathbf{\bm \Lambda}^\star \in \mathbb{R}^{M\times N}$ is a diagonal matrix. Its diagonal entries ${\bm \lambda}^\star = \begin{bmatrix} 
\lambda_1^\star & \ldots & \lambda_{\min\{M,N\}}^\star 
\end{bmatrix}^\top$ and $\xi^\star$ are obtained by the epigraph projection of the $\ell_p$-norm:
\begin{align} \label{eq:S_p_lambda}
(\mathbf{\bm \lambda}^\star,\xi^\star) = \mathcal{P}_{\mathrm{epi}_{\|\cdot\|_{p}}}(\mathbf{\bm \lambda},\xi),
\end{align}
where $\mathbf{\bm \lambda}$ is the vector consisting of the singular values of $\mathbf{X}$.
\subsubsection{Block-wise epigraph \cite{Beck2017}}
Let $\mathbf{x}_\ell \in \mathbb{R}^{N_\ell}$ ($1\leq \ell \leq L$),   $\mathbf{x} = \begin{bmatrix} \mathbf{x}_1^\top & \cdots & \mathbf{x}_{L}^\top \end{bmatrix}^\top \in \mathbb{R}^{\sum^{L}_{\ell = 1}N_\ell}$,  ${\bm \xi} \in \mathbb{R}^L$, and $\|\cdot\|^{(\ell)}: \mathbb{R}^{N_\ell} \rightarrow \mathbb{R}$. Let $\mathrm{epi}_{\{\|\cdot\|^{(\ell)}\}_{\ell=1}^{L}} = \{( \mathbf{x} , {\bm \xi}) \in \mathbb{R}^{\sum^{L}_{\ell = 1}N_\ell} \times\mathbb{R}^{L} \ |\ (\mathbf{x}_\ell,\xi_\ell) \in \mathrm{epi}_{\|\cdot\|^{(\ell)}}\}$ be a block-wise epigraph. If all the norms $\|\cdot\|^{(\ell)}$ are the same, i.e., $\|\cdot\|^{(\ell)} = \|\cdot\|$, we simply denote as $\mathrm{epi}_{\{\|\cdot\|^{(\ell)}\} } = \mathrm{epi}_{\|\cdot\|}$. The block-wise epigraphical projection $(\mathbf{x}^\star,{\bm \xi}^\star) = \mathcal{P}_{\mathrm{epi}_{\{\|\cdot\|^{(\ell)}\}_{\ell=1}^{L}}}(\mathbf{x},{\bm \xi}) $ is given as 
\begin{align}
(\mathbf{x}^\star_{\ell},\xi^\star_{\ell}) = 
\mathcal{P}_{\mathrm{epi}_{\|\cdot\|^{(\ell)}}}(\mathbf{x}_{\ell},{\xi}_{\ell}).
\end{align}
\section{Epigraphical Splitting}\label{appsec:ES}
Let us consider the following convex optimization problem with an inequality constraint:
\begin{align}\label{eq:probES}
\mathbf{x}^\star \in \argmin_{\mathbf{x} } f(\mathbf{x}) \ \ \mathrm{s.t.} \ \ \sum_{\ell=1}^{L} g_\ell (\mathbf{x_\ell}) \leq \eta,
\end{align}
where $\mathbf{x} = \begin{bmatrix} \mathbf{x}_1^\top & \cdots & \mathbf{x}_{L}^\top \end{bmatrix}^\top  \in \mathbb{R}^{\sum_{\ell = 1}^{L} N_\ell }$, $f \in \Gamma_0(\mathbb{R}^{\sum_{\ell = 1}^{L} N_\ell})$, and $g_\ell \in \Gamma_0(\mathbb{R}^{N_\ell})$ ($g_\ell \neq 0$, $1 \leq {}^\exists \ell \leq L$). Epigraphical splitting \cite{Chierchia2014SVM, Ono2019, Chierchia2015, Ono2014, Chierchia2013}  equivalently reformulates the above inequality constraint into multiple epigraphical constraints and a half-plane one by introducing new variable ${\bm \xi} \in \mathbb{R}^{L}$ as follows:
\begin{align}\label{eq:afterES}
\argmin_{\mathbf{x},\  {\bm \xi} } f(\mathbf{x})  \  \mathrm{s.t.}  \ \sum_{\ell=1}^{L} \xi_\ell \leq \eta,\ (\mathbf{x}, {\bm \xi}) \in \mathrm{epi}_{\{ g_\ell  \}_{\ell = 1}^{L}}.
\end{align}
Thus, as long as each epigraphical projection for $\mathrm{epi}_{g_\ell }$ has the closed-form expression, the involved constraint in \eqref{eq:probES} can be handled efficiently by using some proximal splitting algorithm, such as the PDS algorithm \cite{Chambolle2010,Condat2013,Vu2013}.

\section{Proof for Theorem \ref{theo:ER}}\label{ap:poof_ER}
\begin{proof} 
If, for $k=K-1$, $f^{(K-1)}(\mathbf{z}^{\star(K-1)}) \neq \mathbf{z}^{\star(K)}$, then there exist, at least, one index $i_K \in \{1,\ldots,N_K\}$ and $\epsilon_{i_K}  \in (0,1)$ satisfying 
\begin{align}
0 \leq [f^{(K-1)}(\mathbf{z}^{\star(K-1)})]_{i_K} <  \epsilon_{i_K} {z}_{i_K}^{\star(K)} < {z}_{i_K}^{\star(K)}.
\end{align}
Define $\widetilde{\mathbf{z}}^{(K)}$ by replacing the $i_K$-th element of $\mathbf{z}^{\star(K)}$ as
\begin{align}\label{eq:ztild1}
\widetilde{\mathbf{z}}^{(K)} := \begin{bmatrix}  \ldots & z_{i_K-1}^{\star(K)} & \epsilon_{i_K} {z}_{i_K}^{\star(K)} & z_{{i_K}+1}^{\star(K)} & \ldots \end{bmatrix}^\top.
\end{align} 
Then, $f^{(K)}(\widetilde{\mathbf{z}}^{(K)})<f^{(K)}({\mathbf{z}}^{\star(K)})$ $(({\mathbf{z}}^{\star(K-1)},\widetilde{\mathbf{z}}^{(K)})\in \mathrm{epi}_{f^{(K-1)}})$ from the strictly increasing property of $f^{(K)}$. This contradicts that ${\mathbf{z}}^{\star(K)}$ is the minimizer of the problem \eqref{eq:reform4}. 

Next, let us assume that the equalities \eqref{eq:theorem} hold for $k = K-1$, and does not for $k=K-2$, i.e., $f^{(K-1)}(\mathbf{z}^{\star(K-1)}) =  \mathbf{z}^{\star(K)},\ f^{(K-2)}(\mathbf{z}^{\star(K-2)}) \neq  \mathbf{z}^{\star(K-1)}$. Then there exist, at least, one index $i_{K-1} \in \{1,\ldots,N_{K-1}\}$ and $\epsilon_{i_{K-1}}  \in (0,1)$ that satisfy 
\begin{align}
0 \leq [f^{(K-2)}(\mathbf{z}^{\star(K-2)})]_{i_{K-1}} < \epsilon_{i_{K-1}} {z}_{i_{K-1}}^{\star(K-1)} < {z}_{i_{K-1}}^{\star(K-1)}. 
\end{align}
Define $\widetilde{\mathbf{z}}^{(K-1)}$ as $\widetilde{{z}}^{(K-1)}_{i_{K-1}} = \epsilon_{i_{K-1}} {z}_{i_{K-1}}^{\star(K-1)}$ and $\widetilde{{z}}^{(K-1)}_{i} = {z}_{i}^{\star(K-1)}$ ($i\neq i_{K-1}$).
Since $f^{(K-1)}$ is a strictly increasing block-wise vector function, there exist, at least, one index $i_{K} \in \{1,\ldots,N_K\}$ and $\epsilon_{i_{K}}  \in (0,1)$ that satisfy 
\begin{align}
[f^{(K-1)}(\widetilde{\mathbf{z}}^{(K-1)})]_{i_K} < \epsilon_{i_K} {z}_{i_K}^{\star(K)} < {z}_{i_K}^{\star(K)}.
\end{align}
Define $\widetilde{\mathbf{z}}^{(K)}$ as in \eqref{eq:ztild1}, then, by the same discussion, we can derive that
$
f^{(K)}(\widetilde{\mathbf{z}}^{(K)}) < f^{(K)}({\mathbf{z}}^{\star(K)})
$. By repeating the same discussion, we can verify the theorem.\end{proof}
\section{Proof for \textcolor{black}{Proposition} \ref{prop:S2Sinf}}\label{ap:S2Sinf}
\begin{proof} 
It follows trivially for Schatten-2 norm $\|\cdot\|_{S_2}$ since it is equivalent to the Frobenius norm $\|\cdot\|_F$. As for the Schatten-$\infty$ norm  $\|\cdot\|_{S_\infty}$, it is equivalent to the induced norm $\|\cdot\|_{\mathrm{op}}$ based on $\ell_2$ norm as: $\|\mathbf{A}\|_{S_\infty} = \|\mathbf{A}\|_{\mathrm{op}}$ (see Table. \ref{tab:Notations}). 
We first remark that for ${}^\forall \mathbf{A}\in \mathbb{R}_+^{M \times N}$, there exists $\mathbf{x}^{\star} \in \mathbb{R}_+^N$ or $\mathbb{R}_-^N$ ($\mathbb{R}_-$ is the set of non-positive real numbers) achieves the maximum $\mathbf{x}^{\star} \in \argmax_{\|\mathbf{x}\|_2 \leq 1} \|\mathbf{A}\mathbf{x}\|_2$. To prove this we assume the maximizer $\mathbf{x}^{\star}$ includes positive and negative values ${x}^{\star}_{n} \geq 0$ ($n \in \mathcal{N}_\mathrm{p}$) and ${x}^{\star}_{n} < 0$ ($n \in \mathcal{N}_\mathrm{n}$), where $\mathcal{N}_\mathrm{p}\ \cup\  \mathcal{N}_\mathrm{n} = \{1,\ldots,N\}$. Then, from the basic triangle inequality $(a+b)^2 = |a+b|^2 \leq (|a|+|b|)^2$, we derive the above argument as follows:
\begin{align}
\|\mathbf{A}\mathbf{x}^{\star}\|_2^2 =&\ \sum_m \left(\sum_{n\in \mathcal{N}_\mathrm{p}} A_{m,n}x_n^{\star} + \sum_{n\in \mathcal{N}_\mathrm{n}} A_{m,n}x_{n}^{\star}\right)^2\nonumber\\ 
\leq &\ \sum_m \left(\sum_{n\in \mathcal{N}_\mathrm{p}} A_{m,n}x_n^{\star} + \sum_{n\in \mathcal{N}_\mathrm{n}} A_{m,n}(-x_{n}^{\star})\right)^2 \nonumber\\
=&\ \|\mathbf{A}\widetilde{\mathbf{x}}\|_2^2,\\
[\widetilde{\mathbf{x}}]_n =&\
\begin{cases}
x_n^{\star} &  (n\in \mathcal{N}_\mathrm{p}) \\
-x_n^{\star} & (n\in \mathcal{N}_\mathrm{n})
\end{cases},\ \widetilde{\mathbf{x}} \in \mathbb{R}^N_+,\ \|\widetilde{\mathbf{x}}\|_2 \leq 1.
\end{align}
Obviously, if $\widetilde{\mathbf{x}} \in \mathbb{R}^N_+$ is the maximizer, then $-\widetilde{\mathbf{x}} \in \mathbb{R}^N_-$ (the set of the vectors with non-positive entries) is also the maximizer. 

Next, we assume $\mathbf{A},\ \mathbf{B} \in \mathbb{R}^{M\times N}_+$ satisfy $\mathbf{A} \leq \mathbf{B}$ ($A_{m,n} \leq B_{m,n}$) and $\mathbf{x}^{\star} \in \mathbb{R}_+^N$ is such that $\mathbf{x}^{\star} \in \argmax_{\|\mathbf{x}\|_2 \leq 1} \|\mathbf{A}\mathbf{x}\|_2$. From $\mathbf{A}\mathbf{x}^{\star} \leq \mathbf{B}\mathbf{x}^{\star}$ and the non-decreasing property of the $\ell_2$-norm,
\begin{align}
\|\mathbf{A}\|_{S_\infty}^2 =&\ \|\mathbf{A}\mathbf{x}^{\star}\|_2^2 \leq \|\mathbf{B}\mathbf{x}^{\star}\|_2^2 \leq \max_{\|\mathbf{x}\|_2 \leq 1} \|\mathbf{B}\mathbf{x}\|_2^2 = \|\mathbf{B}\|_{S_\infty}^2.
\end{align}
Thus, $\|\cdot\|_{S_\infty}$ is a non-decreasing function. It is, however, not a strictly increasing function as in the following example.
For $\mathbf{A}$ and $\mathbf{B}$ be $
\mathbf{A} = \begin{bmatrix}
1 & 0 \\
0 & 0 
\end{bmatrix}$, $\mathbf{B} = \begin{bmatrix}
1 & 0 \\
0 & 1
\end{bmatrix} 
$, we can easily verify $\mathbf{A} \leq \mathbf{B}$, $A_{1,1} < B_{1,1}$, and $\|\mathbf{A}\|_{S_\infty}=\|\mathbf{B}\|_{S_\infty}$.
\end{proof}
\section{Proof for Proposition \ref{prop:lambda_solution}} \label{ap:lambda_solution}
\begin{proof}
We consider the following two cases. First, we assume that $\lambda^\star$ is greater than the highest absolute value in the input vector, i.e., $\lambda^\star > |x_{\rho(1)}|$. In this case, the solution $\lambda^\star$ should satisfy the following equation:
\begin{align}
\varphi(\lambda^\star) = - \lambda^\star - \textcolor{black}{\xi} = 0\  \Rightarrow \ \lambda^\star = - \textcolor{black}{\xi}\  (>|x_{\rho(1)}|).
\end{align}

Next, let us consider $|x_{\rho(N_0)}| \geq \lambda^\star > |x_{\rho(N_0+1)}|$ ($1 \leq N_0 \leq N$, $|x_{\rho(N+1)}| = 0$). In this case,
$\psi(\lambda^\star) = \sum_{n=1}^{N_0}(|x_{\rho(n)}|-\lambda^\star) - \lambda^\star - \xi = 0
$. Thus, the solution $\lambda^\star$ should satisfy
\begin{align}
|x_{\rho(N_0+1)}| < \lambda^\star = \frac{S_{N_0} - \textcolor{black}{\xi}}{(N_0+1)} \leq |x_{\rho(N_0)}|.
\end{align}
Conversely, we can verify \eqref{eq:lambda_solution} by introducing into $\phi(\lambda)$.
\end{proof}
\section{Vectorial Total Variation Minimization Algorithms With and Without Epigraphical Relaxation} \label{ap:VTVPDS}
Let us consider vectorial TV (VTV) \cite{Bresson2008} $\|\mathbf{x}\|_{\mathrm{VTV}}$ for color images $\mathbf{x} = \begin{bmatrix}
\mathbf{x}_1^\top & \mathbf{x}_2^\top & \mathbf{x}_3^\top
\end{bmatrix}^\top \in \mathbb{R}^{3N}$ which is a mixed norm formulated as 
\begin{align}\label{eq:TVnorm_ori}
\|\mathbf{x}\|_{\mathrm{VTV}} := \sum^{N}_{n=1} \sqrt{\sum^{3}_{c=1}(d_{c,n,1}^2 + d_{c,n,2}^2}),
\end{align}
where $d_{c,n,1}$ and $d_{c,n,2}$ are the $c$-th channel vertical and horizontal differences of the $n$-th pixel, respectively. Then, the definition \eqref{eq:TVnorm_ori} can be equivalently reformulated by using the $\ell_{1,2}$-mixed norm as
\begin{align}
&\ \|\mathbf{x}\|_{\mathrm{VTV}} =\|\mathbf{P}^{(4)}\mathbf{Dx}\|_{\textcolor{black}{2,1}} = f^{(\textcolor{black}{2})} \circ f^{(\textcolor{black}{1})} (\mathbf{PDx}),\nonumber\\
&\ f^{(\textcolor{black}{1})}: =f_{\|\cdot\|_2}:\mathbb{R}^{6N} \rightarrow \mathbb{R}_+^{N}: \nonumber\\
&\quad\begin{bmatrix}
\mathbf{x}_1^\top \  \ldots \  \mathbf{x}_{N}^\top 
\end{bmatrix}^\top  \mapsto \begin{bmatrix}
\|\mathbf{x}_1\|_2 \  \ldots \  \|\mathbf{x}_{N}\|_2 
\end{bmatrix}^\top,\nonumber\\
&\ f^{(\textcolor{black}{2})}: \mathbb{R}^{N} \rightarrow \mathbb{R}_+: \mathbf{x} \mapsto \|\mathbf{x}\|_1,
\end{align}
where $\mathbf{P}^{(4)}$ is the permutation matrix that permutes the vertical and horizontal differences of the R, G, and B channels of each sample consecutively aligned as $\mathbf{P}^{(4)}\mathbf{Dx} = \begin{bmatrix} \mathbf{d}_1^\top & \cdots & \mathbf{d}_{N}^\top \end{bmatrix}^\top$, where $\mathbf{d}_n = \begin{bmatrix} d_{1,n,1},\ d_{1,n,2},\ d_{2,n,1},\ d_{2,n,2},\ d_{3,n,1},\ d_{3,n,2} \end{bmatrix}^\top$. From Proposition \ref{prop:lp}, VTV satisfies the assumption of \eqref{eq:genform}.

Hereafter we show the algorithms for VTV minimization with/without ERx (VTVwERx/VTVwoERx). First, we apply ERx as
\begin{align}\label{eq:VTVepi}
\mathcal{S}_{\mathbf{x}} =&\ \argmin_{\mathbf{x} \in [0,1]^{3N}} \|\mathbf{P}^{(4)}\mathbf{D}\mathbf{x}\|_{\textcolor{black}{2,1}} \ \mathrm{s.t.}\ \mathbf{x} \in \mathcal{B}_2(\mathbf{y}, \epsilon),  \nonumber\\
\xrightarrow{\mathrm{ERx}} {\mathcal{S}}_{\mathbf{x}} \times {\mathcal{S}}_{\mathbf{z}}  
 =&\ \argmin_{\mathbf{x}\in\mathbb{R}^{3N}, \mathbf{z}} \|\mathbf{z}\|_{1}+\iota_{\mathcal{B}_2(\mathbf{y},\epsilon)}(\mathbf{x})\nonumber\\
 & + \iota_{\mathrm{epi}_{\|\cdot\|_2}}(\mathbf{P}^{(4)}\mathbf{D}\mathbf{x},\mathbf{z}) + \iota_{[0,1]^{3N}}(\mathbf{x}),
\end{align}
The algorithm of VTVwERx can be solve by PDS as
\begin{align} 
\mathbf{p} =&\  \begin{bmatrix}
\mathbf{x}^\top & \mathbf{z}^\top 
\end{bmatrix}^\top,\ 	G(\mathbf{p}) = \iota_{[0, 1]^{3N}}(\mathbf{x}), \nonumber\\
	H(\mathbf{q}) =&\  \|\mathbf{q}_1\|_{1} + \iota_{\mathcal{B}_2(\mathbf{y},\epsilon)}(\mathbf{q}_2) + \iota_{\mathrm{epi}_{\|\cdot\|_2}}(\mathbf{q}_3,\mathbf{q}_4),\nonumber\\
	\mathbf{q} =&\  \mathbf{F}\mathbf{p},\  \mathbf{F}=  
	\begin{bmatrix}
		\mathbf{O} & \mathbf{\Phi}^\top & (\mathbf{P}^{(4)}\mathbf{D})^\top & \mathbf{O}\\
		\mathbf{I} & \mathbf{O} & \mathbf{O} & \mathbf{I}\\
	\end{bmatrix}^\top.
\end{align}
The proximity operators of $\iota_{[0, 1]^{3N}}$, $\|\cdot\|_{1}$, and $\iota_{\mathcal{B}_2(\mathbf{y},\epsilon)}$ are in Table \ref{tab:complexity}. For VTVwoERx, we apply PDS as
\begin{align} 
\mathbf{p} =&\  \mathbf{x},\ 	G(\mathbf{p}) = \iota_{[0, 1]^{3N}}(\mathbf{x}), \
	H(\mathbf{q}) =  \|\mathbf{q}_1\|_{\textcolor{black}{2,1}} + \iota_{\mathcal{B}_2(\mathbf{y},\epsilon)}(\mathbf{q}_2),\nonumber\\
	\mathbf{q} =&\  \mathbf{F}\mathbf{p},\  \mathbf{F}=  
	\begin{bmatrix}
		(\mathbf{P}^{(4)}\mathbf{D})^\top&
		\mathbf{\Phi}^\top
	\end{bmatrix}^\top,
\end{align}
where the proximity operator of the $\ell_{\textcolor{black}{2,1}}$-norm is in Table \ref{tab:complexity}.
\bibliographystyle{IEEEtran}
\bibliography{refs}
\end{document}